\documentclass[sn-mathphys,Numbered,iicol]{sn-jnl}
\pdfoutput=1

\usepackage{hyperref}   

\usepackage{graphicx}%
\usepackage{multirow}%
\usepackage{amsmath,amssymb,amsfonts}%
\usepackage{amsthm}%
\usepackage{mathrsfs}%
\usepackage[title]{appendix}%
\usepackage{xcolor}%
\usepackage{textcomp}%
\usepackage{manyfoot}%
\usepackage{booktabs}%
\usepackage{algorithm}%
\usepackage{algorithmicx}%
\usepackage{algpseudocode}%
\usepackage{listings}%
\usepackage{natbib}

\usepackage{tikz,calc}
\usetikzlibrary{shapes,patterns,arrows,positioning,automata}
\usepackage{pgfplots}
\usepackage{units}
\usepackage{ulem}
\usepackage[T1]{fontenc}
\usepackage{enumitem}   

\usepackage{url}

\def\Eq#1{Eq.~(\ref{#1})}
\def\Eqs#1{Eqs.~(\ref{#1})}

\def\app#1{Appendix~\ref{#1}}
\def\Fig#1{Fig.~\ref{#1}}
\def\Sect#1{Section~\ref{#1}}
\def\Sects#1{Sections~\ref{#1}}

\usepackage{bm}
\def\p{\mathbf{p}}
\def\phat{\hat{\mathbf{p}}}

\def\x{\mathbf{x}}

\def\k{\mathbf{k}}

\def\pp{\mathbf{p'}}

\def\kp{\mathbf{k'}}

\def\mf{m^2_q}
\def\mg{m^2_g}
\def\zg{\zeta_g}
\def\zq{\zeta_q}

\usepackage{dcolumn}

\renewcommand{\d}{\text{d}}

\begin{document}


\title[AMY Lorentz invariant parton cascade - the thermal equilibrium case]{AMY Lorentz invariant parton cascade - the thermal equilibrium case}


\author[1]{\fnm{Aleksi} \sur{Kurkela}}\email{aleksi.kurkela@uis.no}

\author[2]{\fnm{Robin} \sur{Törnkvist}}\email{robin.tornkvist@hep.lu.se}

\author[2]{\fnm{Korinna} \sur{Zapp}}\email{korinna.zapp@hep.lu.se}

\affil[1]{\orgdiv{Faculty of Science and Technology}, \orgname{University of Stavanger}, \orgaddress{\street{4036 Stavanger}, \country{Norway}}}

\affil[2]{\orgdiv{Department of Physics
}, \orgname{Lund University}, \orgaddress{\street{Box 118, SE-221 00 Lund}, \country{Sweden}}}


\abstract{We introduce the parton cascade \textsc{Alpaca}, which evolves parton ensembles corresponding to single events according to the effective kinetic theory of QCD at high temperature formulated by Arnold, Moore and Yaffe by explicitly simulating elastic scattering, splitting and merging. By taking the ensemble average over many events the phase space density (as evolved by the Boltzmann equation) is recovered, but the parton cascade can go beyond the evolution of the mean because it can be turned into a complete event generator that produces fully exclusive final states including fluctuations and correlations. The parton cascade does not require the phase space density as input (except for the initial condition at the starting time). Rather, effective masses and temperature, which are functions of time and are defined as integrals over expressions involving the distribution function, are estimated in each event from just the parton ensemble of that event. We validate the framework by showing that ensembles sampled from a thermal distribution stay in thermal equilibrium even after running the simulation for a long time. This is a non-trivial result, because it requires all parts of the simulation to intertwine correctly.}

\maketitle


\section{Introduction}
\label{section:introduction}

According to our current understanding of collisions of heavy nuclei at collider energies the high density in the final state of these collisions leads to strong final state re-scattering that rapidly drives the system towards local thermal equilibrium~\cite{Schlichting:2019abc,Chesler:2015lsa}. At (proper) time $\mathcal{O}(\unit[1]{fm/c})$, the systems enters an extended phase of hydrodynamic evolution, during which it undergoes continued strong longitudinal and moderate transverse expansion. During this phase, it is well described by viscous hydrodynamics~\cite{Song:2011qa, Song:2013qma,Romatschke:2007mq} with a low shear viscosity to entropy density ratio $\eta/s$ and equation of state of a deconfined quark-gluon medium~\cite{ Ratti:2018ksb}. At the pseudo-critical temperature $T_\text{c} \simeq \unit[160]{MeV}$~\cite{HotQCD:2014kol} the partonic medium freezes out into hadrons that continue to interact until the kinetic freeze-out around $\unit[150]{MeV}$. Correlations due the collective expansion of the system are imprinted onto the hadronic final state and give rise to an azimuthal anisotropy (quantified by the flow coefficients $v_n$), a near-side long-range correlation in rapidity known as the ridge, and a hardening of transverse momentum spectra with $\langle p_\perp\rangle$ increasing with hadron mass. The comparatively high temperature at chemical freeze-out also leads to enhanced strangeness production~\cite{STAR:2005gfr,PHENIX:2004vcz,BRAHMS:2004adc,PHOBOS:2004zne,Armesto:2015ioy,Foka:2016vta}.

The interpretation of these observations as signals of final state collectivity is, however, challenged by the observation of these signatures in proton-nucleus and high-multiplicity proton-proton collisions. Naively, these systems are too small and too dilute to develop a sizeable thermalized medium. Still, the question has to be asked to what extent there is final state re-scattering also in these small systems and whether this could lead to a partial equilibration that is visible in hadron distributions (see~\cite{Nagle:2018nvi,Schlichting:2016sqo} for reviews). 

\smallskip

In the case of heavy-ion collisions, the rapid equilibration to an extent that viscous hydrodynamics becomes applicable can be understood both in terms of strong-coupling dynamics \cite{Chesler:2009cy,Heller:2011ju,Chesler:2010bi,Casalderrey-Solana:2013aba}, as well as in terms of weak-coupling dynamics~\cite{Baier:2000sb,Baier:2002bt,Berges:2013eia}. In the latter case it can be studied using the effective kinetic theory of QCD by Arnold, Moore and Yaffe (AMY)~\cite{Arnold:2002zm,Kurkela:2015qoa,Kurkela:2018vqr}, describing the dynamics of quarks of gluons. Kinetic theory is a natural candidate for the description of small collision systems~\cite{Kurkela:2018qeb,Kurkela:2019kip,Kurkela:2021ctp,Ambrus:2021fej}, since it is valid in- and out-of-equilibrium. 

\medskip

Progress in understanding similarities and differences between small and large collision systems will depend critically on detailed apples-to-apples comparisons between theory and experimental data. This is particularly important for the small systems, where the interpretation of data is complicated by biases and auto-correlations. Monte Carlo (MC) event generators are tools that can bridge the gap between theory and experiments and are indispensable for a thorough understanding of collider data in all areas. Compared to the perturbative regime of proton-proton physics, where sophisticated event generators built on the Standard Model (and beyond) are available, event generators for soft physics and heavy-ion physics are far less advanced. It is our aim to fill in this gap by developing a MC event generator representation of the AMY effective kinetic theory that is applicable to small and large collision systems with a focus on soft particle production. \textsc{Alpaca} (AMY Lorentz invariant PArton CAscade) is implemented as a module in the \textsc{Sherpa}~\cite{Sherpa:2019gpd} event generator. The main difference to existing parton cascade codes~\cite{Geiger:1991nj, Geiger:1998fq,Zhang:1997ej,Molnar:2000jh,Bass:2002fh,Xu:2007aa,Borchers:2000wf,Shin:2002fg,Ma:2006fm,Zhang:2019utb} is that it reproduces exactly the AMY kernels for all elastic scattering, splitting and merging processes. Furthermore, it is implemented in a way that is Lorentz-invariant by construction. This avoids problems with residual causality violation~\cite{Molnar:2019yam} and altered correlations and fluctuations  due to particle subdivision~\cite{Lin:2021mdn}. In this paper we focus on the implementation of the effective kinetic theory in a parton cascade code and show that it reproduces AMY in thermal equilibrium. A major challenge in this is to extract the screening masses and effective temperature $T_*$, which are non-local in phase-space, from the parton ensemble.For the validation of the thermal equilibrium we initialise the system in thermal equilibrium and convince ourselves that the distribution function does not change. This will only be the case when detailed balance is respected. In particular, splitting and merging rates have to be the same, which requires a non-trivial interplay between elastic and inelastic processes. The scattering rates depend on the screening masses and the splitting and merging rates on the effective temperature, detailed balance is therefore a critical test of our dynamical extraction of these quantities. Running the simulation for a long time in thermal equilibrium is an important validation exercise showing that the dynamics are correctly encoded. The next step is then to study how a system that is initialised out of equilibrium approaches equilibrium. Since this is a different question we will discuss it in a separate publication.

\smallskip

The paper is organised as follows: in \Sect{section:lorentzinvariance} we briefly review the method used to make the parton cascade Lorentz invariant and \Sect{section:amy} gives a summary of the AMY effective kinetic theory. Details on the implementation in \textsc{Alpaca} are given in \Sect{section:alpaca}, with an introduction to the framework in \ref{subsection:alpaca_timeevolution}, a discussion of the set-up used for the thermal equilibrium simulations in \ref{subsection:alpaca_box}, details on elastic scattering in \ref{subsection:alpaca_elastic}, splitting and merging in \ref{subsection:alpaca_inelastic} and lastly the results of a long run with all the different components included is shown in \ref{subsection:full_run}. We conclude the discussion in \Sect{section:conclusions} and give further technical details in the appendices.

\section{A Lorentz invariant parton cascade}
\label{section:lorentzinvariance}

The no-interaction theorem~\cite{Currie:1963rw} states that $N$ particles moving
in a $6N$ dimensional phase space cannot interact if Lorentz invariance
(instead of Galilei invariance) is required. This theorem can, however, be
circumvented by regarding the particles as moving in a $8N$ dimensional phase
space~\cite{Peter:1994yq}, i.e. formulating the theory in terms of four-positions and four-momenta of the particles. With a physical choice of the quasi-potential
describing the interactions the particles are classically off-shell (i.e.\
$p_i^2 \neq m_i^2$) when they are within the range of the quasi-potential and
on-shell otherwise. The four-vectors can be parameterised by an evolution parameter $\tau$, that is a Lorentz scalar. The equations of motion are then given by
\begin{align}
 \frac{d}{d \tau} p_{i\mu}(\tau) & = \{H,p_{i\mu}\} = - \frac{\partial H}{\partial x_i^\mu}, \\
 \frac{d}{d \tau} x_{i\mu}(\tau) & = \{H,x_{i\mu}\} = + \frac{\partial H}{\partial p_i^\mu}\,,
\end{align}
where $H$ is the classical Hamiltonian.

\smallskip

We here follow the approach of~\cite{Borchers:2000wf} and use a simplified Hamiltonian, in which there are only binary interactions of particles taking place at discrete points in $\tau$\footnote{A similar way of formulating a Lorentz invariant parton cascade has recently been proposed in~\cite{Nara:2023vrq}.}. The interactions thus lead to instantaneous (in $\tau$) changes of the momenta and between interactions the particles travel on free trajectories. We thus require only the free Hamiltonian, which for a system of $N$ particles is the sum of $N$ single particle free Hamiltonians. In \textsc{Alpaca} we neglect the (parametrically small) effective masses and treat all partons as massless. The Hamiltonian is then given by
\begin{equation}
 H = \sum_{i=1}^N \lambda p_i^2 \,,
\end{equation}
where $\lambda$ is an arbitrary Lagrange multiplier \cite{Chagas-Filho:2006cps}. The equations of motion are easily solved to yield
\begin{align}
 p_{i\mu}(\tau) & = \text{const}, \\
 x_{i\mu}(\tau) & = 2 \lambda p_{i\mu} (\tau - \tau_{0i}) + x_{i\mu}(\tau_{0i}) \,.
    \label{eq:x_freestream}
\end{align}
The parameter $\tau$ does not have a physical interpretation. One can, however,
chose an arbitrary but fixed reference frame and calculate the particles'
momentum and position in this frame. Using $x_{i\mu} = (ct, \mathbf{r}_i)$ and $p_{i\mu}
= (\omega_i/c, \mathbf{k}_i)$ with $|\mathbf{k}_i| = \omega_i/c$ one finds
\begin{equation}
 \frac{\d \mathbf{r}_i}{\d t} = \frac{\d \mathbf{r}_i}{\d \tau} \frac{\d \tau}{\d t}
 = 2 \lambda \mathbf{k}_i \frac{c^2}{2 \lambda \omega_i}
 = \frac{c^2 \mathbf{k}_i}{\omega_i} \,,
\end{equation}
where
\begin{equation}
 \tau = \frac{c^2 (t - t(\tau_{0i}))}{2 \lambda \omega_i} + \tau_{0i}
 \label{eq:tauoft}
\end{equation}
was used. This is, of course, the expected result (and independent of $\lambda$).

\smallskip

In order to find the $\{\tau_{0i}\}$ at which the initial conditions
$\{x_i(\tau_{0i})\}$ are specified, the parameter $\tau$ has to be related to
some observer time. This can be done by generalising the procedure introduced in~\cite{Peter:1994yq} for a 2-particle system to $N$ particles\footnote{This is possible since it only requires that the motion of the centre-of-momentum decouples from the rest, which is satisfied for any number of particles.}. The position of the centre-of-momentum $X_\mu$ in the centre-of-momentum frame satisfies
\begin{equation}
 X_\mu^\text{c.m.}(\tau) = \lambda P_\mu^\text{c.m.} \tau
 \qquad \text{with} \qquad
 P_\mu = \sum_{i=1}^N p_{i\mu} \,,
\label{Eq::xincms}
\end{equation}
from which it follows
\begin{equation}
 t^\text{c.m.}(\tau) = \lambda E^\text{c.m.} \tau \,,
 \label{Eq::relttau}
\end{equation}
where, without loss of generality, $\tau_0 = t^\text{c.m.}(\tau_0)=0$ was set.

\medskip

In order to decide whether a given pair of particles interacts, one needs to measure their distance. Following~\cite{Borchers:2000wf}, the Lorentz invariant distance squared between two particles is defined as
\begin{align}
    \label{eq:LIdistance}
 d_{ij}^2 & = - \left( x_\mu - \frac{x_\nu p^\nu}{p^2}p_\mu \right)  
              \left( x^\mu - \frac{x_\nu p^\nu}{p^2}p^\mu \right) \nonumber \\
          & = - \left( x^2 - \frac{(x_\mu p^\mu)^2}{p^2} \right) \,, 
\end{align}
where $x = x_i-x_j$ is the relative four-distance between the particles, and $p =
p_i+p_j$ is the total four-momentum of the pair. In the centre-of-momentum frame of the pair, this reduces to the squared three-distance. When the distance becomes
smaller than $\sqrt{\sigma_\text{scat}/\pi}$, where $\sigma_\text{scat}$ is the scattering cross section, an interaction occurs\footnote{This corresponds to the black disk approximation of the cross section. The generalisation to more sophisticated models is straightforward. We have tested a Gaussian shape for the scattering probability, but did not observe significant differences.}. The interactions are ordered in $\tau$ in a frame-independent way. This avoids the problem of causality violation in parton cascades that appears when ordering interactions in observer time, in which case the ordering is frame dependent.%

In general, when a pair of particles interacts, the condition $d_{ij} < \sqrt{\sigma_\text{scat}/\pi}$ is satisfied during a finite $\tau$ interval and one has to make a choice at which point in $\tau$ the interaction takes place. We here choose the point of closest approach between the particles, i.e. the $\tau$ corresponding to $\min(d_{ij}(\tau))$. The value of $\tau$ for which the closest approach is reached can be calculated from \Eq{eq:LIdistance} and is given by 
\begin{equation}
 \bar \tau = \frac{1}{2} \left(\tau_{0i} + \tau_{0j} \right) + 
\frac{[x_i(\tau_{0i}) - x_j(\tau_{0j})]_\mu [p_i - p_j]^\mu}{4 \lambda p_{i\mu}
p_j^\mu} \,.
\label{Eq::taubar}
\end{equation}
The parton cascade proceeds by moving from one scattering to the next. The $1 \to 2$ splitting processes are easily included by translating the splitting rate in terms of an observer time into a splitting rate in $\tau$ using \Eq{eq:tauoft} (see \Sect{subsection:alpaca_inelastic_splitting}).

\section{The AMY Effective Kinetic Theory}
\label{section:amy}

In~\cite{Arnold:2002zm}, P.~Arnold, G.~Moore, and L.~Yaffe derived an effective kinetic theory of QCD at high temperatures. At sufficiently high $T$, where the QCD coupling $g(T)$ is small, the temperature $T$ sets a hard scale, while $gT$ is a softer scale. It is a starting assumption of the effective-theory setup that these parametrically separated scales are also quantitatively well-separated. Under this assumption the QCD plasma consists of quarks and gluons as well-defined quasi-particles with typical momenta of order $T$. These hard partons propagate as nearly free and nearly massless particles, since thermal masses are $\mathcal{O}(gT)$ and the small-angle scattering rate is $\mathcal{O}(g^2T)$. The rate of large-angle scattering is $\mathcal{O}(g^4T)$. 

In addition to elastic $2\leftrightarrow 2$ scattering, $1 \leftrightarrow 2$ splitting and merging processes have to be taken into account. In vacuum, these are kinematically allowed only when all three momenta are exactly collinear. In plasma, this restriction disappears when a soft scattering is involved. The process is then nearly collinear with a rate $\mathcal{O}(g^4T)$, i.e.,\, comparable to the large-angle elastic-scattering rate. The process is not instantaneous, as it involves an off-shell propagator. The formation time of splitting/merging processes turns out to be $\mathcal{O}(1/g^2T)$, i.e.,\, the same as the small-angle-scattering mean free path. Consequently, additional soft collisions occur during the formation time making it an $N+1 \leftrightarrow N+2$ process. The additional soft collisions are not resolved individually by the radiation and cannot be treated as classically independent. Instead, they act coherently giving rise to the QCD analogue of the Landau-Pomeranchuk-Migdal (LPM) effect \cite{Landau:1953um, Migdal:1956tc}. The splitting/merging processes involving $N$ coherent soft scatterings are denoted as $``1 \leftrightarrow 2"$ processes.

\smallskip

The effective kinetic theory is formulated as a set of Boltzmann equations
\begin{equation}
    \left(\partial_t + \frac{\p}{p_0} \cdot \nabla_\mathbf{x} \right) f_s(\x,\p,t) = - C_s^{2\leftrightarrow 2}[f] - C_s^{``1\leftrightarrow 2"}[f]
\end{equation}
for the phase space densities $f_s(\x,\p,t)$ of partons of species $s$ at position $\x$ with momentum $\p$ at time $t$. The collision kernels $C_s^{2\leftrightarrow 2}$ and $C_s^{``1\leftrightarrow 2"}$ encode the effect of elastic scattering and splitting/merging, respectively.

The elastic scattering kernel includes all $2\leftrightarrow 2$ processes, because all of them have parametrically the same hard scattering rate. It has the usual form
\begin{align}
    & C_a^{2\leftrightarrow2}[f]  =  \frac{1}{4|\p|} \sum_{bcd} \int_{\k\pp\kp} \delta^{(4)}(P+K-P'-K') \nonumber\\ 
    &  \times \nu_b |\mathcal{M}^{ab}_{cd} |^2 (2\pi)^4\big\{ f_a(\p)f_b(\k)[1\pm f_c(\pp)][1\pm f_d(\kp)] \nonumber \\ 
    & - f_c(\pp)f_d(\kp)[1\pm f_a(\p)][1\pm f_b(\k)] \big\} \,, \label{eq:C_elastic}
\end{align}
where the dependence of the phase space densities $f_s$ on position and time has not been written out. Capital letters denote four-vectors and the particles are taken as massless. The factor $\nu_s$ is the number of spin and colour states (i.e.\ $\nu_q = 6$ for quarks and anti-quarks and $\nu_g = 16$  for gluons). The effective in-medium matrix elements squared $|\mathcal{M}^{ab}_{cd} |^2$ are taken as summed over the final state and averaged over the initial state\footnote{Differences in factors of $\nu_s$ compared to~\cite{Arnold:2002zm} arise because in~\cite{Arnold:2002zm} the squared matrix elements are taken as summed over both initial and final state.}. It should be noted that due to screening in the plasma the matrix elements depend on the phase space densities. The short hand notation
\begin{equation}
    \int_\p \dots = \int\frac{\d^3 p}{2|\p| (2\pi)^3} \dots
\end{equation}
for the Lorentz-invariant momentum integration has been used, and the factor in front of the integral contains a factor of $1/2$ for symmetry. The factors $(1 \pm f)$, where the upper sign is for gluons and the lower is for (anti)quarks, encode Bose enhancement or Pauli blocking, depending on the species. The first term in \Eq{eq:C_elastic} is a gain term accounting for partons arriving at momentum $\p$ due to elastic scattering, while the second is the loss term accounting for partons leaving momentum $\p$.

\smallskip

The kernels for the effective splitting/merging processes can be written as
\begin{align}
    & C_a^{``1\leftrightarrow 2"}[f] =  \frac{(2\pi)^3}{2|\p|^2\nu_a}\sum_{b,c}\int_0^\infty \d p'\,\d k'\delta(p-p'-k') \nonumber \\ 
    & \ \times \gamma^a_{bc}(\p;p'\phat,k'\phat)\Big\{f_a(\p)[1\pm f_b(p'\phat)][1\pm f_c(k'\phat)] \nonumber \\ 
    &  \qquad \qquad \qquad \qquad - f_b(p'\phat)f_c(k'\phat)[1\pm f_a(\p)] \Big\} \nonumber \\
    & + \frac{(2\pi)^3}{|\p|^2\nu_a}\sum_{b,c}\int_0^\infty \d k\,\d p'\delta(p+k-p') \nonumber \\ 
    &  \times \gamma^a_{bc}(p'\phat;\p,k\phat) \Big\{f_a(\p)f_b(k\phat)[1\pm f_c(p'\phat)] \nonumber \\
    & \qquad \qquad \qquad \quad - f_c(p'\phat)[1\pm f_a(\p)][1\pm f_b(k\phat)] \Big\},     \label{eq:C_inelastic_collinear}
\end{align}
where the (small) transverse momenta associated with the splitting or merging have been integrated out neglecting the deviations from exact collinearity in the phase space densities. The functions $\gamma^a_{bc}$ are the differential splitting/merging rates for the $a \leftrightarrow b+c$ processes and contain the effect of multiple soft scattering during the formation time. As in the elastic scattering case, they depend implicitly on the phase space densities. The rates $\gamma^a_{bc}$ should be understood as summed (not averaged) over all external partons. 

\smallskip

The effective kinetic theory is leading-order accurate in the coupling $\lambda = g^2 N_c = 4 \pi \alpha_s N_c$ for hard partons. 

In order for any kinetic theory to be applicable the duration of scattering events must be small compared to the mean free path. In the case of the AMY effective kinetic theory in equilibrium this in ensured by the assumption that the temperature is sufficiently high for the theory to be weakly coupled. Out of equilibrium, the question of applicability of the theory requires a more careful discussion, which is provided in~\cite{Arnold:2002zm}. The most important requirements are that the phase space densities support a separation of scales as outlined at the beginning of this section. This means that all relevant excitations should have momenta that are large compared to the screening masses. The effective masses, in turn, have to be much larger than all other mass scales ($\Lambda_\text{QCD}$ and quark masses). Furthermore, the phase space densities have to be sufficiently slowly varying, in particular they should not vary significantly over the length/time of the formation time of near the collinear splitting/merging processes. Lastly, the formulation of the AMY kinetic theory is technically leading-order accurate only for isotropic distributions due to certain plasma instabilities. However, numerical evidence points that the effect of the unstable modes may be quantitatively small \cite{Berges:2013fga}.

\section{The \textsc{Alpaca} framework}
\label{section:alpaca}

\subsection{General remarks}
\label{subsection:alpaca_timeevolution}

\textsc{Alpaca} (AMY Lorentz invariant PArton CAscade) is a parton cascade which solves the Boltzmann equation of the AMY effective kinetic theory indirectly by, in a Lorentz invariant way, evolving a parton ensemble according to the AMY collision kernels. To fully capture the dynamics of these kernels, three types of processes are considered: elastic scattering, quasi-collinear merging and quasi-collinear splitting. 

In addition to being able to evolve finite-sized systems, \textsc{Alpaca} also contains an implementation of infinitely sized ensembles in thermal equilibrium, used  to verify simulations in a simpler setting with analytically known results. The infinite-sized system is implemented through a box with periodic boundary conditions, something that introduces further complications to the Lorentz invariant framework, which is discussed in detail in \Sect{subsection:alpaca_box}.

The evolution of the particle ensemble takes place in $\tau$ as defined in \Sect{section:lorentzinvariance}. Each event starts at $\tau=0$ and is set to end at some $\tau_{\mathrm{max}}$. Since $\tau$ does not have a physical meaning it is convenient to also define an observer time $t_\text{max}$ that the system should reach. To ensure that all particles are evolved to $t_\text{max}$ the simulation will keep running until all particles have $t\geq t_{\mathrm{max}}$ even when this requires going to $\tau > \tau_\text{max}$. There is no need for $t_\text{max}$ and $\tau_\text{max}$ to be the same, but we here choose $t_\text{max} = \tau_\text{max}$. After the initialization of an event, the next possible processes of the three described above are found and ordered in $\tau$. Between any processes taking place partons are treated classically as point particles and are considered free streaming, according to \Eq{eq:x_freestream}.

The possible interactions of particles implemented in \textsc{Alpaca} to fully capture the dynamics given by AMY are elastic scattering and inelastic splitting/merging. However, due to the LPM effect present in the inelastic processes as well as the restriction of collinearity, the inelastic and elastic processes are treated differently. For the elastic case, an effective cross section extracted from the elastic collision kernel is used. In the current implementation we use a black disk approximation for the cross section, i.e.,\ an interaction between a pair of particles occurs if their invariant distance at closest approach satisfies $d_{ij} < \sqrt{\sigma/\pi}$. If several processes are allowed, $\sigma$ is the sum of the cross sections of the individual processes. It is then decided in a second step which processes is going to take place according to the relative contributions to the total cross section. 

In the Lorentz invariant framework a problem arises in a box with periodic boundary conditions. As discussed in \Sect{subsection:alpaca_box} the algorithm for finding the closest approaches between pairs of particles requires the cross section as input. The cross section however, depends on the local quantities $m_{g/q}(\x,\p)$ and $f(\x,\p)$ which are not known a priori. The solution is to find an overestimate of the cross sections that is independent of local quantities and then reject interactions with a probability that is given by the true cross section (evaluated at the point of the potential interaction) divided by the overestimate\footnote{This is a standard Monte Carlo method known as multi-channel accept/reject.}. The cross sections and overestimates for the different processes are given in \Sects{subsection:alpaca_elastic_general}, \ref{subsection:alpaca_inelastic_merging} and \app{appendix:formulae_elasticmatrixelements}.

For the inelastic processes, the effect of multiple coherent scattering during the formation time (LPM effect) is already included in the splitting/merging rates $\gamma^a_{bc}$. We therefore don't explicitly simulate them, but treat the inelastic processes as effective $1\leftrightarrow 2$ processes. However, since we are dealing with an ensemble of discrete particles, a pair of particles will never be exactly collinear. We therefore have to allow for a small relative transverse momentum $k_\perp$ for the merging processes, and consequently also for splitting (the two types of processes have to cover the same phase space in order to preserve detailed balance). The $k_\perp$ is absorbed by a nearby parton, which we call the 'recoil parton'. This recoil parton thus effectively acts as the (possibly multiple) parton(s) with which the soft coherent scattering occurs during the formation time of the splitting/merging. Since merging processes initially consists of a pair they can be treated similarly to elastic scattering, where we extract an effective cross section from the inelastic collision kernels and then evaluate this using a black disk approach.

For collinear splittings, an implementation of the Veto algorithm is utilized, see \Sect{subsection:alpaca_inelastic_splitting} for details. Since the splitting and merging rate is infra-red divergent we have to introduce a small cut-off in $x$. We do this by introducing a minimum momentum for all particles, $p_{\mathrm{min}}$, so that no particle is initialized with $p < p_\text{min}$ or allowed to scatter/split/merge into a momentum lower than this. This does not compromise the theory's accuracy.

Currently, \textsc{Alpaca} is only utilizing the general event handling framework in \textsc{Sherpa}, due to the focus on testing the model in thermal equilibrium. We plan to integrate \textsc{Alpaca} in the event generation in \textsc{Sherpa} in the future to allow for hadronisation and inclusion of hard processes like jet production.

\subsection{Simulating a system in equilibrium}
\label{subsection:alpaca_box}

In numerical studies of systems of particles in thermal equilibrium, one usually considers the system in a box with periodic boundary conditions. The position of a particle leaving the box at one end is then shifted such that the particle re-enters the box at the other end (the momentum remains unchanged). The justification for this procedure is that a system in thermal equilibrium is homogeneous and therefore, if one imagines the space to be divided into equally sized cells, all cells look the same on average. The periodic boundary condition makes all cells look exactly the same.

\smallskip

To simulate the evolution of a system of particles, one has to consider the interactions of each particle with all other particles. Since all cells are identical, it is enough to consider all interactions inside one cell. One therefore has to identify all pairs of particles that interact inside the box within a pre-defined time interval. In a parton cascade considering only binary interactions, one thus has to compute the time and distance of closest approach for each pair of particles. The distance of closest approach decides whether the pair will scatter, and the time at which this happens decides whether the scattering will take place inside the box (or whether one or both particles will have left the box before the encounter). To correctly account for all possible interaction inside the box, one has to consider scattering of a particle inside the box with particles coming from neighbouring cells. Both the time and distance of closest approach depend only on the relative distance between the two particles. One thus captures all possible encounters of particles 1 and 2 and all their copies in other cells by considering the copy of particle 1 located in the box and all copies of particle 2 on other cells.

\smallskip

When the system with periodic boundary conditions evolves in an observer time, e.g.\ the time in the rest frame of the box, a particle in the box can only collide inside the box with a particle coming from inside the box or one of the neighbouring 26 cells\footnote{This can be seen by slicing each side of a solid cube in a grid of 9 evenly sized cells. Slicing through each side (excluding the opposite side of one that has already been sliced) will result in the original cube being divided into 27 evenly sized smaller cubes. The small cube in the middle of the original cube then has 26 neighbours.}, which are copies of the original cell, surrounding the box. If the second particle comes from farther away, the closest approach of the pair will always happen outside the box. Such an evolution is, however, not Lorentz invariant, but frame dependent. As discussed in \Sect{section:lorentzinvariance} the evolution of a multi-particle system can be simulated in a frame independent way by ordering scattering events not in an observer time $t$, but in a generalised time $\tau$, which is a Lorentz scalar.  In this way the evolution of the system becomes Lorentz invariant, but now a particle in the box can scatter inside the box with a copy of another particle coming from any cell. One therefore has to consider a countably infinite number of encounters for this pair. Of these we are only interested in those that bring the particles close enough for a scattering. And of those that lead to a scattering we need to find the first one, i.e.\ the one with the smallest $\tau$ starting from the current $\tau$ of the evolution.

We have developed an algorithm that finds the earliest closest approach for a given pair of particles that is close enough for an interaction to occur. It systematically checks closest approaches between the first particle in the box with copies of the second particle coming from other cells ordered in increasing invariant time $\tau$. This algorithm needs the scattering cross section as input, because when a closest approach is not close enough for a scattering to occur it has to continue checking other copies of the second particles. The algorithm is specified in full detail in \app{appendix:invariantbox}.

\subsection{Elastic scattering}
\label{subsection:alpaca_elastic}

\subsubsection{General considerations}
\label{subsection:alpaca_elastic_general}

As mentioned in \Sect{subsection:alpaca_timeevolution}, to determine if a parton pair scatters elastically, and if so which type of elastic scattering occurs, multichannel accept/reject Monte Carlo sampling is used. The scattering matrix elements are related to the elastic cross section through
\begin{align}
    \label{eq:sigma_2_to_2}
    \sigma^{ab}_{2\rightarrow 2} & = \frac{1}{2}\sum_{cd}\int \frac{d\sigma^{ab}_{cd}}{d\Phi_{\pp,\kp}}d\Phi_{\pp,\kp} \nonumber\\
    & = \frac{1}{s^2} \sum_{cd} \int d^3\pp d^3\kp\left[\frac{s}{2|\pp||\kp|}\right]  \frac{|\mathcal{M}^{ab}_{cd}|^2}{16(2\pi)^6}(2\pi)^4 \nonumber \\ 
    & \times \delta^{(4)}(P+K-P'-K')[1\pm f(\pp)][1\pm f(\kp)],
\end{align}
where 
\begin{align}
    d\Phi_{\pp,\kp} = & \frac{d^4P'}{(2\pi)^3}\frac{d^4K'}{(2\pi)^3} \delta(P'^2)\theta(P_0')\delta(K'^2)\theta(K_0') \nonumber \\ 
    & \times (2\pi)^4 \delta^{(4)}(P+K-P'-K')
\end{align}
is the phase-space measure of the outgoing particles. Note that we pick up a factor of $1/2$ in the first equality due to either gluons being indistinguishable or from double counting of fermionic processes. The factor within the bracket can be expressed using the relative angle $\theta_{\Delta\p\k}$ between $\p$ and $\k$ as $s/2|\p||\k| = 1-\cos(\theta_{\Delta\p\k})$ and enforces a vanishing contribution for particles moving collinearly. In an isotropic system this expression integrates to unity.

For hard scattering screening effects are negligible and the standard vacuum matrix elements can be used. For soft scattering (small $t$ or $u$), however, the matrix elements receive large corrections from in-medium physics. Following the strategy of~\cite{York:2014wja,Kurkela:2018oqw} we replace the infra-red divergent $1/t^2$ terms in the matrix elements by
\begin{equation}
    \frac{1}{t^2} \to \frac{1}{(t - \zeta_s m^2_s)^2} \,,
\end{equation}
where $m_s$ is the effective mass of the parton in the propagator and the prefactors $\zeta_g = e^{5/3}/4$ and $\zeta_q = e^2/4$ contain the conversion factor between (asymptotic) effective mass and screening mass and a scaling factor that ensures that the modified matrix elements reproduce to leading order the HTL results for drag and momentum broadening ~\cite{York:2014wja,Ghiglieri:2015ala}. Strictly speaking, the factors $\zeta_s$ should be re-calculated for our choice of making the replacement, which differs from~\cite{York:2014wja,Kurkela:2018oqw}. We leave this for the future. We list the matrix elements used in \textsc{Alpaca} in \app{appendix:formulae_elasticmatrixelements}.

In addition to the scattering matrix elements the collision kernels contain factors of $(1\pm f)$ for the outgoing partons, which encode Bose enhancement and Pauli blocking. Thus the integral over the outgoing momenta of the matrix elements does not give the standard cross sections, but effective cross sections containing the Bose enhancement/Pauli blocking factors. The effective cross sections can be used for simulating the scatterings in the standard way. However, also the Bose enhancement/Pauli blocking factors depend on the position of the scattering and are not known beforehand. We therefore include them in the rejection step by including an overestimate of these factors when integrating the effective cross sections and later rejecting the scattering with the ratio of the true Bose enhancement/Pauli blocking factor and the overestimate. 

In practice, if the closest approach occurs at some $\bar{\tau}$ and the distance is smaller than $\sqrt{\tilde \sigma/\pi}$ where $\tilde \sigma$ is the overestimated effective cross section, $m_s^2(\x,\p)$ and $f_s(\x,\p)$ are extracted as described in \Sect{subsection:alpaca_elastic_m2}. In combination with the matrix element, as given in \app{appendix:formulae}, the full cross section is found and the process is accepted or rejected based on the ratio $\sigma^{ab}_{2\rightarrow 2}/\tilde{\sigma}^{ab}_{2\rightarrow 2}$.

If the process is accepted, $t$ is sampled from the differential cross section (using accept/reject Monte Carlo with the overestimates of the matrix elements given in \app{appendix:formulae}) and the particles' momenta are updated.

\subsubsection{The effective masses and phase space densities}%
\label{subsection:alpaca_elastic_m2}

The screening mass $m_s$ for a particle of species $s$, represents the medium screening effects and contributes in a highly non-trivial way to the evolution of the parton ensemble. It is responsible for regulating the otherwise divergent matrix elements in a local way and is in turn essential to capture interesting dynamics in the non-equilibrium case. The (position averaged) definition of the effective mass for a gluon is

\begin{equation}
    \label{eq:mg2}
    m_g^2 = \sum_{s} 2\nu_s \frac{g^2 C_s}{d_A} \int_V \frac{d^3\x}{V} \int \frac{d^3\p}{2|\p|(2\pi)^3}f_s(\p,\x)
\end{equation}
and for a fermion of species $s$

\begin{align}
    \label{eq:mq2}
    m^2_{q(s)/\bar{q}(s)} & = 2g^2C_F \int_V \frac{d^3\x}{V}  \int \frac{d^3\p}{2|\p|(2\pi)^3} \nonumber \\ 
    & \times \left[2f_g(\p,\x)+f_s(\p,\x)+f_{\bar{s}}(\p,\x)\right],
\end{align}
where the average is over a spatial volume $V$. Since $d^3\p/|\p|$ and $d^3\x/V$ are Lorentz invariant measures, and $f(\p,\x)$ is a Lorentz invariant scalar \cite{Treumann:2011zb, HendrikVanHees:2023}, it follows that the screening masses are Lorentz invariant quantities.

The screening masses have to be calculated for each scattering and at fixed observer time by integrating/summing over nearby partons. In any reference frame the two colliding partons will in general experience the scattering at different observer times. We choose to calculate the screening mass at the average of these times, which we denote by $t_\text{D}$. A more serious complication is that when evolving the system in $\tau$ (rather than observer time) the partons' positions and momenta are known at fixed $\tau$, which corresponds to a different observer time for every parton. Partons that are known at observer times later than $t_\text{D}$ for some scattering can be propagated back in time, but for partons with times earlier than $t_\text{D}$ this is not possible. We therefore choose to let the earlier partons  free-stream to $t_\text{D}$. This violates causality, as we are ignoring potential scatterings during that time which would change the partons' positions and momenta. This can be mitigated by solving the time evolution iteratively, as discussed in \Sect{subsection:full_run}.

To extract $m^2_{s}$ in a local way we have the following prescription. For any particle pair with their closest approach at some $\bar{\tau}$, we define $t_\text{D}$ to be the average observer time of the particle pair at $\bar{\tau}$. We then identify the $N_{\mathrm{inc}}$ closest (in terms of Euclidean spatial distance in the local rest frame) particles, w.r.t. to the midpoint of the original particle pair at $t_\text{D}$. This gives a volume $V = 4\pi r_{\mathrm{max}}^3/3$, where $r_{\mathrm{max}}$ is the Euclidian distance to the particle furthest away. Due to the semi-classical approach of \textsc{Alpaca} the partons are treated as pointlike and hence the density of particles can be expressed as

\begin{equation}
    \label{eq:dNdV}
    \frac{dN_a}{d^3\x d^3\p}=\frac{\nu_a}{(2\pi)^3}f_a(\x,\p) = \sum_i \delta^3(\x_i-\x)\delta^3(\p_i-\p),
\end{equation}
which gives an expression for $f(\x,\p)$. For a gluon propagator, following \Eq{eq:mg2}, each of the $N_{\mathrm{inc}}$ included particles then adds a factor of

\begin{equation}
     \frac{\alpha_s C_s \pi}{2V|\p|}
\end{equation}
to $m_{g}^2$ depending on the species $s$ of the included particle. Similarly, following \Eq{eq:mq2} for a quark propagator, each surrounding particle adds a factor 

\begin{equation}
    \frac{32\pi}{3V} \frac{\alpha_s}{\nu_g|\p|} \quad \mathrm{or} \quad \frac{16\pi}{3V} \frac{\alpha_s}{\nu_{q/\bar{q}}|\p|}
\end{equation}
to $m_{q/\bar{q}}^2$ depending on whether the included particle is a gluon or quark (of the same species as the virtual quark). 

\smallskip

In order to evaluate the Bose enhancement and Pauli blocking factors, $1\pm f_a(\x,\p)$, that appear in the AMY collision kernels, we also need explicit approximations of $f_a(\x,\p)$. Consequently, we approximate the phase space density by once again treating the particles as pointlike and looking at the phase space volume $V = V_\x V_\p$ where we have a spherical spatial volume around $\x$, the point of interest, and a spherical shell around the origin containing the (absolute) momentum $p$ in momentum space, i.e. 
\begin{equation}
    V_\x = \frac{4\pi r^3}{3} \quad \mathrm{ and } \quad V_\p = \frac{\pi \Delta p}{3}(12p^2 + (\Delta p)^2).
\end{equation}
Here, $r$ is the spatial distance between the included particle and $\x$ and is set to some fixed value, while $\Delta p$ is the allowed differences in absolute momentum $p$ between the original and included particles, also set to some fixed value.  We then simply count the $N_{\mathrm{inc}}$ particles that exist within $V$ and the extracted phase space density follows through the relation given in \Eq{eq:dNdV}, 
\begin{equation}
    f_a(\x,\p) = \frac{(2\pi)^3}{\nu_a} \frac{N_{\mathrm{inc}}}{V}.
\end{equation}
We also set an upper bound for the extracted phase space density, $f_a(p) = \min(f_a(p), T_*/p)$, since the underlying theory is not valid for $f(p)>T_*/p$.

Note that the implementation above is set up for an isotropic distribution, which our test case of thermal equilibrium described later will be. For anisotropic distributions the method is easily generalized by restricting the volume in momentum space to a sphere around $\p$.

The phase space density also has to be extracted at fixed observer time and the same free-streaming procedure as for the effective masses is used (with the same caveat concerning causality violation applying, although less severely since $f$ is local in momentum space and therefore the spread in times is much smaller, cf.~Eq.~(\ref{Eq::relttau})).

\subsubsection{The thermal equilibrium case}
\label{subsection:alpaca_elastic_thermal}

In order to verify that \textsc{Alpaca} reproduces the dynamics of AMY we put our particles in a box with periodic boundary conditions and sample our initial distributions in two different scenarios. First, the Boltzmann  distribution
\begin{equation}
    \label{eq:thermal_dist}
    f_{g/q/\bar{q}\mathrm{,initial}}(p) = f_{\mathrm{C}}(p) = e^{-p/T}
\end{equation}
which has the benefit of providing us with a simple setup that allows for many quantities to be found analytically and compared to, e.g. the scattering rate. This distribution does not give us the full picture though, since it does not factor in quantum effects like Bose enhancement and Pauli blocking. For some cases these effects are not of importance in our verification of \textsc{Alpaca}, and we can just exclude them in the simulations. For splitting and merging however, this is not possible\footnote{The effective temperature will not relax into the actual temperature if the quantum effects are not present in the distribution.} . Hence, we will also study the scenario of Bose-Einstein and Fermi-Dirac distributions,
\begin{equation}
    \label{eq:thermal_dist}
    f_{g\mathrm{,initial}}(p) = f_{\mathrm{BE}}(p) = \frac{1}{e^{p/T}-1},
\end{equation}
and 

\begin{equation}
f_{q/\bar{q}\mathrm{,initial}}(p) = f_{\mathrm{FD}}(p) = \frac{1}{e^{p/T}+1}  \,.  
\end{equation}

To tackle the divergence for vanishing momentum in $f_{\mathrm{BE}}$ (and later also divergent splitting/merging rates for low energy $x$) we introduce a cutoff in momentum, $p_{\mathrm{min}}$, so that no particle is initialized with momentum lower than this, and no particle is allowed to scatter/split/merge into a state with lower absolute momentum than this.

Both of the infinite thermalized systems above provide simplified scenarios where one can find analytical (and simpler numerical) results to compare against. To validate our implementation we show the following:

\begin{enumerate}[label=(\roman*)]
    \item The elastic scattering rate $dN_{\mathrm{coll.}}/dt$ in the case of thermal equilibrium with no Bose enhancement and Pauli blocking, and with a fixed $\sigma^{2\leftrightarrow 2}$, is consistent with the expected results, both for the classical and Bose-Einstein and Fermi-Dirac case.
    \item The  elastic scattering rate $dN_{\mathrm{coll.}}/dt$ with a non-constant $\sigma^{2\leftrightarrow 2}$ (implemented as described in \Sect{subsection:alpaca_elastic_general}) produces the correct rate w.r.t. the elastic collision kernel in AMY EKT. This is also checked both for the classical and Bose-Einstein and Fermi-Dirac case.
    \item The screening masses $m_s^2$ and phase space densities $f_s$ are extracted correctly, i.e., they match the analytical values, both for the classical and Bose-Einstein and Fermi-Dirac case.
\end{enumerate}

\begin{figure*}[h!]
    \centering
    \includegraphics[width=0.495\linewidth]{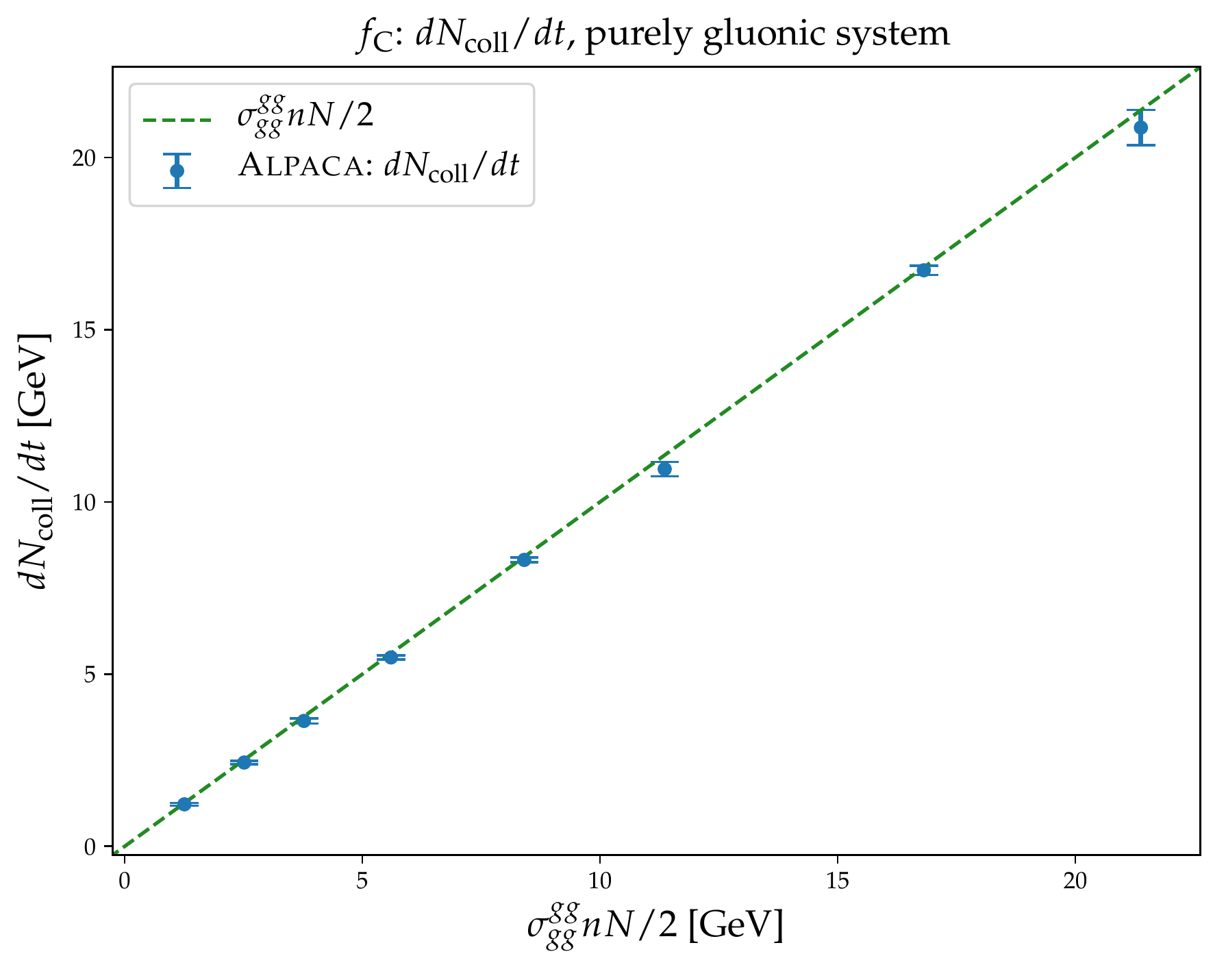}
    \includegraphics[width=0.495\linewidth]{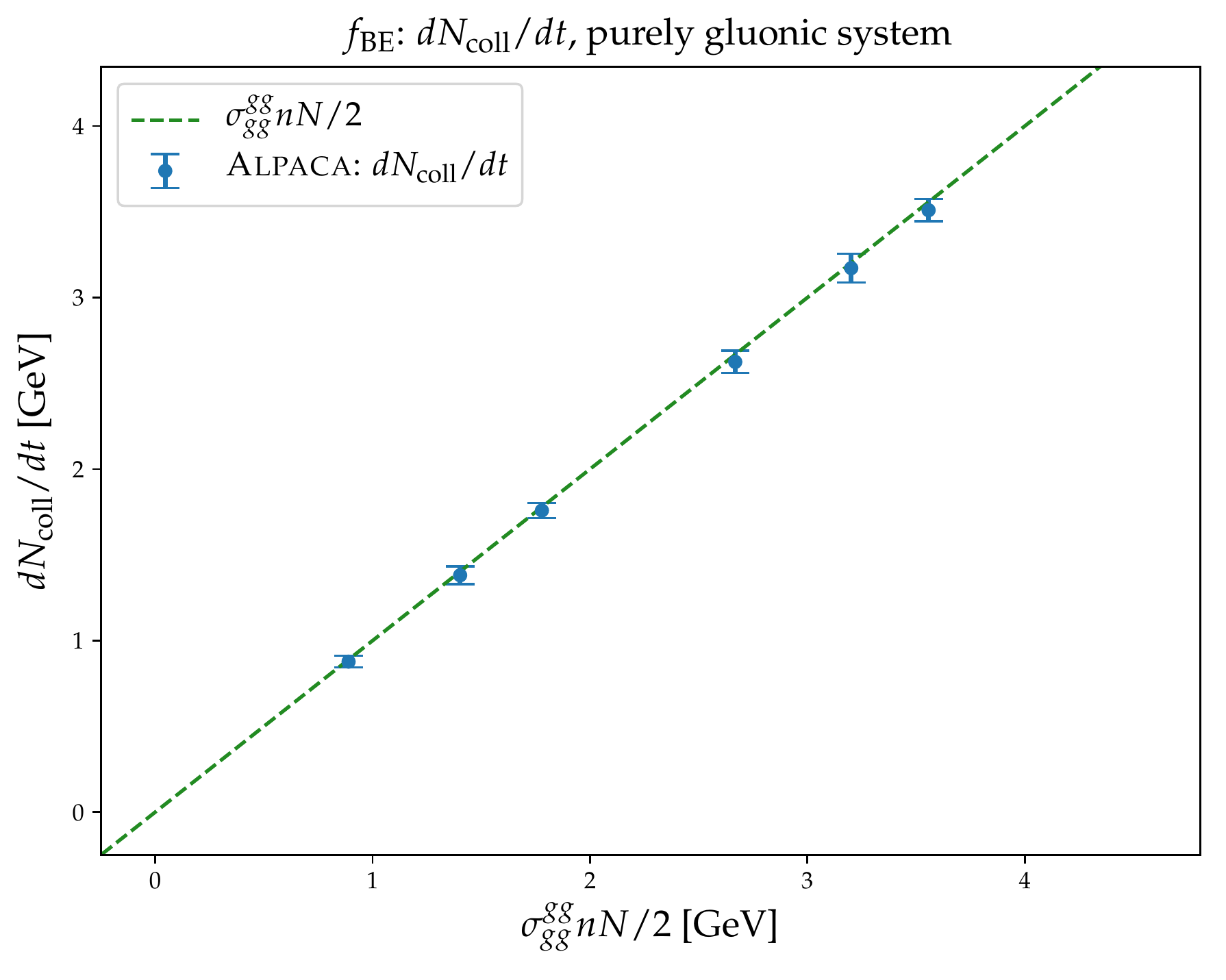}
    \caption{Scaling behaviour in thermal gluonic systems of the elastic scattering rate, $dN_{\mathrm{coll}}/dt$ (without Bose enhancement included) w.r.t. different fixed cross sections $\sigma^{gg}_{gg}$ (varying between $0.5-3$ $\mathrm{GeV}^{-2}$) and temperature (varying between $0.35-0.5$ $\mathrm{GeV}$) and $\tau_{\mathrm{max}} = 1$ $\mathrm{GeV}^{-1}$. Left: $f_{\mathrm{C}}$, box volume $V_x = 10.1^3$ $\mathrm{GeV}^{-3}$. Right: $f_{\mathrm{BE/FD}}$, box volume $V_x = 7.6^3$ $\mathrm{GeV}^{-3}$.}
    \label{fig:dNdt}
\end{figure*}

\noindent \textbf{(i)} The number of particles of species $s$, $dN_s$, contained in an infinitesimal phase-space volume $d^6\xi = d^3\x d^3\p$ around $\x$ and $\p$ can be related to the phase-space density $f_s(\x,\p)$ through
\begin{equation}
    \label{eq:ddN}
    dN_s = \frac{\nu_s}{(2\pi)^3}d^6\xi f_s(\x,\p).
\end{equation}
The collision kernels $C_s[f]$ are related to the rate of change of the number of particles $dN_s$ in $d^6\xi$ between $t$ and $t+dt$, i.e.
\begin{align}
    \frac{d}{dt}dN_s & = \frac{dN_s(t+dt)-dN_s(t)}{dt} \nonumber\\
    & = -\frac{\nu_s}{(2\pi)^3}d^6\xi[\partial_t+\textbf{v}\cdot\nabla]f_s \nonumber \\ 
    & =\frac{\nu_s}{(2\pi)^3}d^6\xi C_s[f]
\end{align}
since $d^6\xi(t+dt) = d^6\xi(t)$ due to the absence of external forces. Separating this into a gain, $dN_s^+$, and loss, $dN_s^-$, term we get that

\begin{align}
    \label{eq:ddNdt}
    \frac{d}{dt}dN_s & = \frac{d}{dt}(dN_s^+-dN_s^-) \nonumber \\ 
    & = \frac{\nu_s}{(2\pi)^3}d^6\xi (C_s^+[f] - C_s^-[f])
\end{align}
with $C_s[f]^+$ and $C_s[f]^-$ corresponding to the gain and loss terms in the collision kernel. The gain (or loss) term of the collision kernel can now be related to the rate of total number of scatterings as

\begin{align}
    \label{eq:total_scattering_rate}
    \frac{dN_{\mathrm{coll.}}}{dt} & = \frac{1}{2} \sum_{s}\frac{d}{dt}\int d(dN_s^\pm) \nonumber \\ 
    & = \frac{1}{2}\sum_s\int d^6\xi \frac{\nu_s}{(2\pi)^3} C_s^\pm[f] \nonumber \\
    & = \sum_s \nu_s \int d^3\x I^\pm_s
\end{align}
where we pick up a factor of $1/2$  due to either integrating over both incoming momentum of indistinguishable particles when $s$ is a gluon, or double counting processes in the sum when $s$ is a quark or antiquark. Here we also define the momentum integrated gain term of the collision kernel for a species $s$ as $I^\pm_s$ (with $+$ and $-$ for the gain and loss part of the collision kernel respectively), which are quantities of interest when evaluating (ii). Considering the case of a purely gluonic system without any Bose-enhancement we can extract a simple scaling behaviour for the rate of total number of collisions, 

\begin{align}
    \frac{dN_{\mathrm{coll.}}}{dt} & = \frac{1}{4} \int d^3\x \int_{\p\k\pp\kp} \nu_g^2 |\mathcal{M}^{gg}_{gg}|_{\mathrm{avg.}}^2(2\pi)^4 \nonumber \\ 
    & \times \delta^{(4)}(P+K-P'-K')f_g(\p) f_g(\k) \nonumber\\
    & = \frac{1}{2}\int d^3\x \int d^3\p d^3\k \frac{\nu_g^2}{(2\pi)^6}\left[\frac{s}{2E_\p E_\k}\right] \nonumber \\
    & \times \sigma^{gg}_{gg} f_g(\p) f_g(\k)
\end{align}
where we have used the definition of the total cross section given in \Eq{eq:sigma_2_to_2}. The expression within the brackets integrates to unity in isotropic systems, as explained in \Sect{subsection:alpaca_elastic_general}, and so for the thermal case with a constant cross section we have the scaling behaviour 

\begin{equation}
    \frac{dN_{\mathrm{coll.}}}{dt} = \frac{1}{2} N_gn_g\sigma^{gg}_{gg},\\
\end{equation}
with $N_g$ being the total number of gluons in the system, and $n_g$ the gluon density. From the equations above we also note that

\begin{equation}
    \label{eq:intC}
    \int \frac{d^3\p}{2(2\pi)^3} C_g^+[f] = \frac{n_g^2 \sigma^{gg}_{gg}}{2\nu_g}.
\end{equation}
The scaling behaviour shown above has been examined in \textsc{Alpaca} through scanning the parameter space, and exhibits the correct scaling behaviour, as shown in \Fig{fig:dNdt}.

\bigskip

\noindent \textbf{(ii)} Including quarks and/or Bose enhancement/Pauli blocking is straight-forward but the collision rate does then not exhibit the same simple scaling behaviour. However, due to the (approximate) time independence of $f(\x,\p)$ in thermal equilibrium we can integrate $C_g^{2\rightarrow 2}[f]+C_{q/\bar{q}}^{2\rightarrow 2}[f]$ numerically, using Monte Carlo methods with stratified sampling, and compare to the results of \textsc{Alpaca} through the relation found in \Eq{eq:ddN}. This numerical integration has been crosschecked against the code used in e.g. \cite{Kurkela:2021ctp}. The results of the comparison of the full numerical integration are shown in Tables~\ref{tab:full_integrals_C} and \ref{tab:full_integrals_BEFD} for Boltzmann and Bose-Einstein/Fermi-Dirac distributions, respectively. A comparison of the rates of each sub-process (for the loss-terms) are shown in Tables~\ref{tab:piecewise_integrals_C} and \ref{tab:piecewise_integrals_BEFD}, again for the different distributions. The explicit expression of the integrated collision kernels are shown in \app{appendix:numerical_solution_of_C}. As can be seen in the tables, the rates coincides well for $f_{\mathrm{C}}$, while for $f_{\mathrm{BE/FD}}$ they are within $\sim 10\%$ of the numerical results. This can be attributed to the cutoff $p_{\mathrm{min}}$ which effectively changes the value of $m_{g/q}^2$ that we extract in \textsc{Alpaca}, while this shift in screening mass is not included in the numerical results.

\begin{table*}[h!]
    \centering
    \begin{tabular}{ c|c|c|c}
         Integral $[\mathrm{GeV}^4]$ & \textsc{Alpaca} & Numerical & Ratio  \\ 
        \hline		\hline 
          $I_g^-$ & $2.16(3) \cdot 10^{-3}$ & $2.15 \cdot 10^{-3}$ & 1.00(2) \\
        $I_g^+$ & $2.18(3) \cdot 10^{-3}$ & $2.17 \cdot 10^{-3}$ & 1.00(2) \\
        \hline
        $I_q^-$ & $5.25(7) \cdot 10^{-4}$ & $5.40 \cdot 10^{-4}$ & 0.97(1) \\
        $I_q^+$ & $5.17(8) \cdot 10^{-4}$ & $5.31 \cdot 10^{-4}$ & 0.97(1) \\
  \end{tabular}
  \caption{The full integrals of the elastic collision kernels $C_g^{\pm,2\rightarrow 2}[f_{\mathrm{C}}]$ ($I_g^\pm$)  and $C_{q/\bar{q}}^{\pm,2\rightarrow 2}[f_{\mathrm{C}}]$ ($I_q^\pm$), see \Eqs{eq:appendix_I_g_loss}-\eqref{eq:appendix_I_q_gain}, are compared between their values produced by \textsc{Alpaca} (through the collision rate, see \Eq{eq:total_scattering_rate}) and by numerically integrating the collision kernels using Monte Carlo methods. The comparison is preformed in the thermal case with $T=0.4$ GeV, $\alpha_s = 0.3$, $\tau_{\mathrm{max}} = 1$ $\mathrm{GeV}^{-1}$ and box volume $V_x = 10.1^3$ $\mathrm{GeV}^{-3}$. The errors of the numerical results are negligible compared to the results from \textsc{Alpaca}, and hence not presented.}
    \label{tab:full_integrals_C}
\end{table*}

\begin{table*}[h!]
    \centering
    \begin{tabular}{ c|c|c|c}
         Integral $[\mathrm{GeV}^4]$ & \textsc{Alpaca} & Numerical & Ratio  \\ 
        \hline		\hline 
        $I_g^-$ & $3.341(1) \cdot 10^{-4}$ & $ 3.140\cdot 10^{-4}$ & 1.0639(4)  \\
        $I_g^+$ & $3.340(1) \cdot 10^{-4}$ & $ 3.140\cdot 10^{-4}$ & 1.0635(4)  \\
        \hline
        $I_q^-$ & $6.559(3) \cdot 10^{-5}$ & $ 7.312\cdot 10^{-5}$ & 0.8970(4)  \\
        $I_q^+$ & $6.573(3) \cdot 10^{-5}$ & $ 7.312\cdot 10^{-5}$ & 0.8989(4)  \\
  \end{tabular}
  \caption{The full integrals of the elastic collision kernels $C_g^{\pm,2\rightarrow 2}[f_{\mathrm{BE/FD}}]$ ($I_g^\pm$)  and $C_{q/\bar{q}}^{\pm,2\rightarrow 2}[f_{\mathrm{BE/FD}}]$ ($I_q^\pm$), see \Eqs{eq:appendix_I_g_loss}-\eqref{eq:appendix_I_q_gain}, are compared between their values produced by \textsc{Alpaca} (through the collision rate, see \Eq{eq:total_scattering_rate}) and by numerically integrating the collision kernels using Monte Carlo methods. The comparison is preformed in the thermal case, including quantum corrections in the initial distribution, with $T=0.4$ GeV, $p_{\mathrm{min}}=0.1$ GeV, $\alpha_s = 0.04$, $\tau_{\mathrm{max}} = 1$ $\mathrm{GeV}^{-1}$ and box volume $V_x = 7.6^3$ $\mathrm{GeV}^{-3}$. The errors of the numerical results are negligible compared to the results from \textsc{Alpaca}, and hence not presented.}
    \label{tab:full_integrals_BEFD}
\end{table*}

\begin{table*}[h!]
    \centering
    \begin{tabular}{ c|c|c|c}
         Integral $[\mathrm{GeV}^4]$  & \textsc{Alpaca} & Numerical & Ratio  \\ 
        \hline		\hline 
          $I_g^{-, gg\leftrightarrow gg}$ & $1.44(6) \cdot 10^{-3}$ & $1.43 \cdot 10^{-3}$ & 1.00(5) \\
        $I_g^{-, gq\leftrightarrow gq}$ & $0.69(3) \cdot 10^{-3}$ & $0.69 \cdot 10^{-3}$ & 1.00(4) \\
        $I_g^{-, gg\leftrightarrow q\bar{q}}$ & $0.026(4) \cdot 10^{-3}$ & $0.0027 \cdot 10^{-3}$ & 0.97(16) \\
        \hline
        $I_q^{-, q_iq_j\leftrightarrow q_iq_j}$ & $1.35(10)\cdot 10^{-4}$ & $1.39\cdot 10^{-4}$ & 0.97(7) \\
        $I_q^{-, q_iq_i\leftrightarrow q_iq_i}$ & $0.30(4)\cdot 10^{-4}$ & $0.31\cdot 10^{-4}$ & 0.96(11) \\
        $I_q^{-, q_i\bar{q}_i\leftrightarrow q_i\bar{q}_i}$ & $0.38(4)\cdot 10^{-4}$ & $0.39\cdot 10^{-4}$ & 0.98(10) \\
        $I_q^{-, q_i\bar{q}_i\leftrightarrow q_j\bar{q}_j}$ & $0.04(1)\cdot 10^{-4}$ & $0.04\cdot 10^{-4}$ & 1.01(25) \\
        $I_q^{-, qg\leftrightarrow qg}$ & $2.96(16)\cdot 10^{-4}$ & $3.0\cdot 10^{-4}$ & 0.97(6) \\
        $I_q^{-, q\bar{q}\leftrightarrow gg}$ & $0.20(3)\cdot 10^{-4}$ & $0.21\cdot 10^{-4}$ & 0.95(12) \\
  \end{tabular}
  \caption{The full integrals of the different subprocesses in the loss terms of the elastic collision kernels $C_g^{-,2\rightarrow 2}[f_{\mathrm{C}}]$ ($I_g^-$)  and $C_{q/\bar{q}}^{-,2\rightarrow 2}[f_{\mathrm{C}}]$ ($I_q^-$), see \Eqs{eq:appendix_I_g_loss} and \eqref{eq:appendix_I_q_loss}, are compared between their values produced by \textsc{Alpaca} (through the collision rate of each subprocess, see \Eq{eq:total_scattering_rate}) and by numerically integrating the collision kernels using Monte Carlo methods. The comparison is preformed in the thermal case with $T=0.4$ GeV, $\alpha_s = 0.3$, $\tau_{\mathrm{max}} = 1$ $\mathrm{GeV}^{-1}$ and box volume $V_x = 10.1^3$ $\mathrm{GeV}^{-3}$. The errors of the numerical results are negligible compared to the results from \textsc{Alpaca}, and hence not presented.}
    \label{tab:piecewise_integrals_C}
\end{table*}

\begin{table*}[h!]
    \centering
    \begin{tabular}{ c|c|c|c}
         Integral $[\mathrm{GeV}^4]$  & \textsc{Alpaca} & Numerical & Ratio  \\ 
        \hline		\hline 
        $I_g^{-, gg\leftrightarrow gg}$ & $2.2(2) \cdot 10^{-4}$ & $ 2.08\cdot 10^{-4}$ & 1.06(9)  \\
        $I_g^{-, gq\leftrightarrow gq}$ & $1.0(1) \cdot 10^{-4}$ & $ 1.04\cdot 10^{-4}$ & 1.05(10) \\
        $I_g^{-, gg\leftrightarrow q\bar{q}}$ & $0.024(9) \cdot 10^{-4}$ & $ 0.022\cdot 10^{-4}$ & 1.1(4) \\
        \hline
        $I_q^{-, q_iq_j\leftrightarrow q_iq_j}$ & $1.5(2) \cdot 10^{-5}$ & $ 1.7\cdot 10^{-5}$ & 0.85(13)  \\
        $I_q^{-, q_iq_i\leftrightarrow q_iq_i}$ & $0.36(9) \cdot 10^{-5}$ & $ 0.41\cdot 10^{-5}$ & 0.87(23)  \\
        $I_q^{-, q_i\bar{q}_i\leftrightarrow q_i\bar{q}_i}$ & $0.38(9) \cdot 10^{-5}$ & $ 0.43\cdot 10^{-5}$ & 0.86(22) \\
        $I_q^{-, q_i\bar{q}_i\leftrightarrow q_j\bar{q}_j}$ & $0.006(7) \cdot 10^{-5}$ & $ 0.01\cdot 10^{-5}$ & 0.6(7)  \\
        $I_q^{-, qg\leftrightarrow qg}$ & $4.2(5) \cdot 10^{-5}$ & $ 4.6\cdot 10^{-5}$ & 0.9(1) \\
        $I_q^{-, q\bar{q}\leftrightarrow gg}$ & $0.08(4) \cdot 10^{-5}$ & $ 0.1\cdot 10^{-5}$ & 0.8(4)  \\
  \end{tabular}
  \caption{The full integrals of the different subprocesses in the loss terms of the elastic collision kernels $C_g^{-,2\rightarrow 2}[f_{\mathrm{BE/FD}}]$ ($I_g^-$) and $C_{q/\bar{q}}^{-,2\rightarrow 2}[f_{\mathrm{BE/FD}}]$ ($I_q^-$), see \Eqs{eq:appendix_I_g_loss} and \eqref{eq:appendix_I_q_loss}, are compared between their values produced by \textsc{Alpaca} (through the collision rate of each subprocess, see \Eq{eq:total_scattering_rate}) and by numerically integrating the collision kernels using Monte Carlo methods. The comparison is preformed in the thermal, including quantum corrections in the initial distribution, with $T=0.4$ GeV, $p_{\mathrm{min}}=0.1$ GeV, $\alpha_s = 0.04$, $\tau_{\mathrm{max}} = 1$ $\mathrm{GeV}^{-1}$ and box volume $V_x = 7.6^3$ $\mathrm{GeV}^{-3}$. The errors of the numerical results are negligible compared to the results from \textsc{Alpaca}, and hence not presented.}
    \label{tab:piecewise_integrals_BEFD}
\end{table*}

\bigskip

\begin{figure*}[h!]
    \centering
    \includegraphics[width=0.495\linewidth]{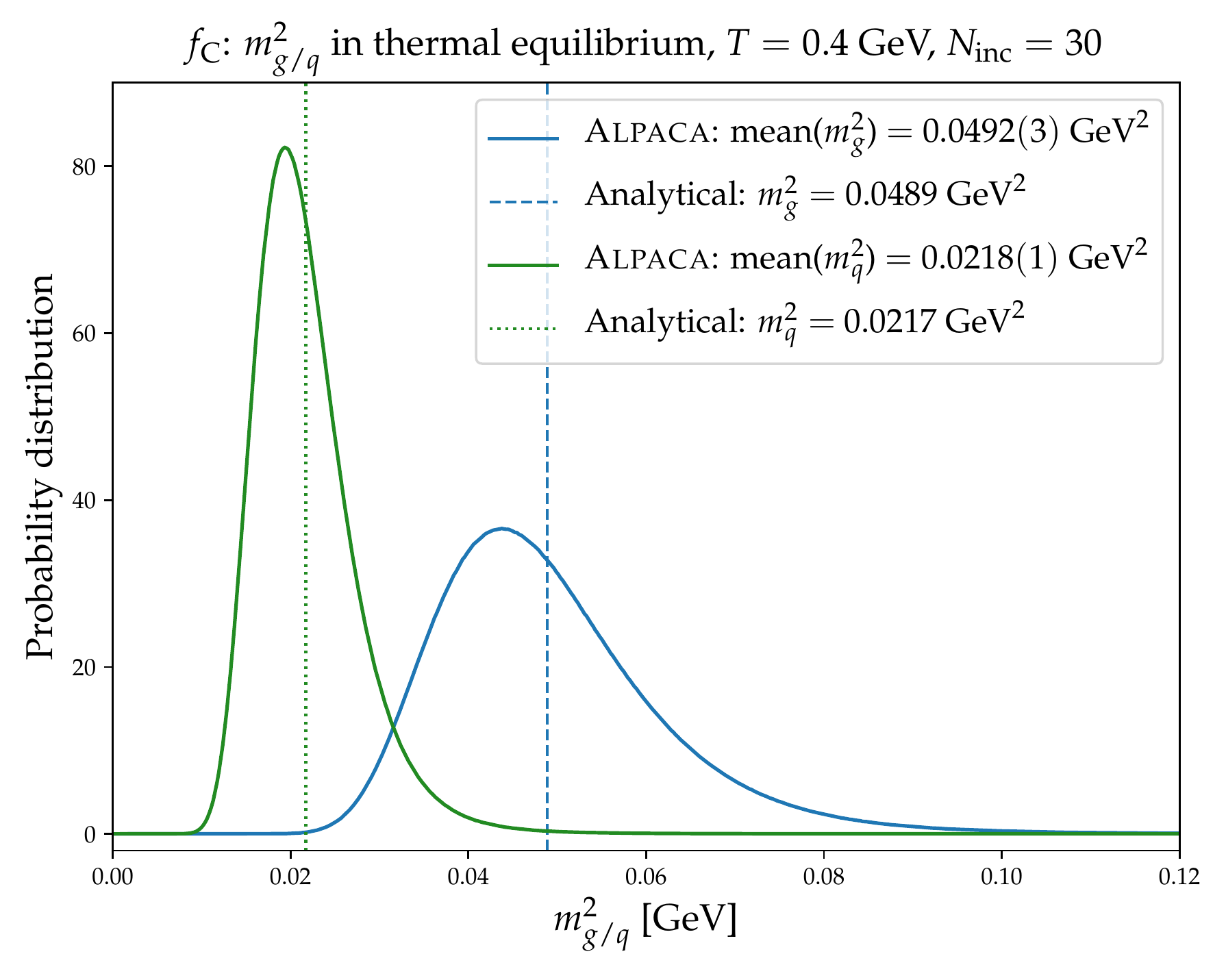}
    \includegraphics[width=0.495\linewidth]{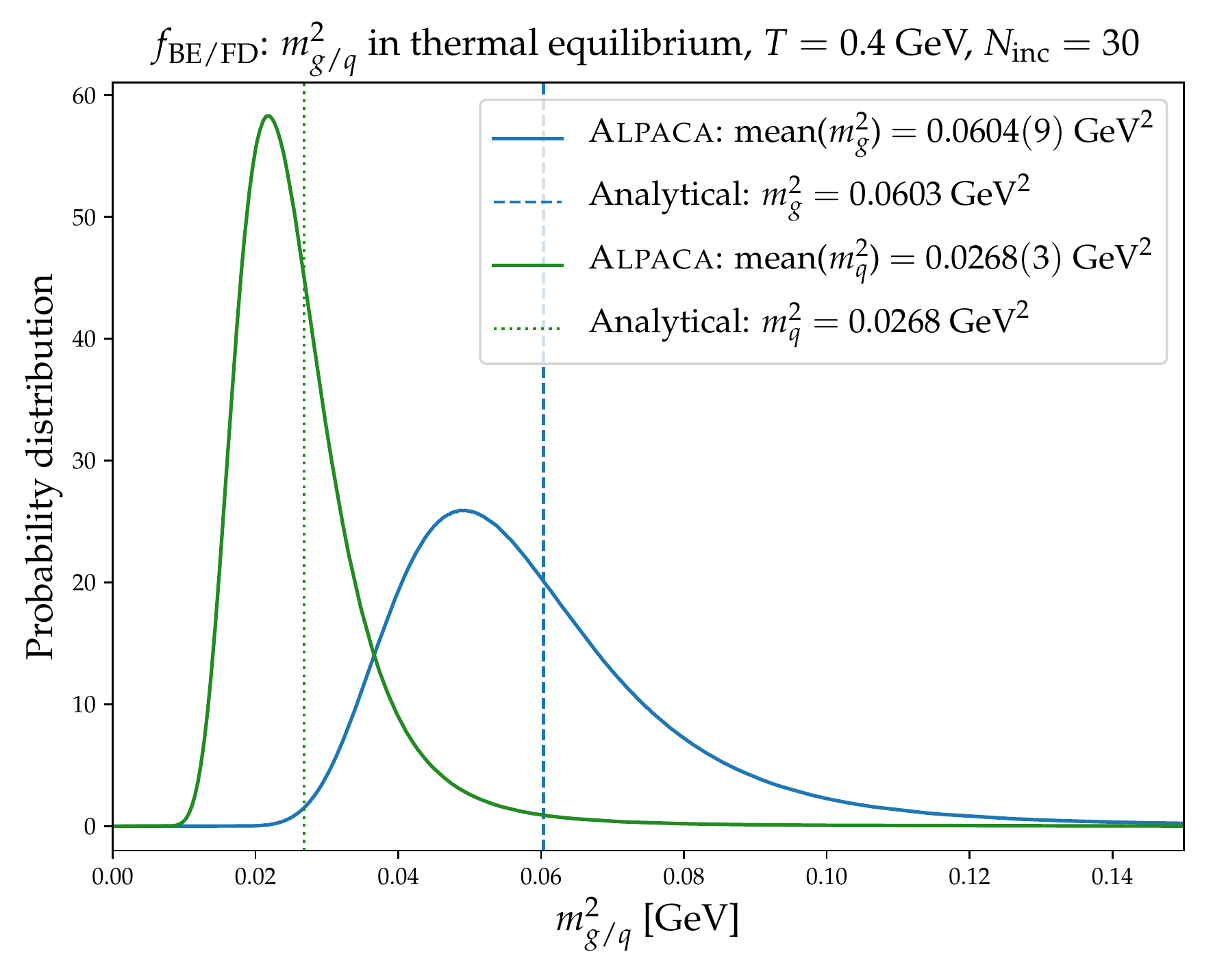}
    \caption{Probability distribution of extracted $m_{g/q}^2$ in thermal equilibrium, for $T=0.4$ GeV, and $30$ included particles. The errors presented for the mean values correspond to the error of the mean of the distribution  (in contrast, the width of the distribution is caused by including a finite number of particles in the extraction and does not translate into an uncertainty on the mean). Left: $f_{\mathrm{C}}$. Right: $f_{\mathrm{BE/FD}}$.}
    \label{fig:m2_N50}
\end{figure*}

\begin{figure*}[h!]
    \centering
    \includegraphics[width=0.495\linewidth]{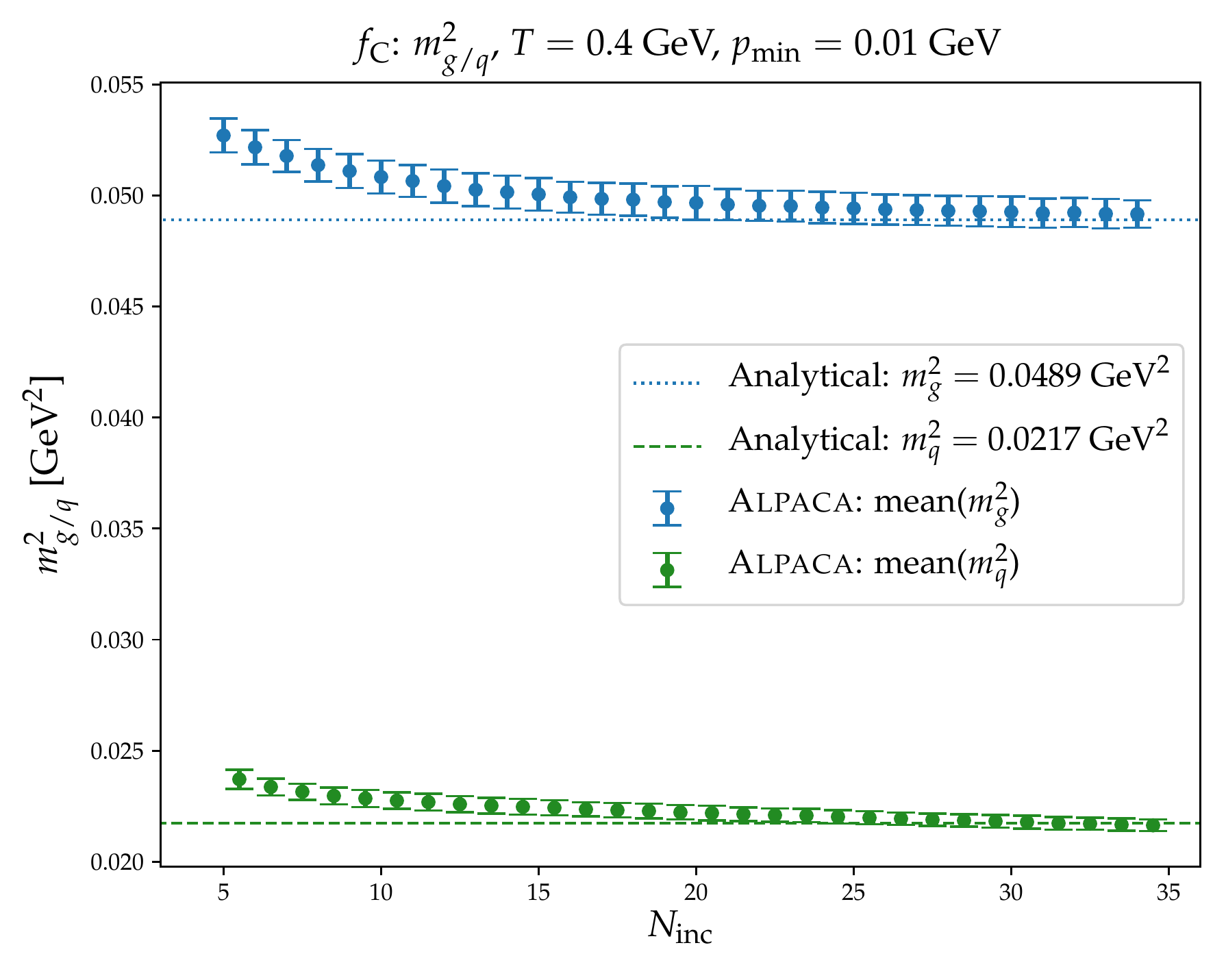}
    \includegraphics[width=0.495\linewidth]{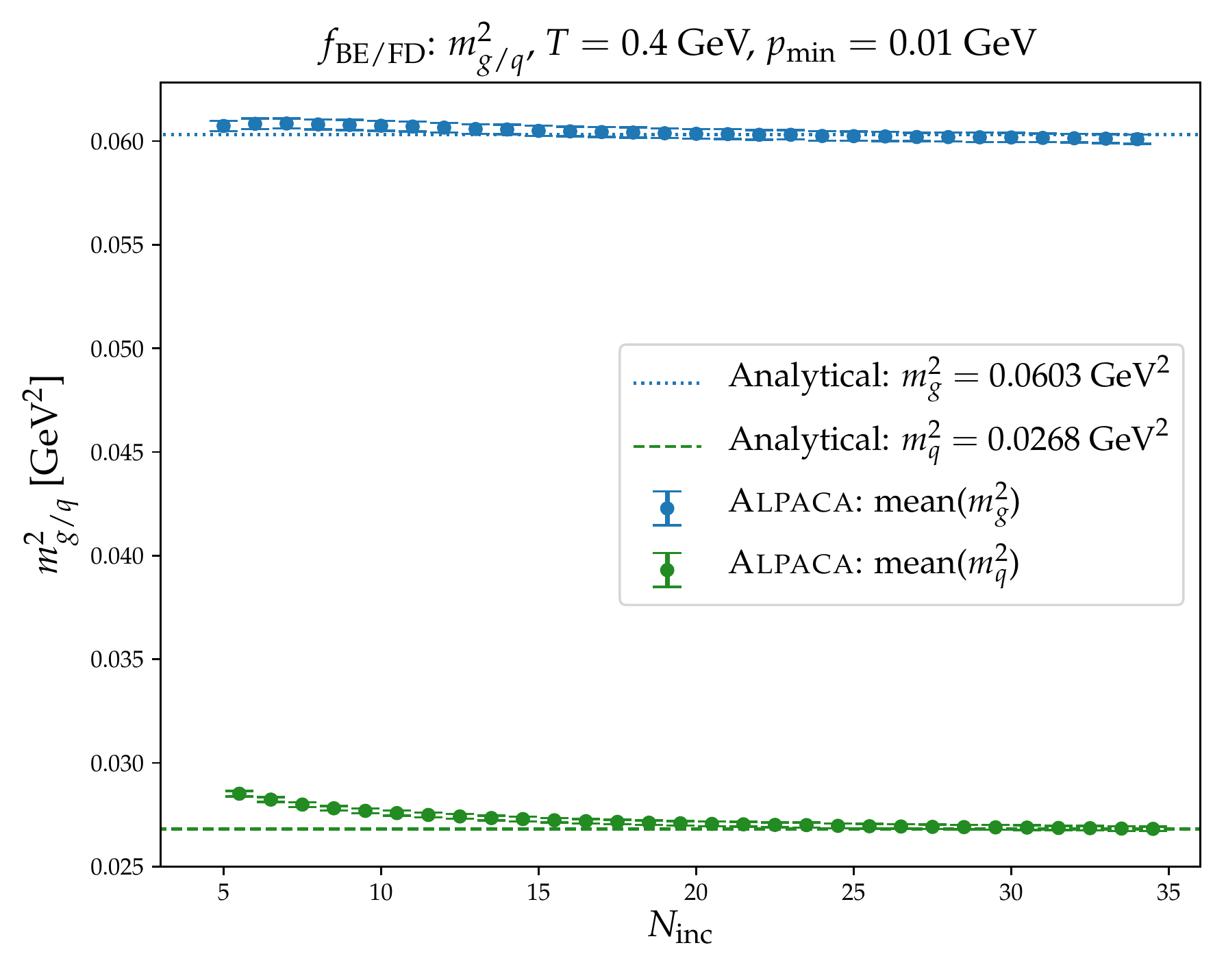}
    \caption{Extracted $m_{g/q}^2$ in thermal equilibrium, as a function of number of included particles, $N_{\mathrm{inc}}$. Left: $f_{\mathrm{C}}$. Right: $f_{\mathrm{BE/FD}}$.}
    \label{fig:m2_vs_Ninc}
\end{figure*}

\begin{figure*}[h!]
    \centering
    \includegraphics[width=0.495\linewidth]{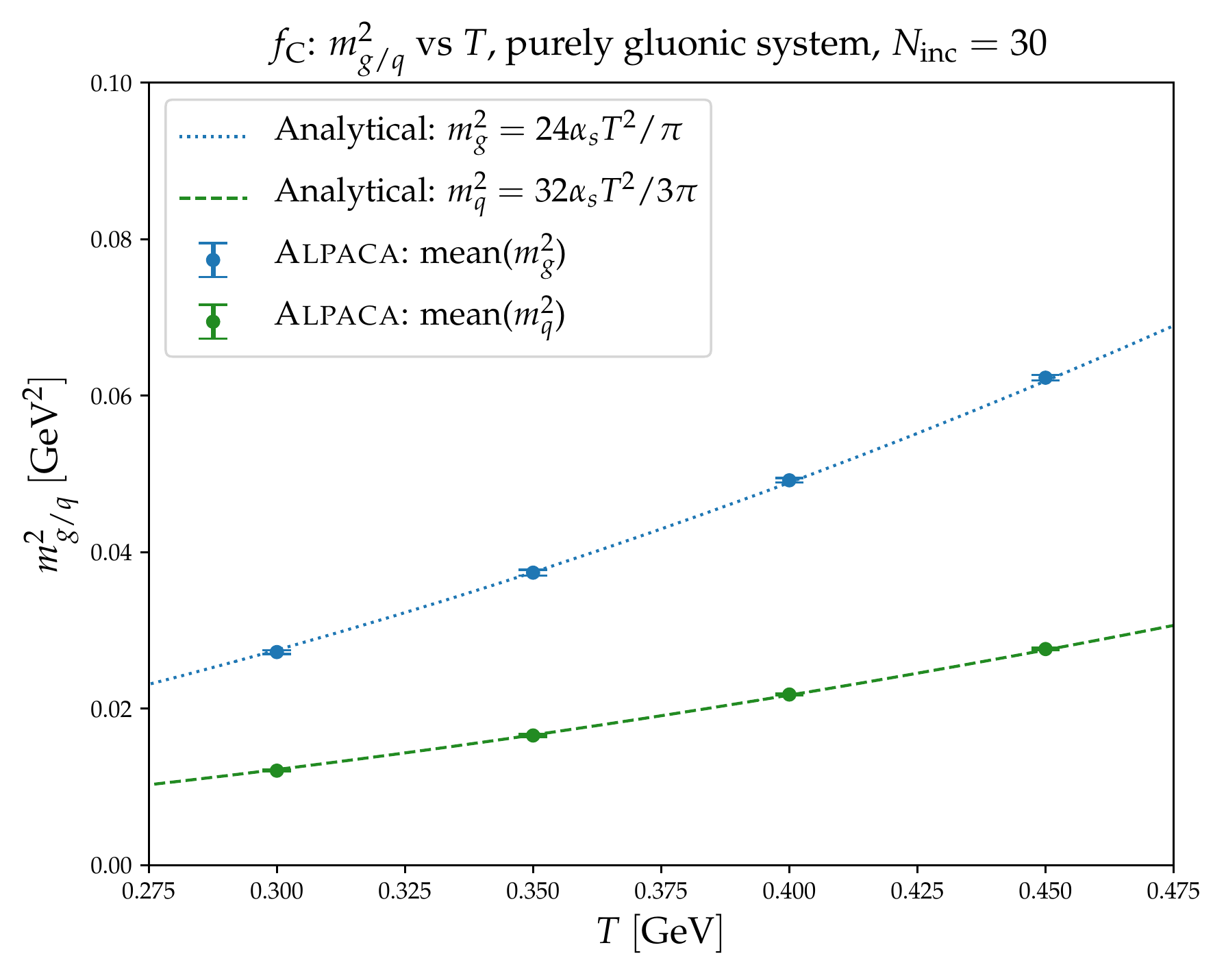}
    \includegraphics[width=0.495\linewidth]{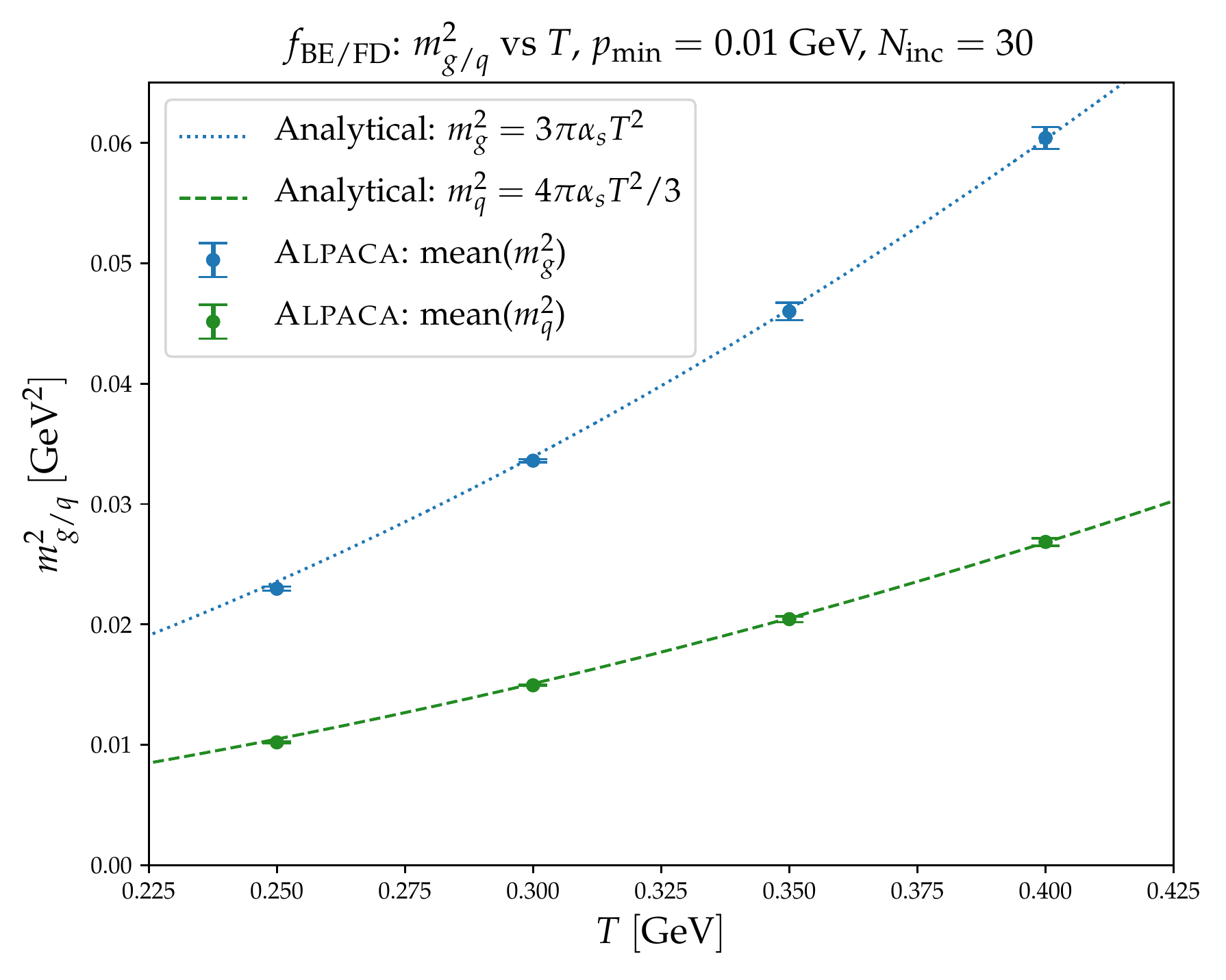}
    \caption{Scaling behaviour of the mean value of extracted $m_{g/q}^2$ in thermal systems, w.r.t different temperatures of the system. Number of included particles in each calculation is fixed to $30$. Left: $f_{\mathrm{C}}$. Right: $f_{\mathrm{BE/FD}}$.}
    \label{fig:m2_vs_T}
\end{figure*}

\noindent \textbf{(iii)} Lastly, we compare the implementation of $m_s^2(\x,\p)$, described in Section~\ref{subsection:alpaca_elastic_m2} to the analytical expectations for thermal equilibrium, which for a system with both quarks and gluons are

\begin{equation}
    m^2_{g}[f_{\mathrm{C}}] = \frac{24\alpha_s}{\pi}T^2, \quad m^2_{q/\bar{q}}[f_{\mathrm{C}}] = \frac{32\alpha_s}{3\pi}T^2
\end{equation}
and

\begin{equation}
    m^2_{g}[f_{\mathrm{BE/FD}}] = 3\pi \alpha_s T^2, \quad m^2_{q/\bar{q}}[f_{\mathrm{BE/FD}}] = \frac{4\pi \alpha_s}{3} T^2.
\end{equation}
These comparisons are shown in \Fig{fig:m2_N50} and \ref{fig:m2_vs_Ninc}. The numerical values are extracted directly after initialisation when the partons all have the same observer time and the complication due to causality violation mentioned above is absent. As shown in \Fig{fig:m2_vs_Ninc}, the mean value converges to the analytical expectation for both quarks and gluons, however, it requires a rather large number of particles to be included in the calculation. Approximating partons using Gaussian distributions (see \Sect{subsection:alpaca_inelastic_tstar}) instead of Dirac delta functions might improve on this convergence rate. In \Fig{fig:m2_vs_T} the scaling behaviour of the extracted $m_s^2$ against different temperatures can be seen.

\begin{figure*}[h!]
    \centering
    \includegraphics[width=0.495\linewidth]{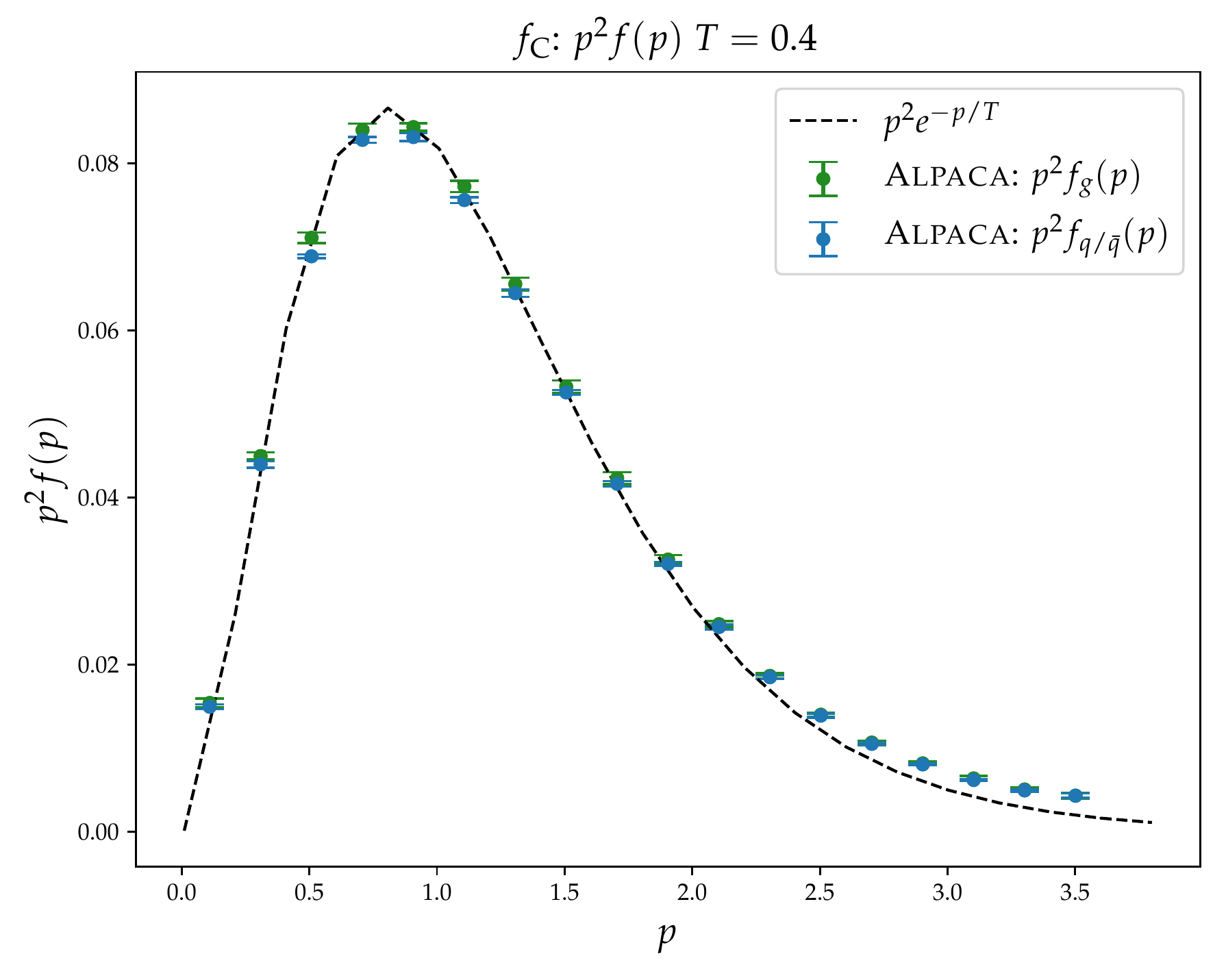}
    \includegraphics[width=0.495\linewidth]{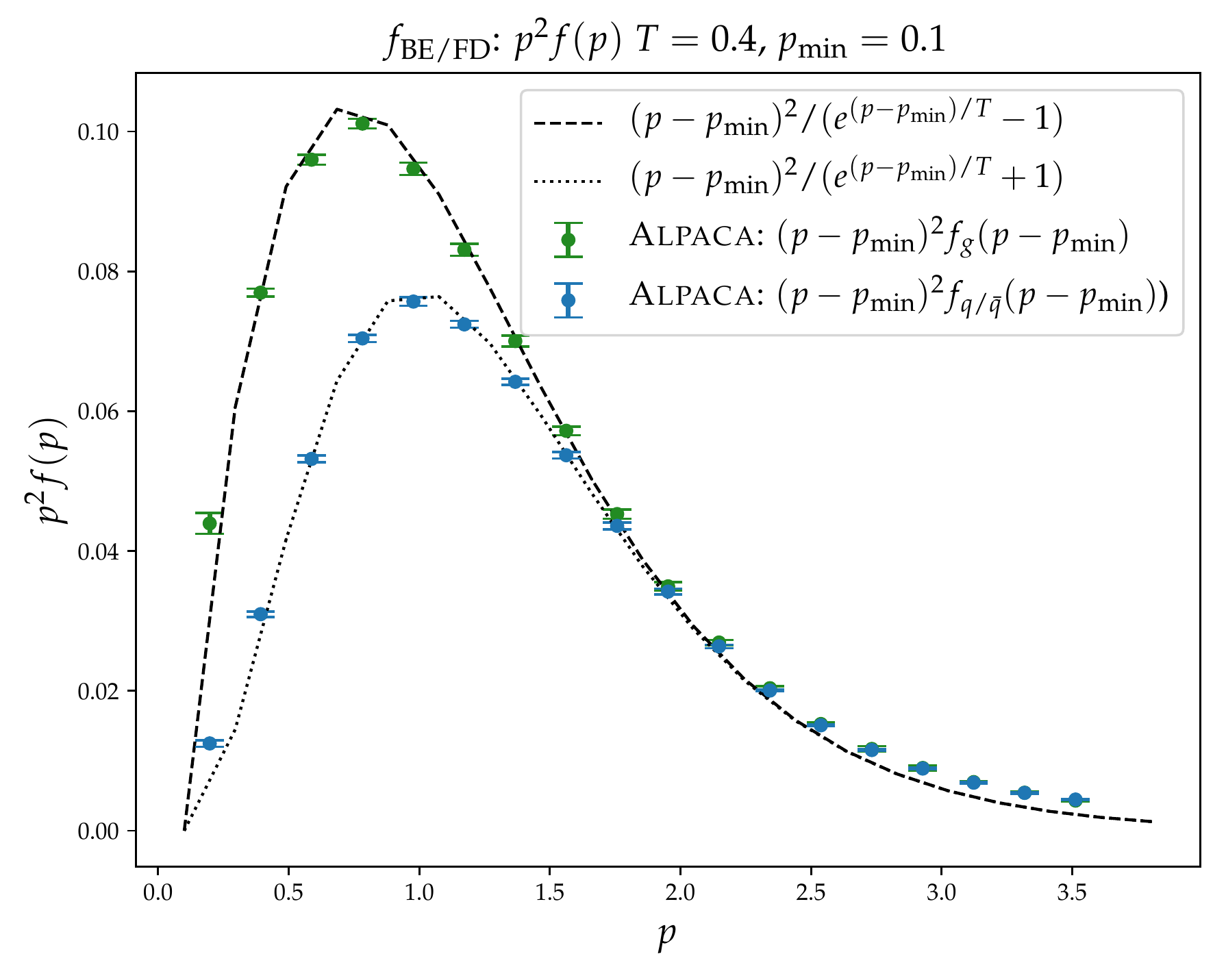}
    \caption{Extracted phase space density, $f(\x,\p)$, for thermal systems with temperature $T=0.4$ GeV. The values are extracted at initial $t$. Left: $f_{\mathrm{C}}$. Right: $f_{\mathrm{BE/FD}}$.}
    \label{fig:f_p}
\end{figure*}

Finally, a comparison of $f(\x,\p)$ can be seen in \Fig{fig:f_p}, which is shown to converge well to the analytical expectation. 

All runs in this section are initialized in a box with volume $V_x=7.6^3$ $\mathrm{GeV}^{-3}$ and $\alpha_s = 0.04$.

\bigskip

\subsection{Quasi-collinear splitting and merging processes}
\label{subsection:alpaca_inelastic}

The effective $``1\leftrightarrow 2"$ splitting/merging rates have been obtained to leading order in~\cite{Arnold:2002ja} in terms of solutions to a linear integral equation. An explicit analytical solution has been found in~\cite{Arnold:2008zu} for very energetic partons, where the LPM effect dominates the dynamics. In the incoherent Bethe-Heitler limit it is also possible to find an analytical solution. Between these  two limiting cases no analytical solution is known. We here use an interpolation that reproduces the analytical results in the Bethe-Heitler and LPM regimes and agrees well with numerical solutions between them. Details are given in \app{appendix:gamma}.

As mentioned in \Sect{section:amy}, the splitting/merging rates describe the splitting (merging) of a (two) parton(s) in conjunction with an arbitrary number of coherent soft scatterings. They thus depend on the density of partons generating soft fluctuations of the background gauge field. It turns out that the phase space densities enter via a single parameter $T_*$, which is the temperature of a system in thermal equilibrium that has the same infra-red behaviour as the system under consideration. In thermal equilibrium it thus reduces to the temperature. The extraction of $T_*$ in \textsc{Alpaca} is discussed in \Sect{subsection:alpaca_inelastic_tstar}.

\subsubsection{Splitting}
\label{subsection:alpaca_inelastic_splitting}

Soft scatterings within the medium can bring partons slightly off-shell which kinematically allows for them to split nearly collinearly into two new partons. The average formation time of this process coincides with the mean free time for soft elastic collisions within the medium and so all soft scatterings occurring during the formation time of the splitting have to be considered coherently. Destructive interference leads to a suppression of the splitting rate compared to the incoherent case, which is known as the LPM effect in QCD. 

We here assume that subsequent splittings are independent and the time evolution of a parton undergoing repeated splitting can be formulated as a Markov chain\footnote{Possible interference effects in subsequent radiation processes with overlapping formation times are discussed in~\cite{Arnold:2023qwi} (and references therein)}. This is a well studied problem in event generators, and a common approach to sample the occurrence of particle branchings is the Veto Algorithm. 

This algorithm can be employed when the splitting probability is of the form
\begin{equation}
    \label{eq:sudakov_prob}
    P_a(t,x) = -\frac{dN_a}{dt} = h_a(t,x)N_a(t)
\end{equation}
for a particle of species $a$, where $x$ (and $1-x$) correspond to the fractions of energy going to the two new particles. Here, $P_a(t,x)$ is the probability for the particle to split at time $t$ with energy fraction $x$, $N_a(t)$ is the probability that a particle of species $a$ has not split yet at $t$, also known as the Sudakov factor, and $h_a(t,x)$ is the instantaneous splitting probability (density) at $t$ and $x$. We solve the splittings in \textsc{Alpaca} by using an overestimate version of the Veto Algorithm. This means the following: We find an overestimate $\tilde{h}_a(t,x)\geq h_a(t,x)$ of the instantaneous splitting probability and then use $\tilde{h}_a(t) = \int \tilde{h}_a(t,x)dx$ to sample $t_{\mathrm{split}}$. The energy fraction $x$ is then sampled uniformly between $x_\mathrm{min}$ and $x_\mathrm{max}$. The draw is accepted or rejected based on the ratio $h_a(t,x)/\tilde{h}_a(t,x)$. If the splitting is rejected we do not reset in time, but set the previously sampled $t_{\mathrm{split}}$ to be our new initial time, and redo the sampling of splitting time and energy fraction until we find a splitting that is accepted. When multiple decay channels are available, $t_\mathrm{split}$ is sampled from the sum of the corresponding splitting probabilities, and the correct decay channel is selected based on the ratio of the splitting probabilities, once the splitting has been accepted.

To implement this algorithm, two quantities are needed: $h_a(t,x)$ and an overestimate $\tilde{h}_a(t,x)\geq h_a(t,x)$. We have previously defined the number of particles of species $a$ in a phase space volume $d^6\xi$ at time $dt$ in \Eq{eq:ddN}, and the number of particles of species $a$ leaving $d^6\xi$ during $dt$ in \Eq{eq:ddNdt}. Hence, the (total) instantaneous splitting probability for a particle in $d^6\xi$ of species $a$ during time $dt$ is 

\begin{equation}
    h_a =  \frac{\frac{d(dN_a^-)}{dt}}{dN} = \frac{C_a^{-, ``1\rightarrow 2"}[f(\x,\p)]}{f(\x,\p)}.
\end{equation}
The term in the splitting/merging kernel, \Eq{eq:C_inelastic_collinear}, which corresponds to a particle of species $a$ with momentum $\p$ splitting to two particles with momentum $\pp$ and $\kp$ in terms of the quasi-collinear branching rate $\gamma^a_{bc}$ is

\begin{align}
    & C_a^{-,``1\leftrightarrow 2"} = \frac{(2\pi)^3}{2|\p|^2\nu_a}\sum_{b,c}\int_0^\infty dp'dk'\delta(p-p'-k')\nonumber\\ 
    & \times \gamma^a_{bc}(\p;p'\phat,k'\phat)  f_a(\p)[1\pm f_b(p'\phat)][1\pm f_c(k'\phat)]
\end{align}
and so

\begin{align}
    \label{eq:h_a}
    h_a = & \int_0^1 dx\frac{(2\pi)^3}{2|\p|\nu_a}\sum_{b,c} \gamma^a_{bc}(\p;xp\phat,(1-x)p\phat) \nonumber \\ 
    & \times [1\pm f_b(xp\phat)][1\pm f_c((1-x)p\phat)].
\end{align}
The instantaneous splitting probability as a function of $x$ is then extracted from \Eq{eq:h_a} as

\begin{align}
    \label{eq:splitting_h}
    h_a(x) = & \frac{(2\pi)^3}{2|\p|\nu_a}\sum_{b,c} \gamma^a_{bc}(\p;xp\phat,(1-x)p\phat) \nonumber\\ 
    & \times[1\pm f_b(xp\phat)][1\pm f_c((1-x)p\phat)].
\end{align}
In \app{appendix:gamma} a detailed account of how to calculate $\gamma^a_{bc}$ is given, and an overestimate of the instantaneous splitting probability given above can be found in \app{appendix:splitting_probability_overestimate}. Note that the overestimates of $h_a(x)$ depend on $T_*$, which we assume does not change significantly over our relevant timescales. Hence, for the overestimates we sample $T_*$ at current $\tau$ (again taking the system to constant observer time using the same procedure as for the effective masses) and multiply with some fixed numerical factor $c>1$ such that our overestimate effective temperature is $cT_*$. The effective temperature is then re-calculated at $\tau_{\mathrm{split}}$ when evaluating $h_a(x)$.

The quantities $h_a(x)$ and $\tilde{h}_a(x)$ are used in the Veto Algorithm to sample $x$ and $t_{\mathrm{split}}$, as described in the beginning of this section, and this gives us a corresponding $\tau_{\mathrm{split}}$. Once a $\tau_{\mathrm{split}}$ is found, a formation time $t_{\mathrm{form}}$ is calculated according to the corresponding expression given in \cite{Arnold:2002zm},
\begin{equation}
    t_{\mathrm{form}} = \frac{p}{m_g^2}\left[1+\frac{g^4T_*p}{m_g^2} \right]^{-\frac{1}{2}}.
\end{equation}
A $\Delta \tau$ interval of length $\tau_{\mathrm{form}}$ corresponding to $t_{\mathrm{form}}$ is placed around $\tau_{\mathrm{split}}$ such that $\tau_{\mathrm{split}}$ is distributed uniformly in the interval. During this formation time the particle cannot scatter elastically or quasi-collinearly merge because elastic scattering during the formation time is re-summed in the splitting kernel and hard elastic scattering and merging are parametrically rare so that neglecting them does not affect the theory's accuracy. When the parton later splits at $\tau_{\mathrm{split}}$ two new partons are created with kinematics following the massless case of final-state emitter and final-state spectator of \textsc{Sherpa}'s Catani-Seymour shower described in \cite{Schumann:2007mg}, and they are assigned the remaining formation time of the original particle. The splitting probabilities $h_a(x)$ does not factor in any relative transverse momentum $k_\perp$ for the new parton pair, since they are derived from the AMY collision kernel which assumes a nearly exactly collinear process. Deviating slightly from this assumption, we allow for a small $k_\perp$ for the new parton pair sampled from the soft gluon spectrum $dk_\perp^2/(k_\perp^2+k_{\perp, \mathrm{reg}}^2)$ for some fixed $k_{\perp, \mathrm{reg}}^2 \ll 1$. The recoil is absorbed by an additional parton such that all partons remain on-shell. This recoil parton is chosen as the (spatially) closest located parton to the parton splitting (which in addition allows the outgoing partons to all have a momentum larger than our cutoff $p_{\mathrm{min}}$).

\subsubsection{Merging}
\label{subsection:alpaca_inelastic_merging}
In the case of evaluating if two particles merge our initial state is a pair, and so we cannot utilize the same methods as we have for particles splitting. Instead, we derive an effective cross section from the merging rate and follow the same approach that we use for elastic scatterings, i.e. to evaluate if two particles merge we consider their Lorentz invariant distance $d_ {ij}^2$ at their closest approach, and if it falls within the effective cross section of merging. In \app{section:relation_between_gamma_M} it is shown that the quasi-collinear splitting rate, $\gamma^a_{bc}$, is related to the effective matrix element $|\mathcal{M}^a_{bc}|^2$ through
\begin{equation}
    \gamma_{ab}^c = \frac{\nu_a\nu_b}{8(2\pi)^4}|\mathcal{M}_{ab}^c|^2.
\end{equation}
It follows, by definition, that for two particles of momentum $\p$ and $\k$ merging into a particle $\pp$ the differential cross section can be expressed as
\begin{equation}
    \frac{d\sigma^{2\rightarrow 1}_{ab}}{d\Phi_\pp} = \frac{|\mathcal{M}_{ab}^c|^2}{2s}[1\pm f(\pp)] = \frac{4(2\pi)^4}{\nu_a\nu_b}\frac{\gamma^c_{ab}}{s}[1\pm f(\pp)]
\end{equation}
with
\begin{equation}
    d\Phi_\pp = \frac{d^4P'}{(2\pi)^3} \delta(P'^2)\theta(P_0') (2\pi)^4 \delta^{(4)}(P'-P-K)
\end{equation}
being the phase space measure for the outgoing particle. The resulting cross section is then
\begin{align}
    \sigma^{2\rightarrow 1} & = \int \frac{d\sigma^{2\rightarrow 1}}{d\Phi_\pp}d\Phi_\pp  \nonumber\\
    & = \frac{4(2\pi)^5}{\nu_a\nu_b} \frac{\delta(s)}{s}\gamma^c_{ab}(t,x,p,k)[1\pm f(\pp)] \nonumber\\
    & = \frac{4(2\pi)^5}{\nu_a\nu_b} \frac{1}{s} \frac{\delta(k_\perp^2)\gamma^c_{ab}(t,x,p,k)}{\left(\frac{E_\p}{E_\k} + \frac{E_\k}{E_\p} + 2\right)}[1\pm f(\pp)]
\end{align}
where $k_\perp^2$ is the relative transverse momentum between $\p$ and $\k$. The Dirac delta present in the cross section enforces the exact collinearity of the merging processes, which would be the only kinematically allowed merging in vacuum. As with the splitting however, we circumvent this condition by introducing a recoil parton to absorb the transverse momentum. Hence, our merging process will allow partons with a non-zero $k_\perp$ to merge. We relate the Dirac-delta to the $k_\perp^2$-distribution used for sampling splittings as $\delta(k_\perp^2) \rightarrow f(k_\perp^2)/2$ where

\begin{equation}
     f(k_\perp^2 )=  \left[(k_\perp^2 + k_{\perp, \mathrm{reg}}^2)\log\left(\frac{k_{\perp\mathrm{max}}^2+k_{\perp, \mathrm{reg}}^2}{k_{\perp, \mathrm{reg}}^2}\right)\right]^{-1}
\end{equation}
is the $k_\perp^2$-distribution normalised to unity 
 and we pick up a factor of $1/2$ since $\delta(k_\perp^2)$ is not defined for $k_\perp^2 < 0$ and so only integrates to $1/2$. This gives us our final expression of the merging cross section,

\begin{align}
    & \sigma^{2\rightarrow 1} = \frac{2(2\pi)^5}{\nu_a\nu_b} \Bigg[s \left(\frac{E_\p}{E_\k} + \frac{E_\k}{E_\p} + 2\right)(k_\perp^2 + k_{\perp, \mathrm{reg}}^2) \nonumber  \\ 
    & \times\log\left(\frac{k_{\perp\mathrm{max}}^2+k_{\perp, \mathrm{reg}}^2}{k_{\perp, \mathrm{reg}}^2}\right)\Bigg]^{-1} \gamma^c_{ab}(t,x,p,k)[1\pm f(\pp)].
\end{align}

As in the case of elastic scattering the merging rates $\gamma$ depend on the local quantity $T_*$ and the kernels also contain Bose enhancement/Pauli blocking factors, which we deal with in the same way as before: when integrating the effective cross section we overestimate the merging rate and the Bose enhancement/Pauli blocking factors and later reject mergings with the ratio of the true to and the overestimated effective cross section.

To calculate the cross section at the closest approach for a parton pair the energy fraction $x$ is also needed, for which we use the Lorentz-invariant definition
\begin{equation}
    x = \frac{P_\mu R^\mu}{P_\mu R^\mu + K_\mu R^\mu }
\end{equation}
where $P^\mu$ and $K^\mu$ are the 4-momenta of the incoming partons, and $R^\mu$ is the 4-momentum of the recoiler, as in \cite{Schumann:2007mg}. 

As mentioned earlier, since the splitting rate is infra-red divergent we have to introduce a small cut-off in $x$. For consistency we apply the same cut-off to the merging rate. This does not compromise the theory's accuracy.\footnote{As pointed out in~\cite{Arnold:2002zm} in the infra-red the divergent rates cancel between gain and loss terms. However, in the parton cascade splitting and merging are algorithmically different processes and we can therefore not rely on this cancellation and have to regularise the rates in the infra-red instead.}

Once a closest approach which falls within the cross section of the merging is found, a formation time is assigned to the particle pair in the same manner as for a splitting parton, see \Sect{subsection:alpaca_inelastic_splitting}. If the merging is accepted in the rejection step to correct to the true cross section, the momenta are updated following (inversely) the massless case of final-state emitter and final-state spectator in \cite{Schumann:2007mg}, where the final-state spectator, or reocil parton, is chosen to be the parton with the closest spatial distance to the merging pair (that also allow for kinematics where no parton ends up with $p<p_{\mathrm{min}}$). The remainder of the formation time of the original pair is also assigned to the new parton.

\subsubsection{The effective temperature $T_*$}
\label{subsection:alpaca_inelastic_tstar}

In the quasi-collinear splitting/merging rate $\gamma$ the quantity $T_*$ appears as a local variable (see \app{appendix:gamma}), this is the effective temperature at which an equilibrium system would have the same soft elastic scattering rate as the system under consideration. The definition of $T_*$ averaged over a spatial volume $V$ is given by 
    
\begin{align}
    \label{eq:Tstar}
    T_* & = \frac{\frac{1}{2}g^2 \sum_s\frac{\nu_s C_s}{d_A}\frac{1}{V}\int d^3\x \int \frac{d^3\p}{(2\pi)^3}f_s(p)[1+f_s(p)]}{g^2\sum_s\frac{ \nu_s C_s}{d_A}\frac{1}{V}\int d^3\x \int\frac{d^3\p}{(2\pi)^3}f_s(p)/p}  
\end{align}
where we note that the denominator is equal to $m_g^2$, and the integrals again have to be calculated at fixed observed time and in the local rest frame. Our prescription to extract $T_*$ locally follows along the lines of the method used to find $m_{g/q}^2$, see \Sect{subsection:alpaca_elastic_m2}, though with a key difference. In our treatment of $m_{g/q}^2$ we considered all particles to be point-like, an assumption that for $T_*$ introduces problems with squared Dirac delta terms of the same variables. To remedy this we instead consider partons as Gaussian distributions with width $\sigma_\p$ and $\sigma_\x$ in the phase space density, i.e. 

\begin{equation}
    f_a(\x,\p) = \frac{(2\pi)^3}{\nu_a} \sum_i \frac{e^{-\frac{(\x_{i}-\x)^2}{2\sigma_{i,\x}^2}} e^{-\frac{(\p_{i}-\p)^2}{2\sigma_{i,\p}^2}}}{(2\pi\sigma_{i,\x}\sigma_{i,\p})^3} 
\end{equation}
where the sum is over all partons $i$ of species $a$. Assuming $\sigma_{i,\x}=\sigma_\x$ and $\sigma_{i,\p}=\sigma_\p$ for all partons $i$, we have for $m_g^2T_*$ that
\begin{align}
    m_g^2T_* & = \frac{\pi\alpha_s}{4V} \sum_{s} C_s \sum_{i(s)} \Bigg[1 + \frac{1}{\nu_{s}(2\sigma_{\x}\sigma_{\p})^3} \nonumber  \\
    & + \sum_{j(s) > i(s)} \frac{1}{\nu_{s}4(2\sigma_\x\sigma_\p)^3} e^{-\frac{(\x_i-\x_j)^2}{4\sigma_\x^2}}e^{-\frac{(\p_i-\p_j)^2}{4\sigma_\p^2}} \Bigg]
\end{align}
with $\sigma_{\x}\sigma_{\p}\geq 1/2$ due to the uncertainty principle. With $T_*$ expressed as above we can proceed to extract it locally in the same manner as in \Sect{subsection:alpaca_elastic_m2}. We include the $N_{\mathrm{inc}}$ closest particles, where each particle $i$ contributes with a term

\begin{equation}
    \frac{\pi\alpha_sC_s}{4V}\left[1 + \frac{1}{\nu_{s(i)}(2\sigma_{\x}\sigma_{\p})^3} \right]
\end{equation}
and each particle pair $i$, $j$ of the same species $s$ adds a term
\begin{equation}
    \frac{\pi\alpha_sC_s}{16V\nu_{s(i)}(\sigma_\x\sigma_\p)^3} e^{-\frac{(\x_i-\x_j)^2}{4\sigma_\x^2}}e^{-\frac{(\p_i-\p_j)^2}{4\sigma_\p^2}}.
\end{equation}

\subsubsection{The thermal equilibrium case}
\label{subsection:alpaca_inelastic_thermal}

To verify the implementation of the splitting/merging kernel in \textsc{Alpaca} we once again utilize the simplicity of an infinite thermalized system. Since the splitting and merging rates are controlled by $T_*$, which only equals $T$ for $f_{BE/FD}$, we will mainly focus our analysis to this initial distribution in the rest of this section, with the exception of looking at fixed $\gamma$. We will verify the following:

\begin{enumerate}[label=(\roman*)]
    \item The total splitting and merging rates $dN_{\mathrm{split}}/dt$ and $dN_{\mathrm{merge}}/dt$ are equal, and correspond to independently extracted numerical values.
    \item The effective temperature $T_*$ is extracted so that it matches the analytical expected value.
\end{enumerate}

The effects of formation time are not of interest for these verifications, since we do not run inelastic and elastic processes simultaneously here, and so formation time for both splitting and merging have been set to zero.

\bigskip

\noindent \textbf{(i)} It follows from \Eq{eq:ddNdt} that for an isotropic system the splitting rate can be related to the collision kernel as 

\begin{align}
    \label{eq:splitting_rate_derivation}
    \frac{dN_{\mathrm{split}}}{dt} & = \sum_s \frac{d}{dt}\int d(dN_s^-) \nonumber\\
    & = \sum_s \int d^6 \xi\frac{\nu_s}{(2\pi)^3} C_s^{-, ``1\rightarrow 2"}[f] \nonumber \\
    & = 2\pi V \sum_{sbc} \int_0^\infty dpdp'dk'\delta(p-p'-k')\gamma_{bc}^s \nonumber \\  
    & \times f_s(\p)[1\pm f_b(p'\hat{\mathbf{p}})][1\pm f_c(k'\hat{\mathbf{p}})].
\end{align}

For the case of a purely gluonic system with constant $\gamma^g_{gg}$ and no Bose enhancement it follows that the splitting and merging rates will not be the same for $f_{\mathrm{BE/FD}}$. Hence, we instead consider the same scenario for $f=f_{\mathrm{C}}$, where it follows that
\begin{align}
    \label{eq:splitting_rate}
    \frac{dN_{\mathrm{split}}}{dt} & = 2\pi V\gamma T^2.
\end{align}

\begin{figure}[h!]
    \centering
    \includegraphics[width=1\linewidth]{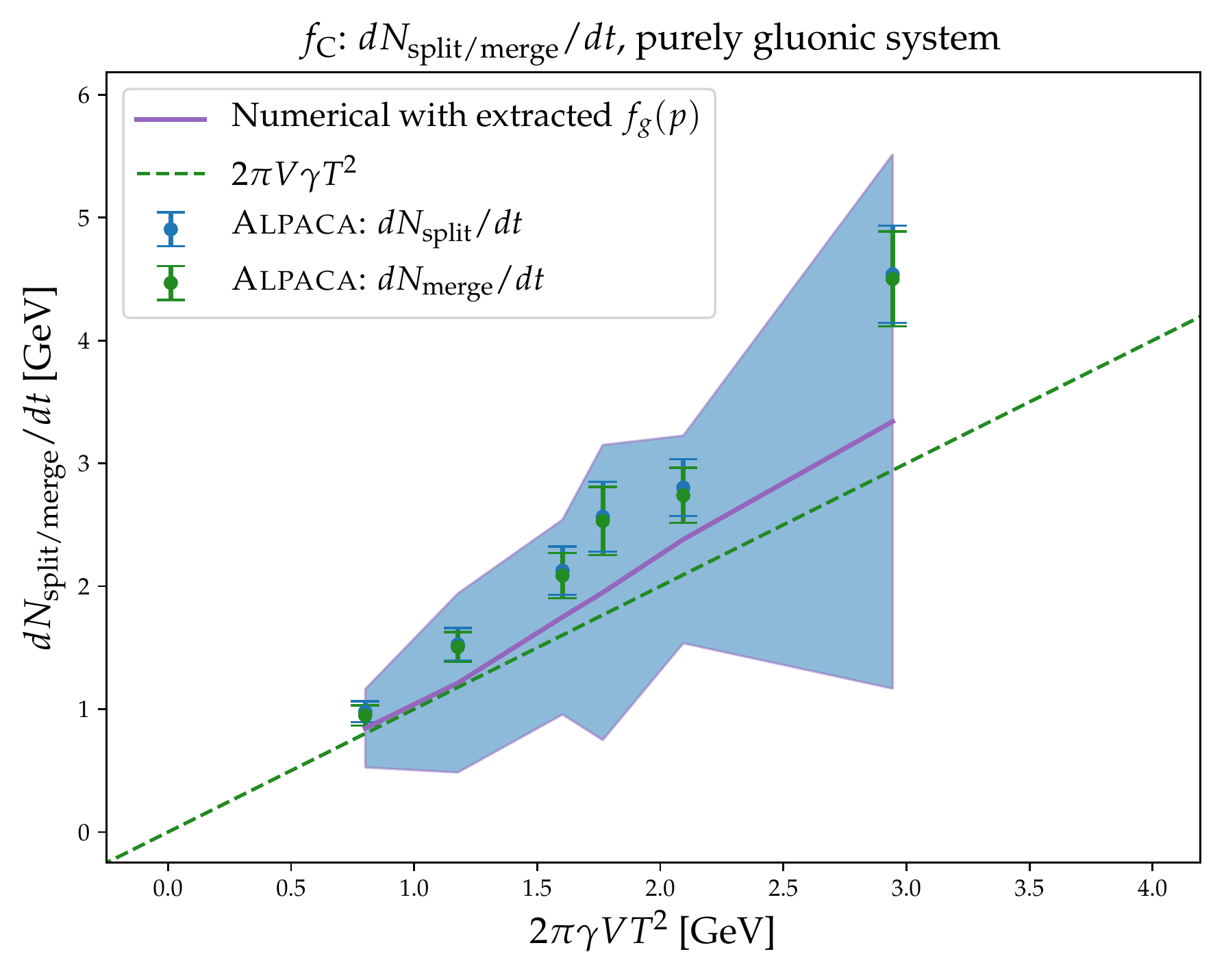}
    \caption{Scaling behaviour of the splitting and merging rates in a gluonic system with fixed $\gamma$ and no Bose enhancement factors are included. The system relaxes to a different phase space distribution, which is extracted at the end of the run. The resulting splitting/merging rate using the extracted distribution in \Eq{eq:splitting_rate_derivation} is shown as a purple line, with error bands in blue. The run parameters vary $T$ between $0.3-0.4$ GeV and $\gamma$ between $0.001-0.005$ $\mathrm{GeV}^2$. Fixed run parameters are $p_{\mathrm{min}}=0.01$ GeV, $\Delta t=250$~$\mathrm{GeV}^{-1}$ and box volume $V_x = 10.1^3$ $\mathrm{GeV}^{-3}$.}
    \label{fig:Nsplit_vs_gamma}
\end{figure}

The thermal equilibrium splitting and merging rates have been examined in \textsc{Alpaca} for fixed $\gamma$ in a gluonic system with no Bose-terms through a scan of the parameter space and the results are shown in \Fig{fig:Nsplit_vs_gamma}. As can be seen, the rates deviate from the analytical value given above. This follows from the fact that the system relaxes to a different distribution when no Bose enhancement is included. The new distribution that the system relaxes into is extracted at the end of the run and used in \Eq{eq:splitting_rate_derivation} to find the corresponding expected splitting/merging rate, which our results fall within errors of.

\begin{figure}[h!]
    \centering
    \includegraphics[width=1\linewidth]{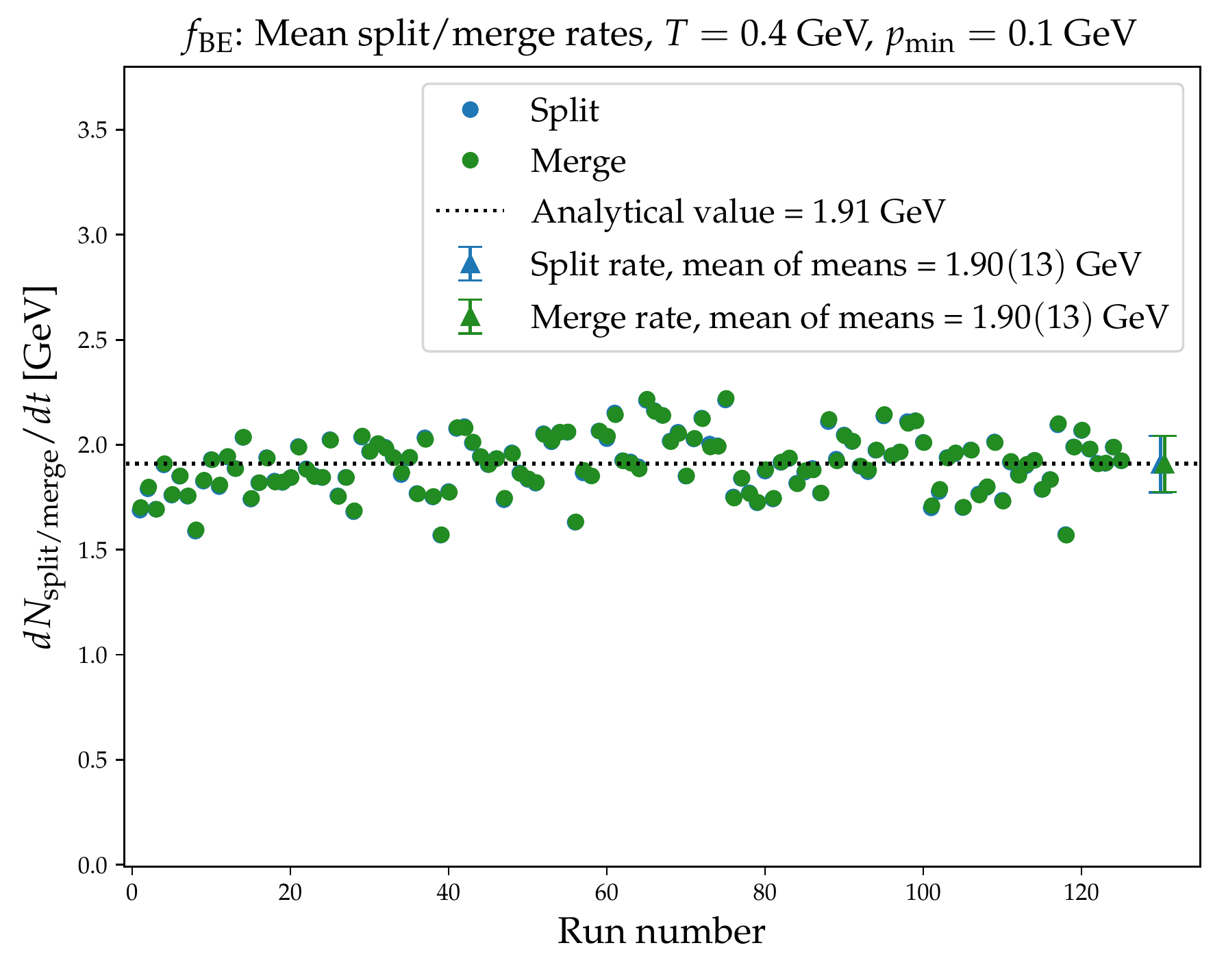}
    \caption{The splitting and mering rate for a gluonic thermal system with $\gamma$ calculated as described in \app{appendix:gamma}. The figure shows results from 125 runs with different initial configurations sampled from a thermal distribution with $\tau_{\mathrm{max}} = 2500$ $\mathrm{GeV}^{-1}$, $\alpha_s = 0.04$ and box volume $V_x = 10.1^3$ $\mathrm{GeV}^{-3}$.}
    \label{fig:Nsplit_dynamic_gamma}
\end{figure}

 We have also looked at the case of a dynamic $\gamma$, extracted as described in \app{appendix:gamma} for a purely gluonic system with $f=f_{\mathrm{BE}}$. The result of this is found in \Fig{fig:Nsplit_dynamic_gamma} and shown to correspond well with the (independently) numerically extracted rate.

\bigskip

\begin{figure}[h!]
    \centering
    \includegraphics[width=1\linewidth]{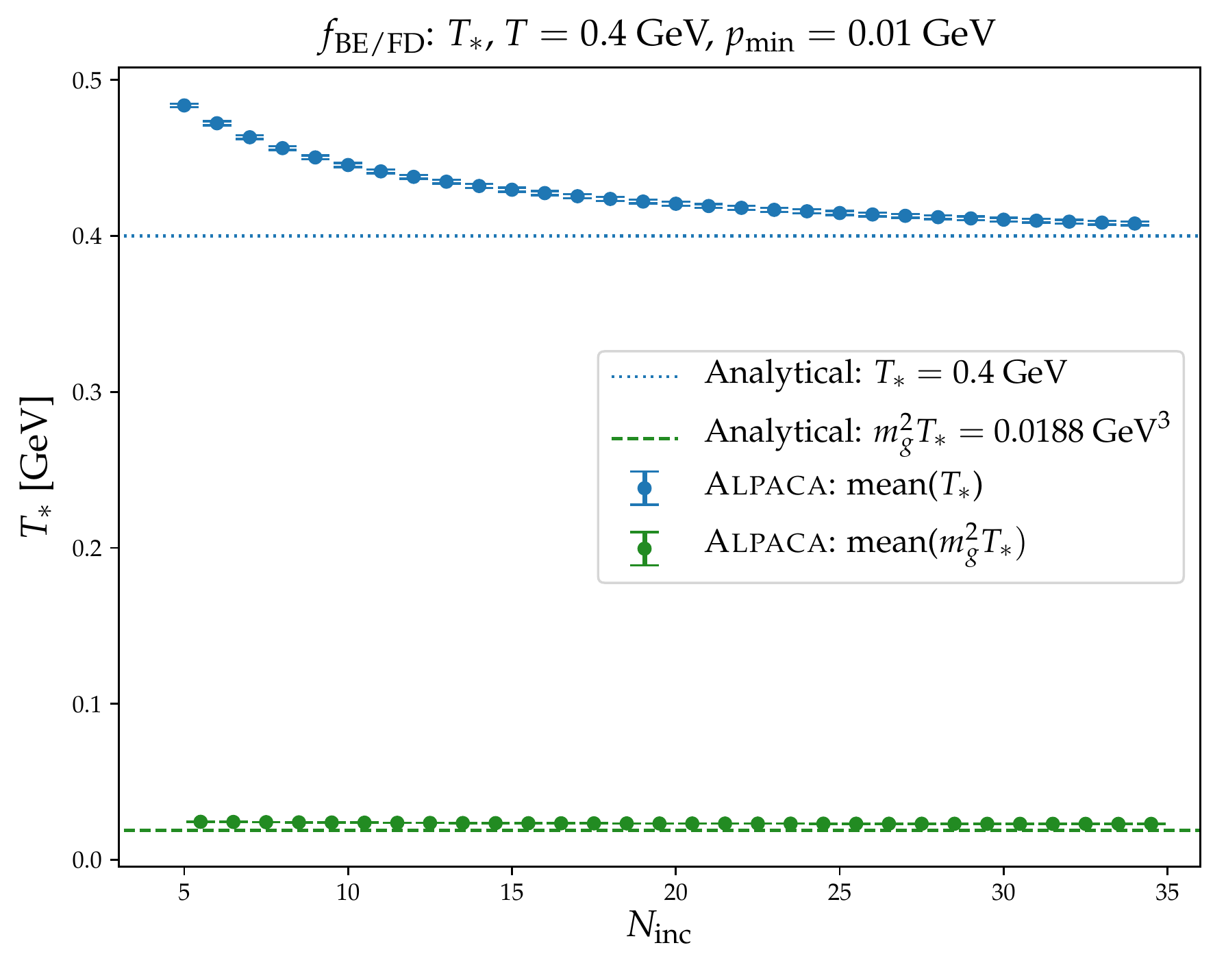}
    \caption{Scaling behaviour of $T_*$ and $m_g^2T_*$ (see \Eq{eq:Tstar}) as a function of number of particles included, $N_{\mathrm{inc}}.$}
    \label{fig:Tstar_vs_Ninc}
\end{figure}

\begin{figure}[h!]
    \centering
    \includegraphics[width=1\linewidth]{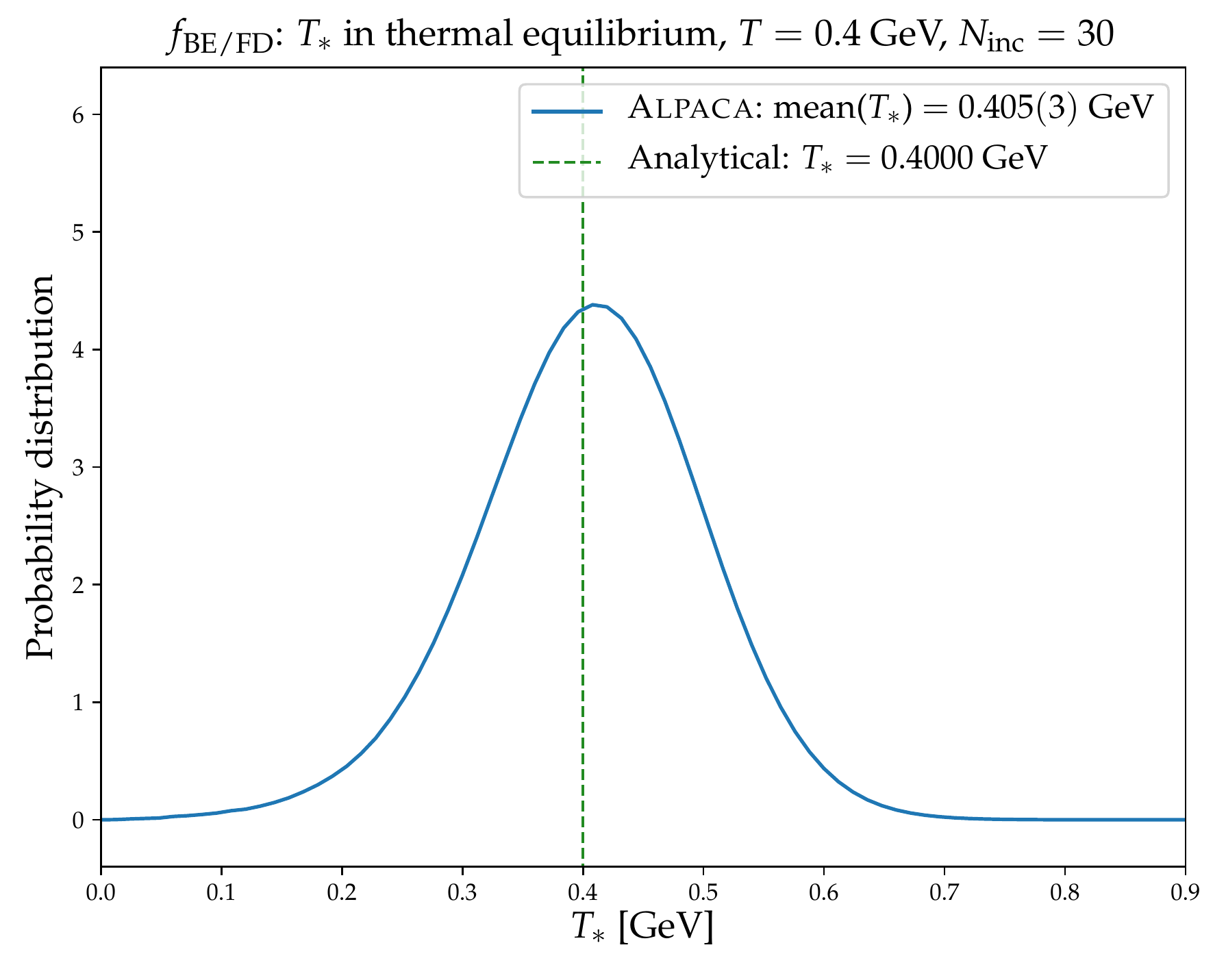}
    \caption{Distribution of extracted $T_*$ in a thermalized system with temperature $T=0.4$ GeV and $N_{\mathrm{inc}} = 30$.}
    \label{fig:Tstar_N30}
\end{figure}

\begin{figure}[h!]
    \centering
    \includegraphics[width=1\linewidth]{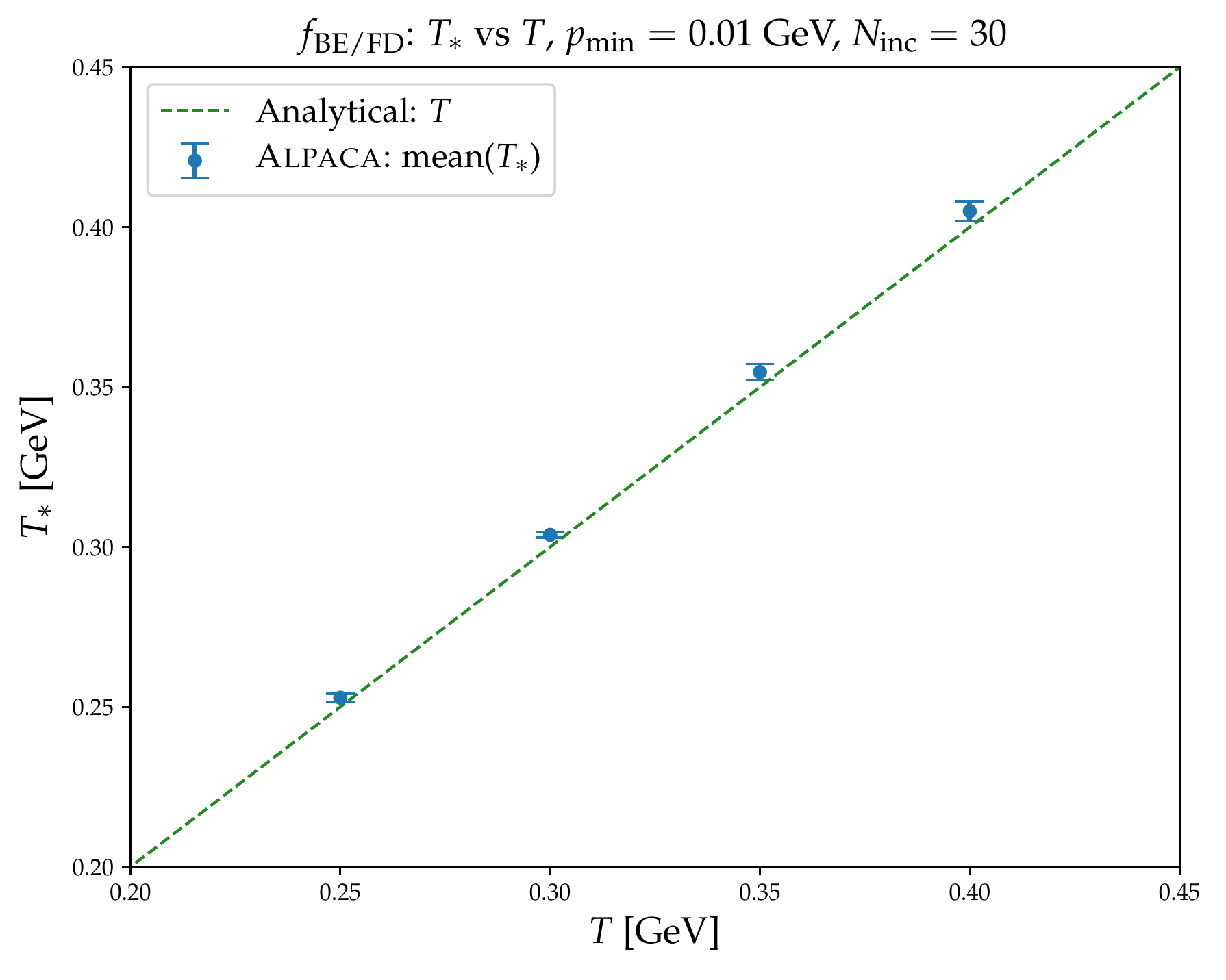}
    \caption{Scaling behaviour of $T_*$ as a function of different initialization temperatures $T$ for $f_{\mathrm{BE/FD}}$.}
    \label{fig:Tstar_vs_T}
\end{figure}

\noindent \textbf{(ii)} The thermal equilibrium provides us with a simple analytical solution to $T_*$ for $f_{\mathrm{BE/FD}}$, which is 
\begin{align}
    T_* & = T.
\end{align}
The convergence of the extracted $T_*$ towards the analytical expectation w.r.t. the number of included particles is shown in \Fig{fig:Tstar_vs_Ninc}. Note that it converges slower than $m_s^2$. This is due to the different implementation for extracting the numerator of $T_*$ compared to the denominator $m_g^2$.
The probability distribution of the extracted $T_*$, including a fixed number of particles, is shown in \Fig{fig:Tstar_N30}. The width of the distribution is due to the discrete sampling of $f_{\mathrm{BE/FD}}$, and will not decrease by increasing number of events. Lastly, a scan over different temperatures has been done to ensure that the effective temperature exhibits the correct scaling behaviour, as it is shown to do in \Fig{fig:Tstar_vs_T}. All values for $T_*$ shown in this section are extracted at initial $\tau$ in systems with box volume $V_x = 7.6^3$ $\mathrm{GeV}^{-3}$.

\subsection{Putting everything together}
\label{subsection:full_run}

Having verified all relevant parts of our implementation separately, we now proceed to the final verification, which is to combine all the pieces and let it run over longer timescales to ensure the dynamics of the system are correct, allowing it to remain in a thermal equilibrium.

There are two caveats to this long run though. Firstly, we have introduced a global cutoff in momentum, $p_{\mathrm{min}}$, which no particle is initialized below, and each process which produces outgoing momenta $k<p_{\mathrm{min}}$ is resampled until $k \geq p_{\mathrm{min}}$. This means that the we cannot expect our system to relax exactly into $f_{\mathrm{BE/FD}}$. Hence, we initialize our system using a shifted initial distribution,

\begin{align}
    p^2f_{\mathrm{BE/FD}} & \rightarrow \frac{(p-p_{\mathrm{min}})^2}{e^{(p-p_{\mathrm{min}})/T}\pm1} \nonumber \\
    & = (p-p_{\mathrm{min}})^2 f_{\mathrm{BE/FD, shift}}.
\end{align}

This causes the analytical values of $m_{g/q}^2$ and $T_*$ to shift compared to the regular Bose-Einstein and Fermi-Dirac case as well, and we extract them separately numerically to compare to our results in \textsc{Alpaca}.

Secondly, as mentioned in \Sect{subsection:alpaca_elastic_m2}, during the evolution the state of the system is known at fixed $\tau$, which means that  all particles have different times. But the effective masses $m_{g/q}^2$ and temperature $T_*$ have to be calculated at fixed time. The evolution of particles with later times than the required time can be backtraced. Since the future of particles with times earlier than the required time is still unknown, we let these particles free stream to the required time. This does, however, introduce a bias since splittings tend to take place at earlier $\tau$ than merging\footnote{Since high energy particles are more likely to split while low energy particles are more likely to merge, we will at any fixed tau on average have had more splittings than mergings following \Eq{Eq::relttau}. For any fixed $t$, the number of particles is the same on average, but at fixed $\tau$ there are on average more particles in the system. This is not fully corrected by letting earlier particles free-stream and hence, when calculating screening masses and effective temperature in this way we have slightly more particles in the system than we should have. This increases the value of $m_{g/q}^2$ which in turn decreases $T_*$.}.

The problem outlined above can be solved in an iterative fashion, by first running an event with $m_{g/q}^2$, $T_*$ and $f(p)$ extracted with the currently available particles at any given $\tau$. When the times $t_i$ for all particles $i$ have passed some $t_{\mathrm{max}}$, the run ends, and the history of position, momenta and flavour for each particle over all $t$ is saved. The system is then restarted and initialized with the same initial distribution that was sampled during the first event. During the course of the second event, $m_{g/q}^2$, $T_*$ and $f(p)$ at any given $t$ is then extracted at fixed time $t$ using the history of the particles from the previous run. The increase/decrease of the dynamic quantities in the first run will affect the dynamics of the evolution, and so the values extracted in the second run will not be exactly correct, though it will iteratively converge toward the correct values by repeating the same procedure, always extracting the dynamic values from the previous run. However, though this procedure should converge in an equilibrium setting there is no guarantee that it will for a system initialized out of equilibrium. A possible solution to remedy this for non-equilibrium systems could instead be to utilize the Lorentz invariant parton cascade setup presented in \cite{Nara:2023vrq}, which is a modified version of what is described in \Sect{section:lorentzinvariance}. The study of non-equilibrium distributions is however left for future publications.

In this chapter we present the results side by side from a run without iteration and a run with one iteration step. We define these cases as $N_{\mathrm{iter}}=0$ and $N_{\mathrm{iter}}=1$ respectively.

For the full run presented in this chapter the following is set/included.

\begin{itemize}
    \item Initialized with both gluons and quarks using $f_{\mathrm{BE/FD, shift}}$, $T = 0.4$ GeV, $p_{\mathrm{min}}=0.1$ GeV, $\alpha_s = 0.04$, box volume $V_x = 7.6^3$ $\mathrm{GeV}^{-3}$ , $t_{\mathrm{min}} = 0$ $\mathrm{GeV}^{-1}$ and $t_{\mathrm{max}} = 1000$ $\mathrm{GeV}^{-1}$, for $100$ separate runs with different initial configurations.
    \item All channels in $2\rightarrow2$ elastic scattering are allowed and evaluated including Bose/Pauli factors.
    \item All channels in $1\rightarrow2$ inelastic splitting/merging are allowed and evaluated including Bose/Pauli factors.
    \item The quantities $m_{g/q}^2$, $T_*$ and $f(p)$ are extracted dynamically from the parton ensemble in each run. In the run where $N_{\mathrm{iter}}=1$, the quantities are extracted using the parton position and momentum history from a previous run.
    \item Formation times are included for splitting and merging.
\end{itemize}

\begin{figure*}[h!]
    \centering
    \includegraphics[width=0.495\linewidth]{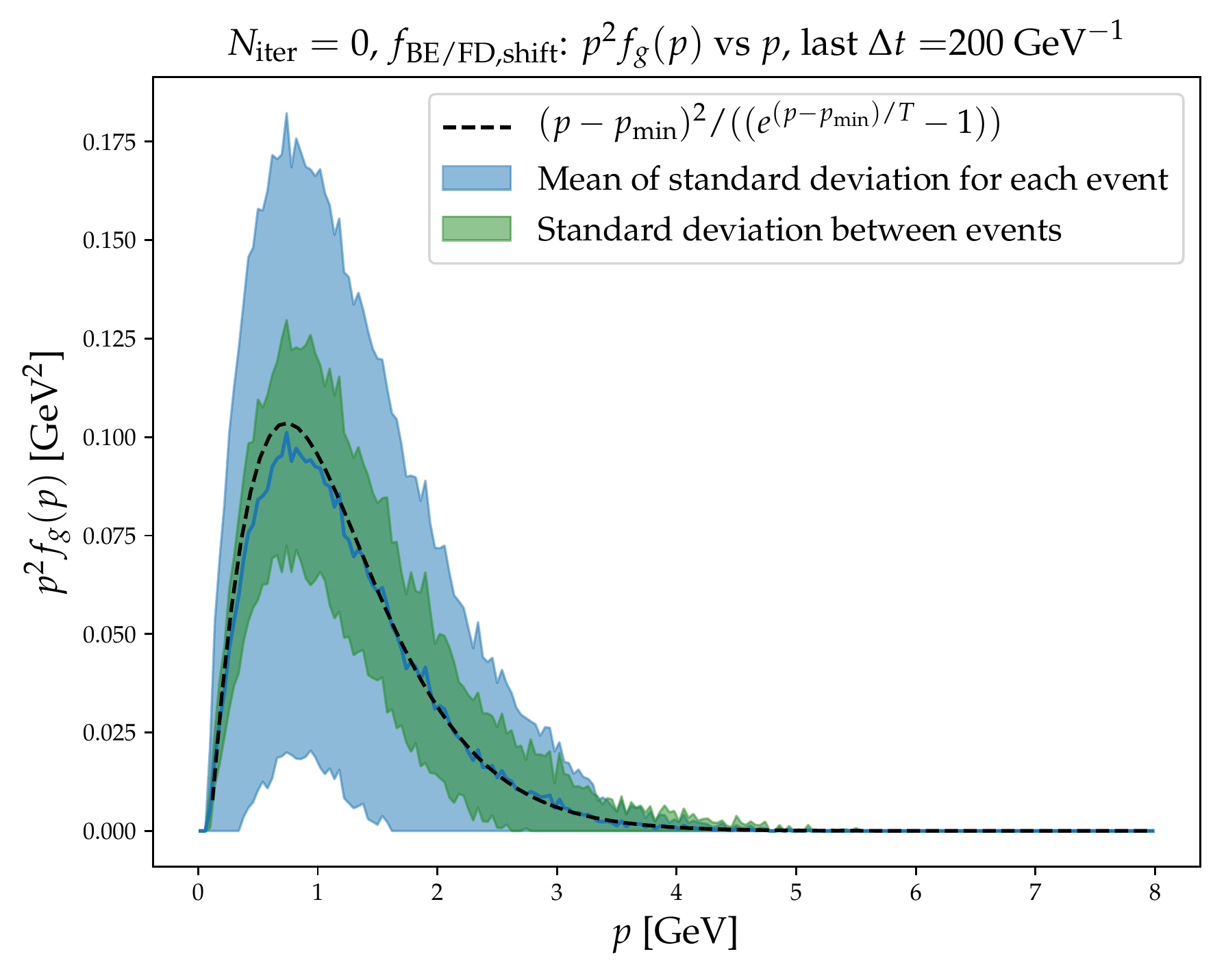}
    \includegraphics[width=0.495\linewidth]{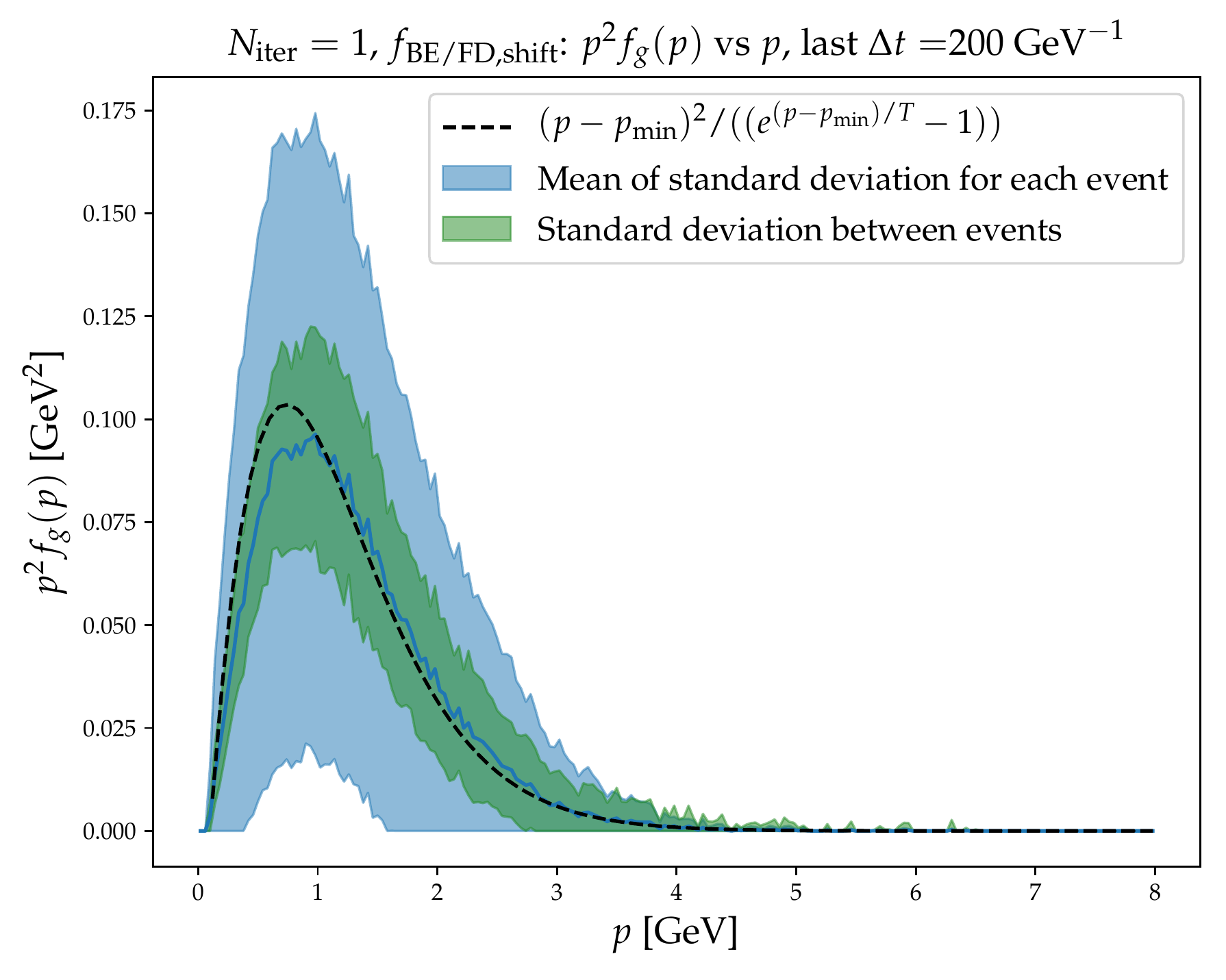}
    \includegraphics[width=0.495\linewidth]{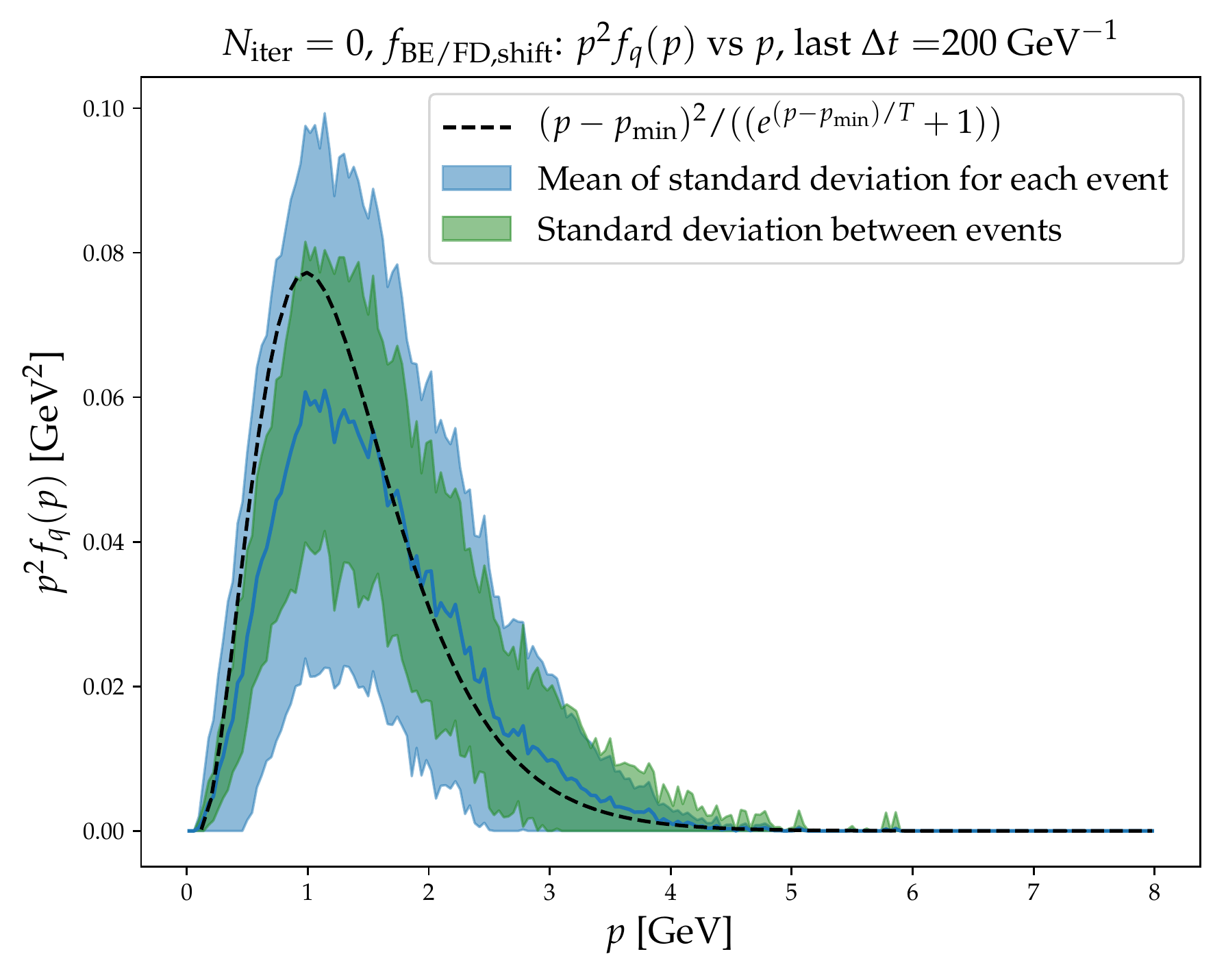}
    \includegraphics[width=0.495\linewidth]{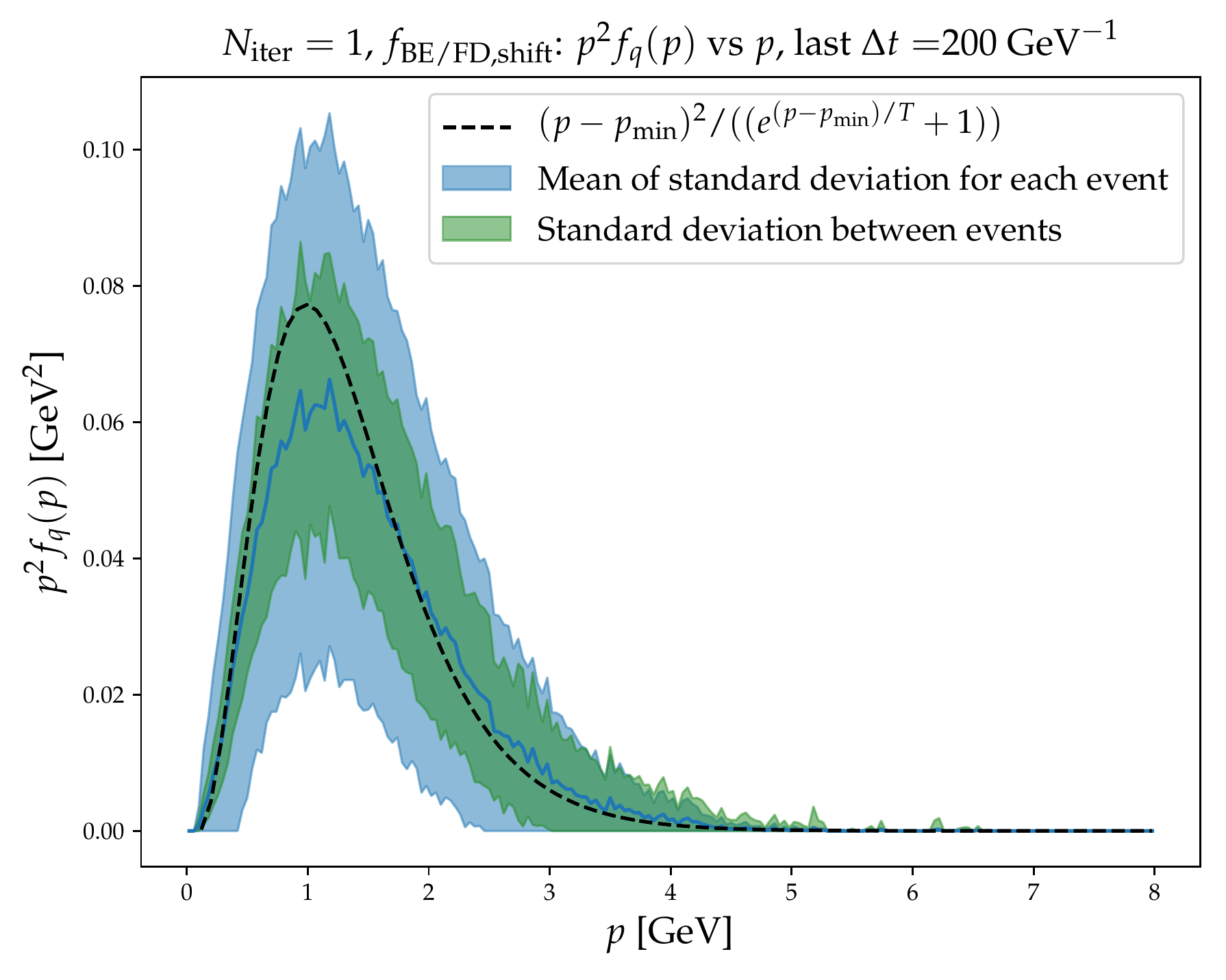}
    \caption{Distribution of the system extracted after the run is finished, averaged over the last $\Delta t = 200$ $\mathrm{GeV}^{-1}$. The blue error bands correspond to mean of the fluctuations within each event over time, while the green error bands correspond to the standard deviation between events. Left: $N_{\mathrm{iter}}=0$. Right: $N_{\mathrm{iter}}=1$. Top: $f_g(p)$. Bottom: $f_q(p)$.}
    \label{fig:fullrun_fg_fq}
\end{figure*}

\begin{figure*}[h!]
    \centering
    \includegraphics[width=0.495\linewidth]{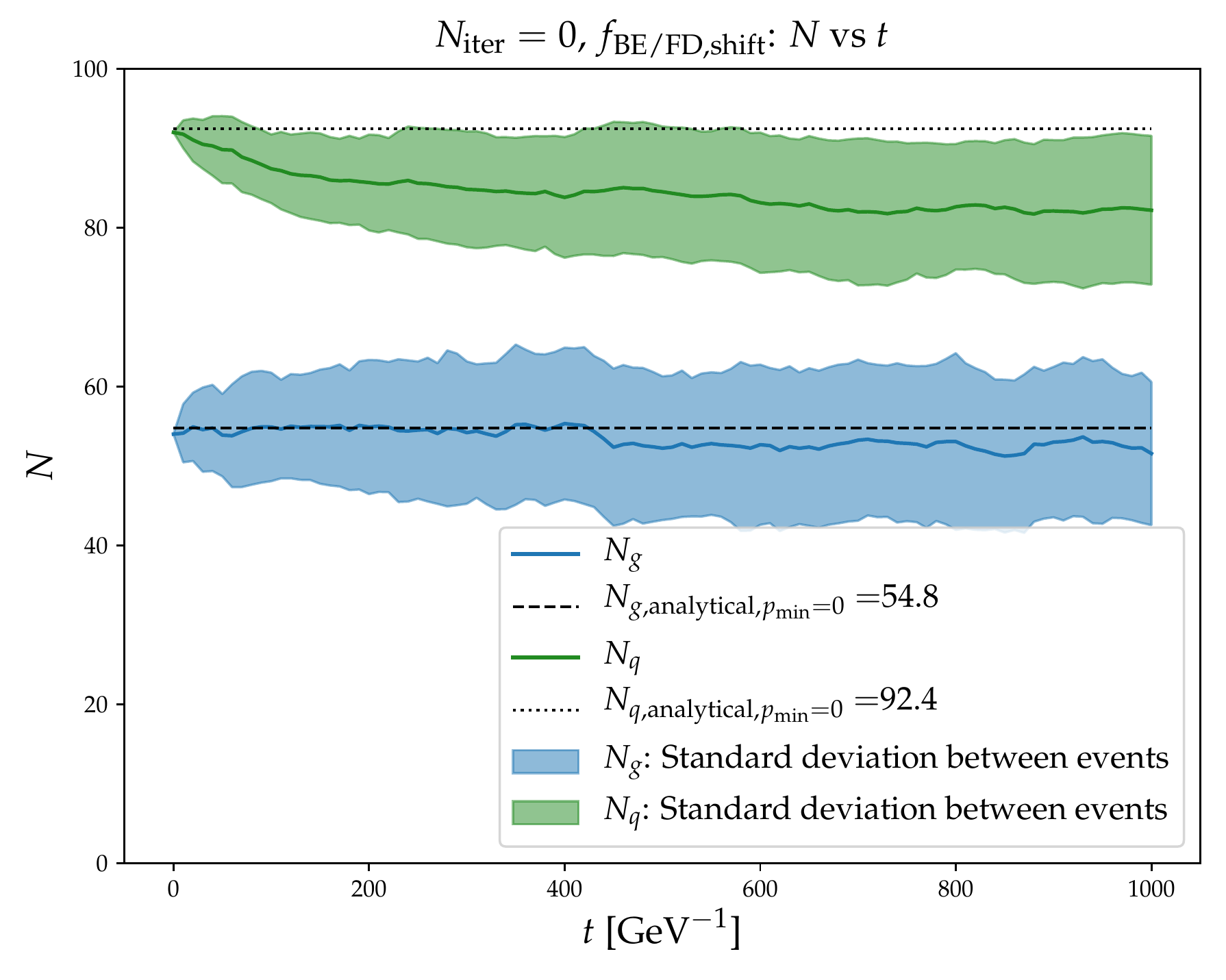}
    \includegraphics[width=0.495\linewidth]{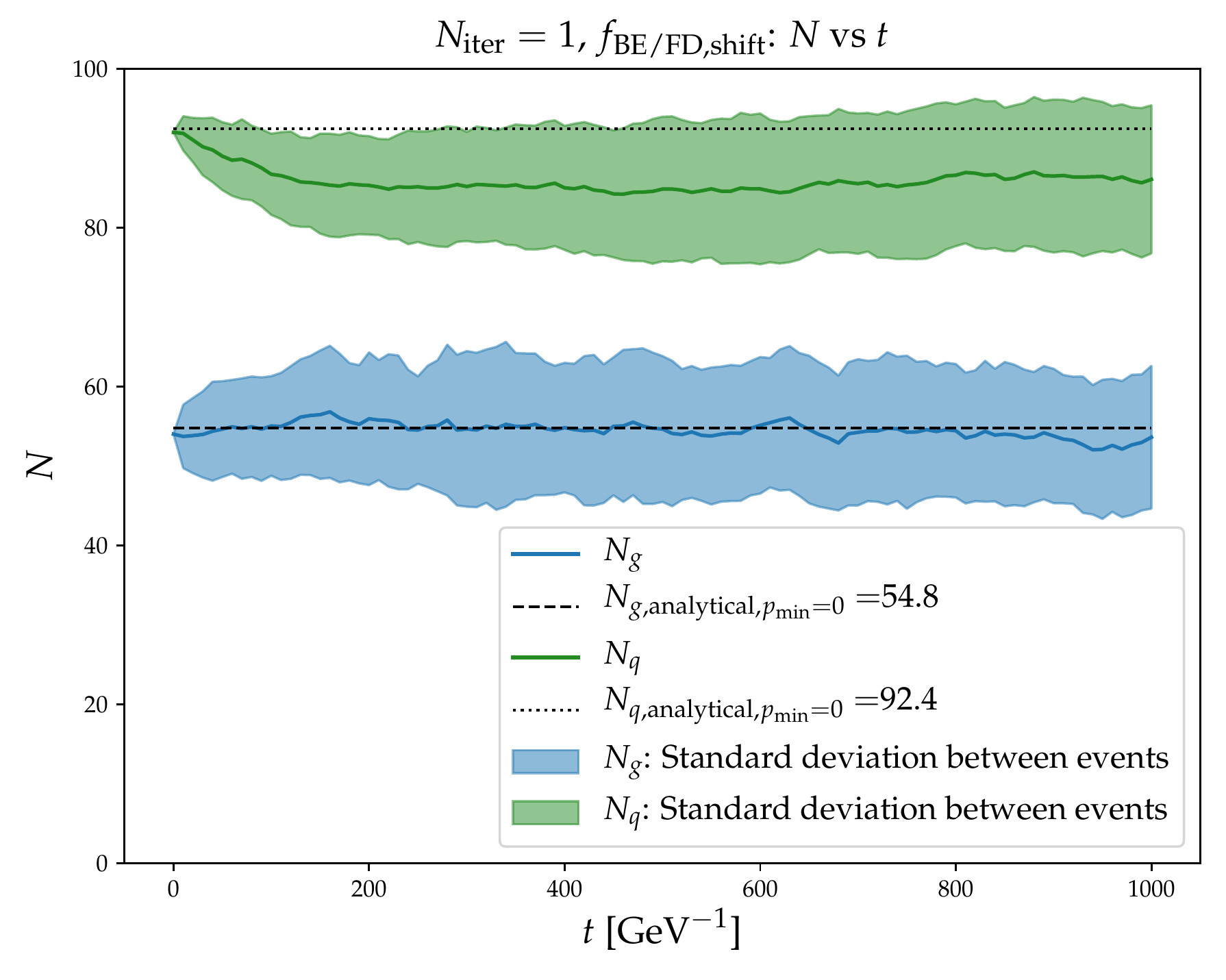}
    \caption{Mean number of particles in the system as a function of $t$. The error bands correspond to the standard deviation between events. Left: $N_{\mathrm{iter}}=0$. Right: $N_{\mathrm{iter}}=1$.}
    \label{fig:fullrun_N}
\end{figure*}

\begin{figure*}[h!]
    \centering
    \includegraphics[width=0.495\linewidth]{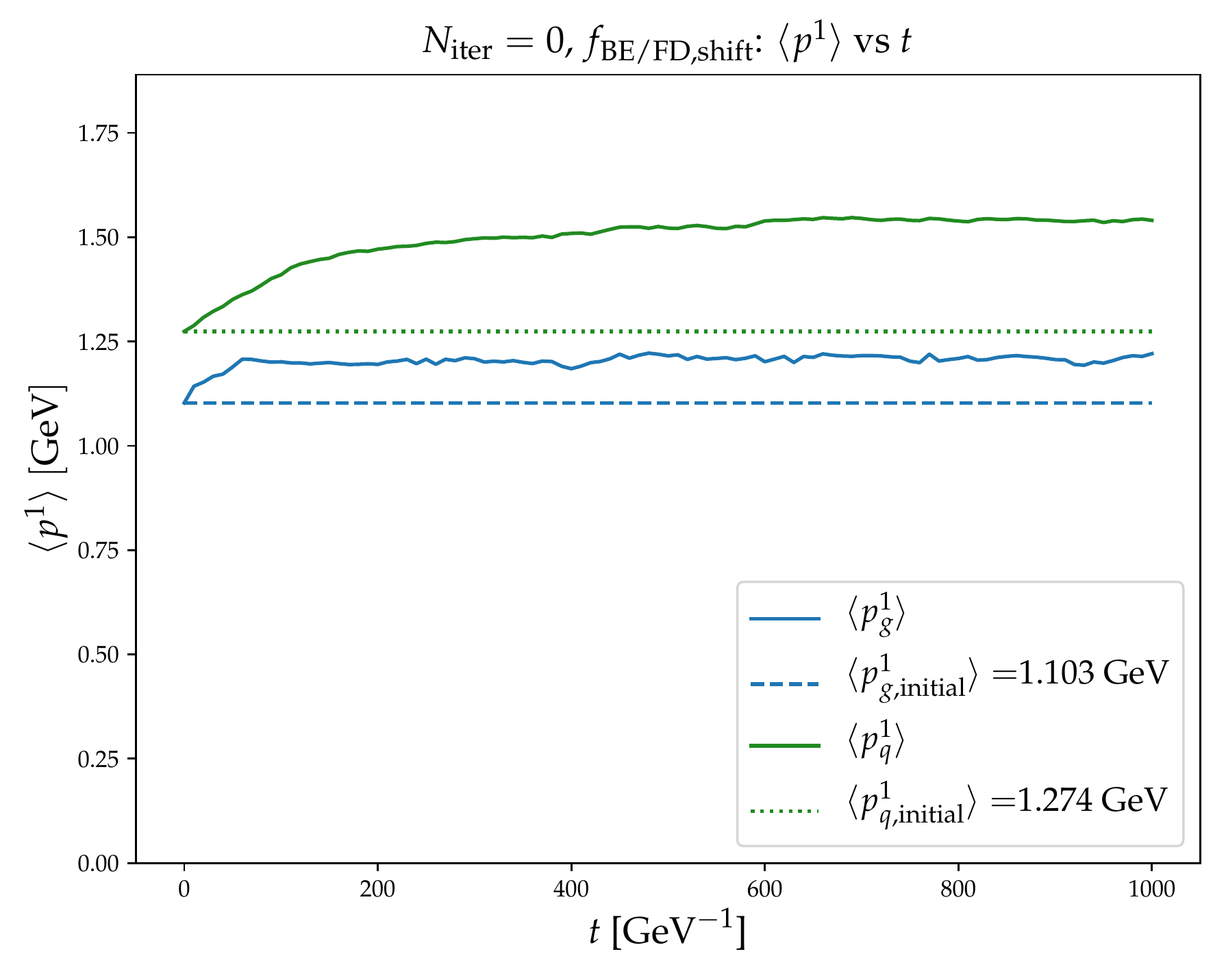}
    \includegraphics[width=0.495\linewidth]{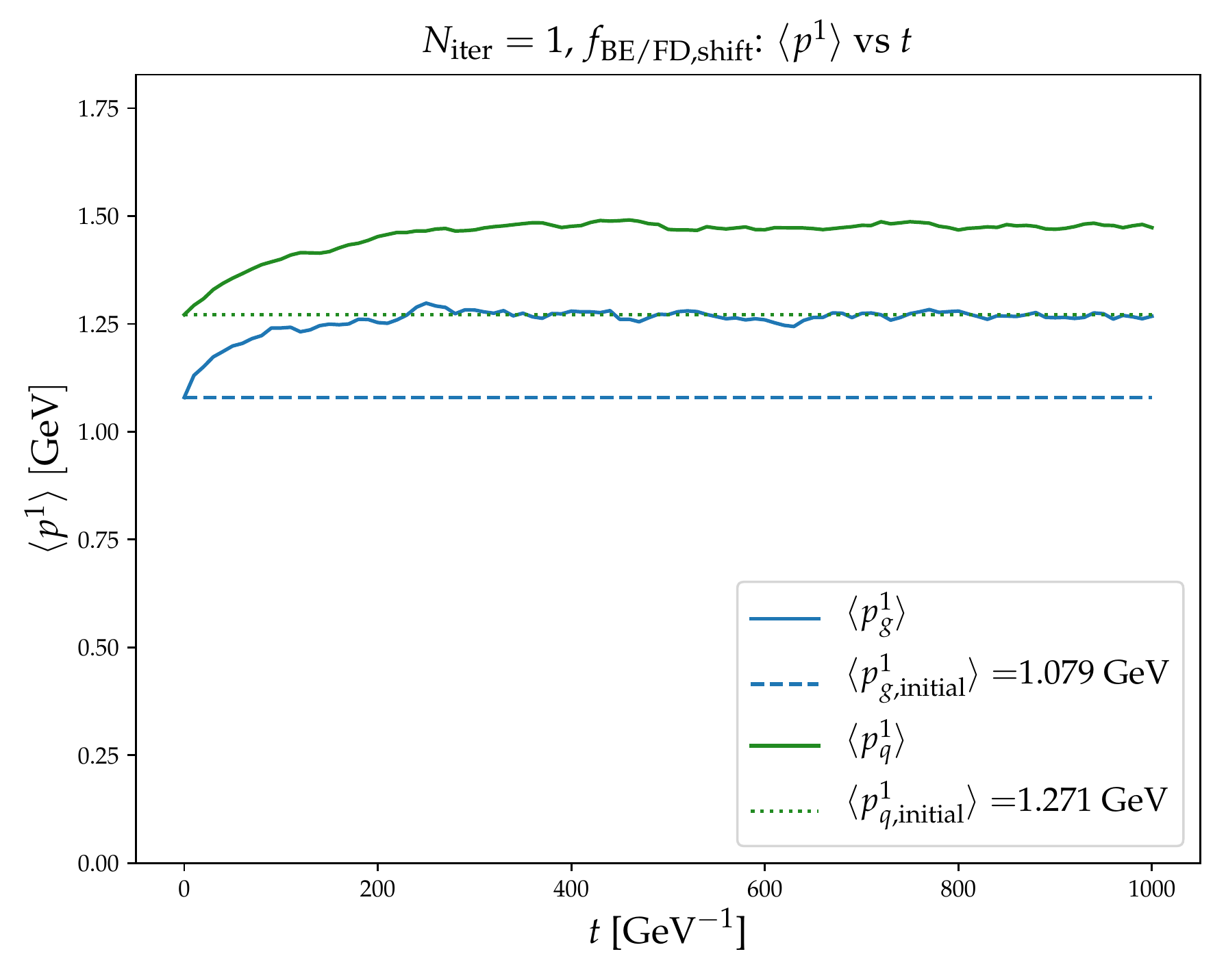}
    \caption{Mean between events of average energy per particle as a function of $t$. Left: $N_{\mathrm{iter}}=0$. Right: $N_{\mathrm{iter}}=1$.}
    \label{fig:fullrun_p1}
\end{figure*}

The resulting distributions that the system relaxes into for gluons and quarks is shown in \Fig{fig:fullrun_fg_fq}. As can be seen there, for gluons there is no significant difference in distribution at the end compared to the start, apart from a small drop in amplitude which can be attributed to the number of particles oscillating around the initial value, see \Fig{fig:fullrun_N}. For quarks the distribution is shifted slightly to larger energies, as can also be observed \Fig{fig:fullrun_p1}, but is still within error bars of the original distribution.

In Table~\ref{tab:fullrun_split_merge_rates} we see the mean splitting/merging rate over time. The total rates for inelastic splitting and merging are consistent with each other, for both the non-iterative and iterative runs, with a slight decrease in the rate for the latter (which can be attributed to a shift in $T_*$, see below). We also see that detailed balance is preserved as the average splitting and merging rates for each channel overlap within errors. The total scattering rates for the two different runs can be found in Table~\ref{tab:fullrun_elastic_rates}, and we see an increase for $N_{\mathrm{iter}}=1$ as expected, since too large $m_{g/q}^2$ (see below) will suppress the scattering rate.

\begin{figure*}[h!]
    \centering
    \includegraphics[width=0.495\linewidth]{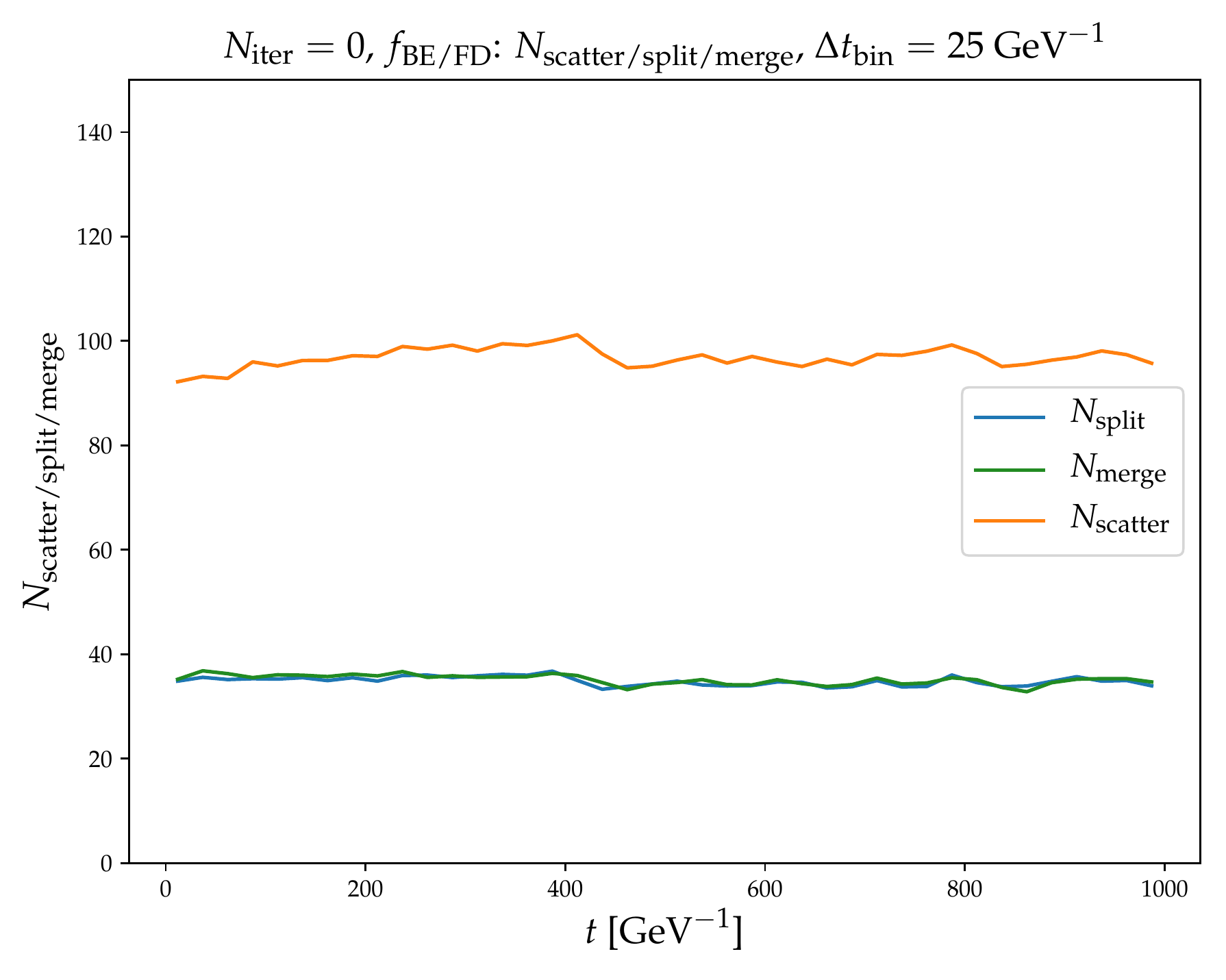}
    \includegraphics[width=0.495\linewidth]{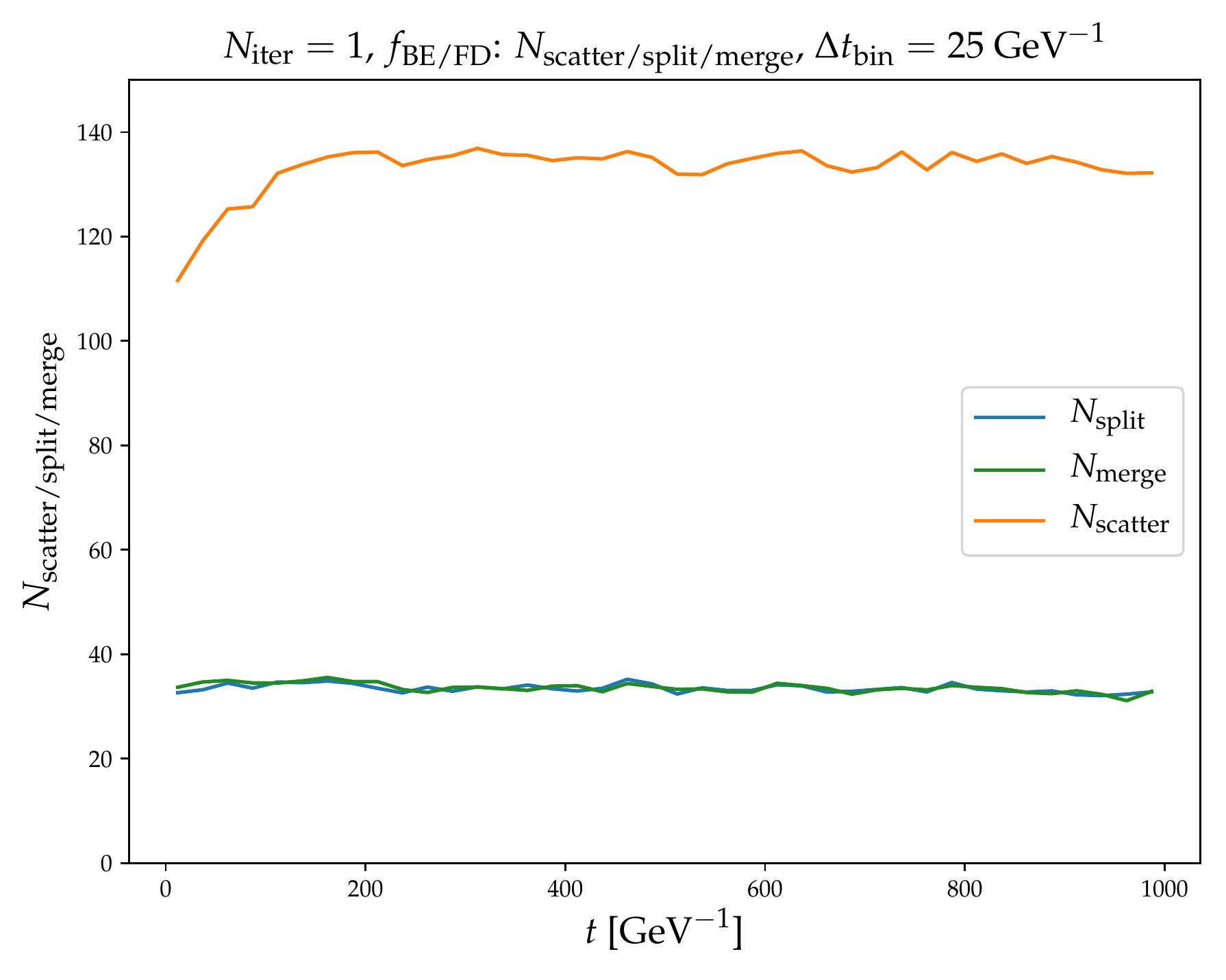}
    \caption{Number of elastic scatterings and inelastic splittings/merging as functions of time, bin width $\Delta t_{\mathrm{bin}}=25$ $\mathrm{GeV}^{-1}$. Left: $N_{\mathrm{iter}}=0$. Right: $N_{\mathrm{iter}}=1$.}
    \label{fig:fullrun_N_vs_tbin}
\end{figure*}

\begin{figure*}[h!]
    \centering
    \includegraphics[width=0.495\linewidth]{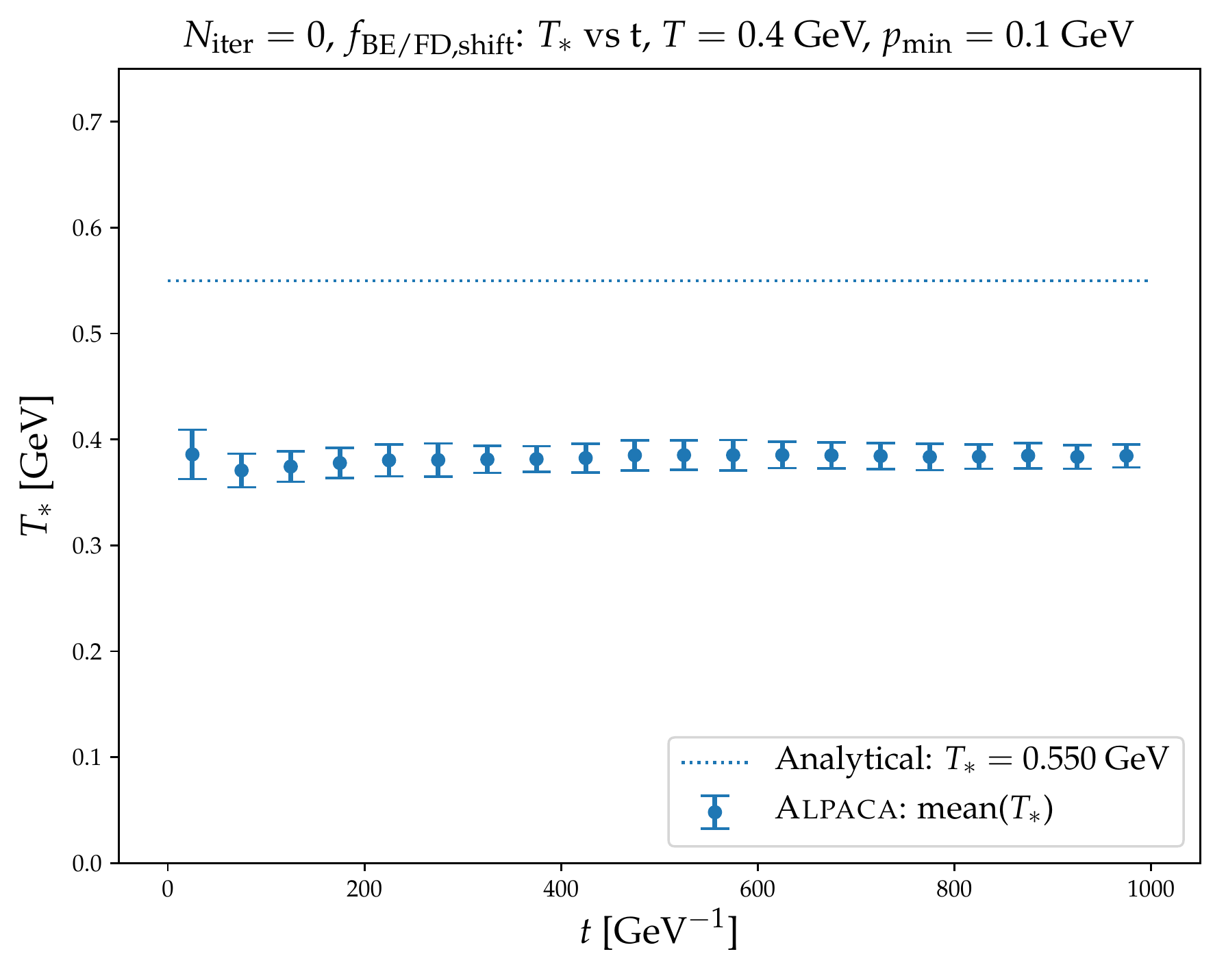}
    \includegraphics[width=0.495\linewidth]{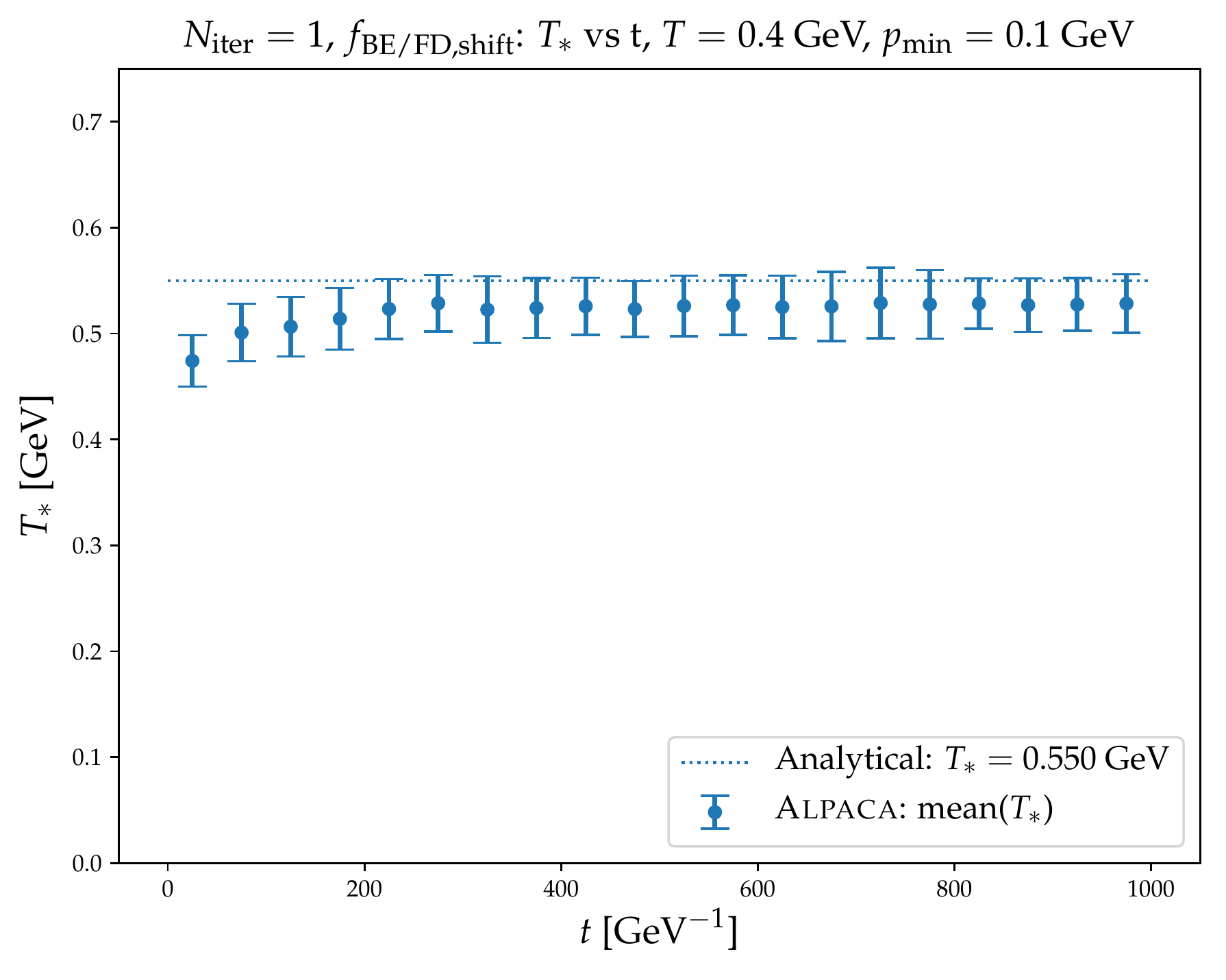}
    \caption{Mean of $T_*$ as a function of time. The error bars correspond to the standard deviation between events. Left: $N_{\mathrm{iter}}=0$. Right: $N_{\mathrm{iter}}=1$.}
    \label{fig:fullrun_Tstar}
\end{figure*}

The rates are also visualized in \Fig{fig:fullrun_N_vs_tbin} as number of scatterings, splittings and mergings as functions of time. No significant variation over time is observed, with the exception of a moderate increase of the scattering rate at early times, which could be due to the system settling into its actual equilibrium configuration (something similar can be seen in $T_*$, cf. Fig.~\ref{fig:fullrun_Tstar}). The corresponding mean free time per particle for elastic scattering is $t_{\mathrm{free, scatter}} = 38(3)$ $\mathrm{GeV}^{-1}$ and $t_{\mathrm{free, split/merge}} = 53 (4)$ $\mathrm{GeV}^{-1}$ for the non-iterative case, and $t_{\mathrm{free, scatter}} = 28 (2)$ $\mathrm{GeV}^{-1}$ and $t_{\mathrm{free, split/merge}} = 57 (4)$ $\mathrm{GeV}^{-1}$ for the case of one iteration.

\Fig{fig:fullrun_N} shows that the total number of particles stays reasonably constant around the initialized values. In \Fig{fig:fullrun_p1} the average energy per particle is shown to increase slightly for both the iterative and non-iterative runs, though seemingly remaining more constant in the latter case once the new equilibrium has been found.

\begin{figure*}[h!]
    \centering
    \includegraphics[width=0.495\linewidth]{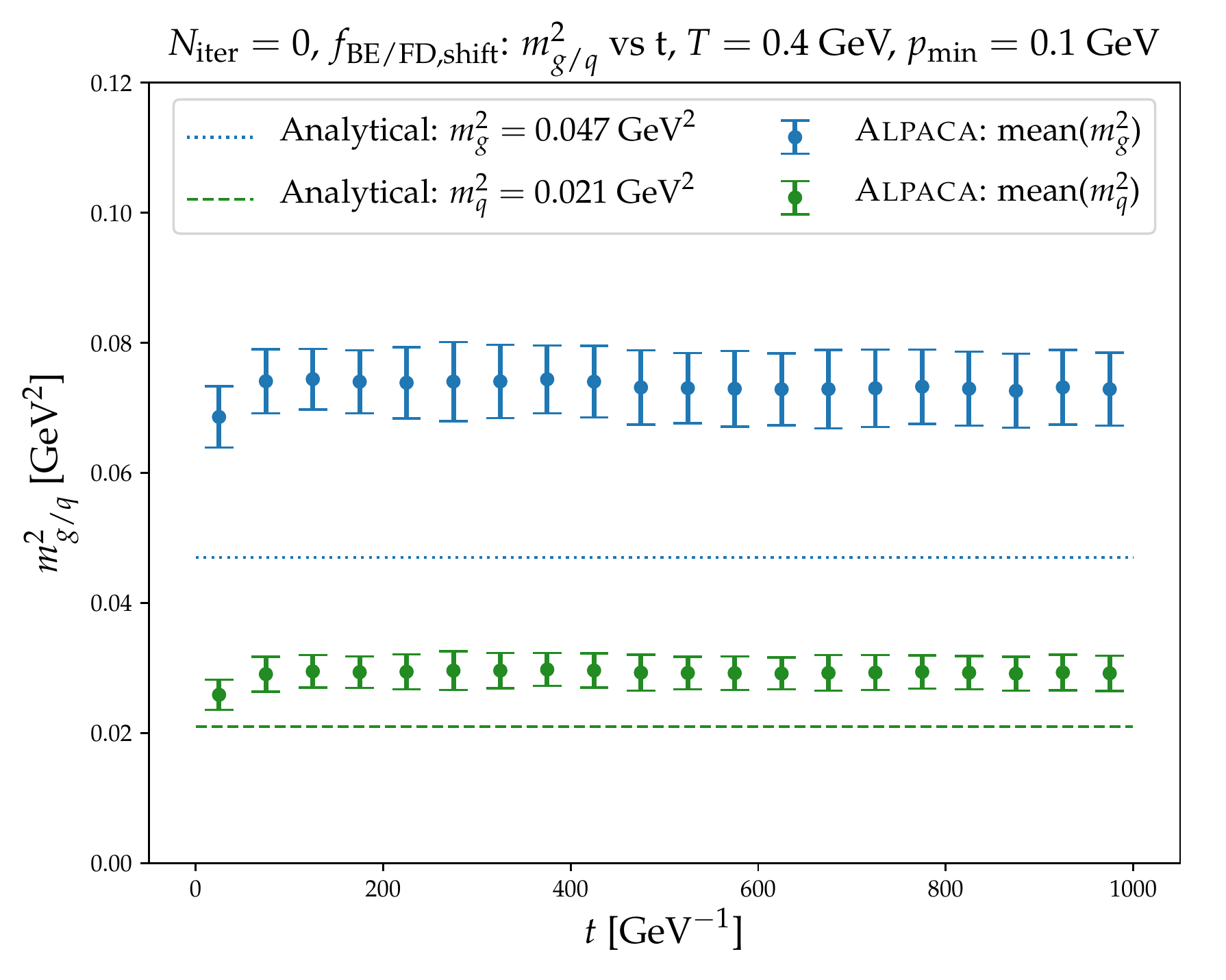}
    \includegraphics[width=0.495\linewidth]{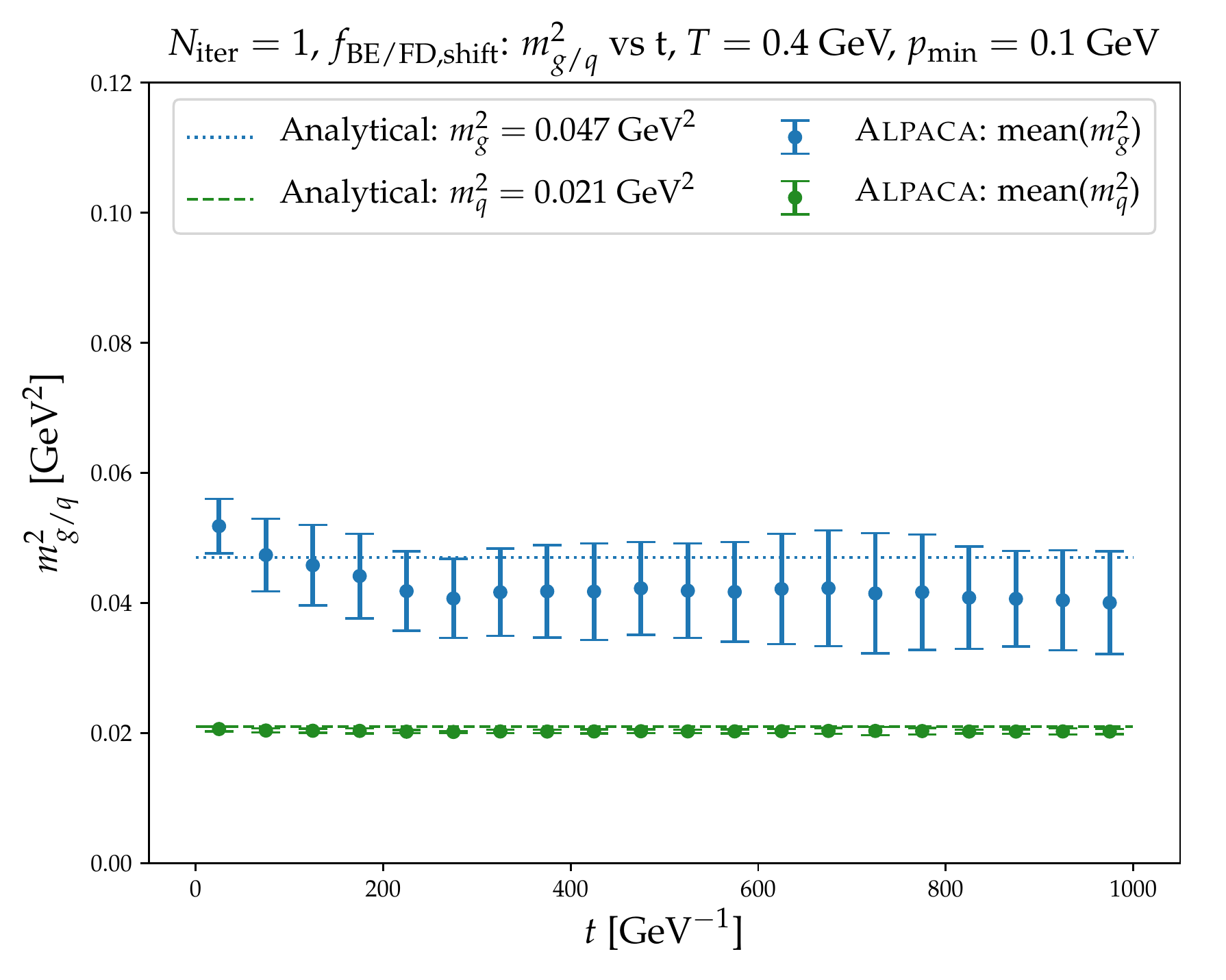}
    \caption{Mean of extracted $m_{g/q}^2$ as a function of time. The error bars correspond to the standard deviation between events. Left: $N_{\mathrm{iter}}=0$. Right: $N_{\mathrm{iter}}=1$.}
    \label{fig:fullrun_m2}
\end{figure*}

Lastly, as discussed in the beginning of this section, we observe a drastic change in $m_{g/q}^2$ and $T_*$ within the first $\Delta t = 100$ $\mathrm{GeV}^{-1}$ in Figs.~\ref{fig:fullrun_m2} and \ref{fig:fullrun_Tstar}. As the particles spread out more in $t$ the system settles into extracting too large values for $m_{g/q}^2$ and consequently too low values $T_*$ for the non-iterative case. For the case of just one iteration, we see how the correct value of both $m_{g/q}^2$ and $T_*$ is extracted, within oscillations. Since the effective masses and temperature are consistent with the true values after just one iteration, it can be expected that further iterations will not lead to significant changes in the evolution. When the true values are not known, one has to find out how many iterations are needed by comparing the results from subsequent iterations.

To summarize, when running our system initialized in an infinite thermal equilibrium, with all relevant quantities needed to faithfully reproduce the AMY Kinetic Theory extracted dynamically during the run, the system remains in this equilibrium (up to small oscillations in particle number and minor shifts in average energy per particle).

\begin{table*}[h!]
    \centering
    \begin{tabular}{ c c|c|c|c|c}
         $N_{\mathrm{iter}}$ & Type & $dN_{1\leftrightarrow2, \mathrm{tot}}/dt$ GeV  & $dN_{g\leftrightarrow gg}/dt$ GeV & $dN_{q\leftrightarrow gq}/dt$ GeV & $dN_{g\leftrightarrow q\bar{q}}/dt$ GeV \\ 
        \hline		\hline 

        0 & \begin{tabular}{@{}c@{}}Split \\ Merge\end{tabular} & \begin{tabular}{@{}c@{}} 1.4(1) \\ 1.4(1)\end{tabular} & \begin{tabular}{@{}c@{}}1.0(1) \\ 1.0(1)\end{tabular} & \begin{tabular}{@{}c@{}}0.40(3) \\ 0.40(3)\end{tabular} & \begin{tabular}{@{}c@{}}0.013(3) \\ 0.009(3)\end{tabular} \\
        \hline
        1 & \begin{tabular}{@{}c@{}}Split \\ Merge\end{tabular} & \begin{tabular}{@{}c@{}}1.3(1) \\ 1.3(1)\end{tabular} & \begin{tabular}{@{}c@{}}0.88(9) \\ 0.84(8)\end{tabular} & \begin{tabular}{@{}c@{}}0.44(4) \\ 0.49(4)\end{tabular} & \begin{tabular}{@{}c@{}}0.020(5) \\ 0.014(3)\end{tabular}
  \end{tabular}
  \caption{Comparison of the splitting and merging rates for the long run with $N_{\mathrm{iter}}\in\{0, 1\}$.}
    \label{tab:fullrun_split_merge_rates}
\end{table*}

\begin{table}[h!]
    \centering
    \begin{tabular}{ c|c}
         $N_{\mathrm{iter}}$ & $dN_{\mathrm{2\rightarrow2}, \mathrm{tot}}/dt$ GeV \\
        \hline		\hline 
        0 & 3.9(3) \\
        \hline
        1 & 5.3(4)
  \end{tabular}
  \caption{Comparison of the total elastic scattering rate for the long run with $N_{\mathrm{iter}}\in\{0, 1\}$.}
    \label{tab:fullrun_elastic_rates}
\end{table}

\clearpage

\section{Conclusions}
\label{section:conclusions}

The AMY effective kinetic theory of QCD at high temperatures has been successfully applied to various aspects of heavy ion collisions and is a natural candidate for the description of small collisions systems. We therefore believe it is time to turn it into a phenomenology tool by constructing a Monte Carlo event generator that solves the AMY Boltzmann equations by explicit simulation. Monte Carlo event generators have the advantage that in principle any observable can be calculated from an event sample (instead of having to do a new calculation for each observable) and that results can be directly compared to experimental data.

We have introduced the parton cascade \textsc{Alpaca} as a part of the multi-purpose event generator \textsc{Sherpa}. \textsc{Alpaca} encodes the AMY collision kernels in a Lorentz invariant framework for simulating the dynamics of multi-particle systems. The ensemble average is thus a solution of the Boltzmann equation. Frame independence is achieved by ordering collision and splitting events not in a frame dependent observer time but in a frame independent generalised time~\cite{Peter:1994yq}.

The partons can undergo elastic scattering and effective quasi-collinear splitting or merging, where in the latter coherent multiple soft scattering during the formation has to be taken into account. Both types of processes, elastic and inelastic, depend on the phase space densities of partons through the effective masses $m_g$ and $m_q$ related to screening effects in the medium and the effective temperature $T_*$. These quantities are local in position space but require integration over the momentum. We show that an estimate of the effective masses and temperature can be obtained from a single event without the need to provide further information about the phase space densities. In thermal equilibrium the analytical values are recovered with increasing precision as more and more partons are included in the estimate. However, in order to get an accurate value a large number of up to 30 particles has to be included. It remains to be seen whether this is good enough for out-of-equilibrium systems. In a similar way estimates of the local phase space density needed for Bose enhancement and Pauli blocking factors can be obtained.

To further validate the \textsc{Alpaca} framework we show that in thermal equilibrium
\begin{itemize}
    \item the elastic scattering rate for fixed cross section exhibits the expected scaling with temperature, volume and cross section
    \item the effective masses have the correct value and the expected dependence on temperature
    \item the elastic scattering rates with dynamical matrix elements reproduce the expected results within a few percent
    \item for fixed $\gamma$ the splitting/merging rate exhibits the expected scaling with $\gamma$, temperature and volume
    \item the effective temperature scales as expected with temperature
    \item with dynamical $\gamma$ the splitting and merging rates are the same and agree well with numerical results
    \item when putting everything together and running over longer time scales the system remains in thermal equilibrium
\end{itemize}
This gives us confidence that \textsc{Alpaca} indeed faithfully reproduces the AMY dynamics. Going from thermal equilibrium to out-of-equilibrium systems requires only small extensions. The next step will thus be to apply \textsc{Alpaca} to out-of-equilibrium systems to study equilibration and to the modeling of heavy ion collisions.

{\bf Acknowledgements.}

The authors thank Aleksas Mazeliauskas for useful discussions. This work is part of a project that has received funding from the European Research Council (ERC) under the European Union's Horizon 2020 research and innovation programme  (Grant agreement No. 803183, collectiveQCD).

\bibliography{references}

\appendix

\section{Box with periodic boundary conditions in Lorentz invariant framework}
\label{appendix:invariantbox}

The Lorentz invariant distance (squared) between to particles $i$ and $j$ is given by
\begin{align}
	d_{ij}^2(\tau) = & - (x_i(\tau)-x_j(\tau))^2 \nonumber \\
    & +\frac{[(x_i(\tau)-x_j(\tau))_\mu(p_i+p_j)^\mu]^2}{(p_i+p_j)^2}\,,
\end{align}
where the positions of the particles are functions of $\tau$
\begin{equation}
 x_\mu(\tau) = 2 \lambda p_\mu (\tau - \tau_0) + x_{0\mu}
\end{equation}
with $x_{0\mu}=x_\mu(\tau_0) = (t_0; \mathbf{x}_0)$ and the arbitrary but fixed Lagrange multiplier  $\lambda$.

The closest approach (i.e.\ the minimum of $d_{ij}^2$) is reached for $\tau = \bar \tau_{ij}$, which is given by
\begin{equation}
 \bar \tau_{ij} = \frac{1}{2} \left(\tau_{0i} + \tau_{0j} \right) + 
\frac{[x_{0i} - x_{0j}]_\mu [p_i - p_j]^\mu}{4 \lambda p_{i\mu}
p_j^\mu} \,.
\end{equation}
The distance of closest approach is then simply given by $d_{ij}^2(\bar \tau_{ij})$, which reduces to 
\begin{align}
	d_{ij}^2(\bar \tau_{ij}) = & 2 \frac{[(x_{0i}-x_{0j})_\mu p_i^\mu] [(x_{0i}-x_{0j})_\mu p_j^\mu]}{2 p_{i\mu}p_j^\mu} \nonumber \\
 & - (x_{0i} - x_{0j})^2 \,.
\end{align}

\medskip

From now on we drop the index $\mu$ from four-vectors and introduce the notation $[ab]$ for the scalar product of two four-vectors $a$ and $b$. Three-vectors are denoted by boldface symbols.

We consider here a cubic box\footnote{Other rectangular box shapes can be obtained in the same way, but have no advantage above the cubic box.} with side length $L$. A copy of a particle $j$ in another cell is generated by shifting the initial position $\mathbf{x}_{0}$ along the coordinate axes by multiples of the side length $L$
\begin{equation}
	x_{0j} \longrightarrow x_{0j} + (0; k,l,m) L = x_{0j} + s
\end{equation}
with integer $k$, $l$, and $m$.

The ``invariant time'' of closest approach $\bar \tau_{ij}$ with shifted particle $j$ is then given by
\begin{align}
	\bar \tau_{ij} & = \frac{1}{2} \left(\tau_{0i} + \tau_{0j} \right) + 
	\frac{[(x_{0i} - x_{0j} - s)(p_i - p_j)]}{4 \lambda [p_i p_j]} \nonumber\\
	 & = \frac{1}{2} \left(\tau_{0i} + \tau_{0j} \right) + 
	 \frac{[(x_{0i} - x_{0j})(p_i - p_j)]}{4 \lambda [p_i p_j]} \nonumber \\
  & \quad - \frac{[s(p_i - p_j)]}{4 \lambda [p_i p_j]} \nonumber\\
 	& = \bar \tau_{ij,0} - \frac{[s(p_i - p_j)]}{4 \lambda [p_i p_j]} \label{Eq:taubar} \,,
\end{align}
where $\bar \tau_{ij,0}$ is the invariant time of closest approach without shift (i.e. $k=l=m=0$). The invariant squared distance becomes
\begin{align}
	d_{ij}^2(\tau) = & - (x_i(\tau)-x_j(\tau)-s)^2 \nonumber \\
 & \quad +\frac{[(x_i(\tau)-x_j(\tau)-s)(p_i+p_j)]^2}{2[p_i p_j]} \nonumber\\
	 = & d_{ij,0}^2(\tau) + 2[s(x_i(\tau)-x_j(\tau))] - s^2 \nonumber \\ 
	& - \frac{[(x_i(\tau) - x_j(\tau))(p_i + p_j)]}{[p_i p_j][s(p_i + p_j)]} \nonumber \\
 & \quad + \frac{[s(p_i + p_j)]^2}{2[p_i p_j]} \,,
\end{align}
where $x_{i/j}(\tau)$ refers to the particles' positions without shift and $d_{ij,0}^2(\tau)$ denotes the squared invariant distance without shift. Similarly, the invariant distance squared at closest approach becomes
\begin{align}
	d_{ij}^2(\bar \tau_{ij}) = & 2 \frac{[(x_{0i}-x_{0j}-s) p_i] [(x_{0i}-x_{0j}-s) p_j]}{2 [p_ip_j]} \nonumber \\
    & - (x_{0i} - x_{0j}-s)^2  \nonumber\\
	= & d_{ij,0}^2(\bar \tau_{ij}) + 2 [s(x_{0i} - x_{0j})] - s^2 \nonumber \\
	  & - 2\frac{[(x_{0i} - x_{0j})p_i][sp_j]}{[p_i p_j]} \nonumber \\
   & - 2\frac{[(x_{0i} - x_{0j})p_j][sp_i] - [sp_i][sp_j]}{[p_i p_j]}\,.  
	  \label{Eq:dij2}
\end{align}

Our task is to find the triple $(k,l,m)$ that produces the first scattering, i.e.\ the first encounter with $d_{ij}^2(\bar \tau_{ij}) < \sqrt{\sigma/\pi}$. To achieve this, it is helpful to first consider $k$, $l$, and $m$ as continuous variables that span a three-dimensional space ($\mathbb{R}^3$). In this space $\bar \tau_{ij}$ is constant on planes, whose position and orientation are determined by \Eq{Eq:taubar}. Equating this to some $\tau_0$ and solving for $m$ yields the value of $m$ for which, given $k$ and $l$, the invariant time of closest approach equals $\tau_0$.
\begin{align}
	m_0(k,l,\tau_0) = & \frac{1}{p_{i,z} - p_{j,z}} \Bigg\{ \frac{4 \lambda [p_i p_j]}{L} (\tau_0 - \bar \tau_{ij}) \nonumber \\
 & - k(p_{i,x} - p_{j,x}) - l(p_{i,y} - p_{j,y})\Bigg\}.
\label{Eq:m0plane}
\end{align}
It is also worth noting that the direction of fastest increase of $\bar \tau_{ij}$ is given by $\mathbf{p}_i - \mathbf{p}_j$ (the gradient of $\bar \tau_{ij}$ w.r.t. $(k,l,m)$).

Furthermore, in the $(k,l,m)$ space, the invariant distance of closest approach $d_{ij}(\bar \tau_{ij})$ vanishes on a line. It can be found from the condition that the gradient (again w.r.t. $(k,l,m)$) vanishes. The partial derivatives of \Eq{Eq:dij2} are given by
%
    \begin{align}
    	& \frac{\partial }{\partial k} d_{ij}^2(\bar \tau_{ij}) = \nonumber \\  
    	& \underbrace{\frac{2L}{[p_i p_j]}[(x_{0i} - x_{0j})(p_{i,x}p_j + p_{j,x}p_i)] - 2L(x_{0i,x} - x_{0j,x})}_{=: a_k} \nonumber \\
     & - \frac{2L}{[p_ip_j]}[s(p_{i,x}p_j + p_{j,x}p_i)] + 2L^2k \nonumber\\
    	= & a_k + 2L^2k \left(\frac{2p_{i,x}p_{j,x}}{[p_ip_j]} + 1 \right) \nonumber \\
    	 & + 2L^2 l \frac{p_{i,x}p_{j,y} + p_{j,x}p_{i,y}}{[p_ip_j]} \nonumber \\
    	 & + 2L^2 m \frac{p_{i,x}p_{j,z} + p_{j,x}p_{i,z}}{[p_ip_j]}, \label{Eq:ddijdk}
    \end{align}

\begin{align}
	& \frac{\partial }{\partial l} d_{ij}^2(\bar \tau_{ij}) = \nonumber \\
    & \underbrace{\frac{2L}{[p_i p_j]}[(x_{0i} - x_{0j})(p_{i,y}p_j + p_{j,y}p_i)] - 2L(x_{0i,y} - x_{0j,y})}_{=: a_l} \nonumber \\
    & - \frac{2L}{[p_ip_j]}[s(p_{i,y}p_j + p_{j,y}p_i)] + 2L^2l \nonumber \\
	= & a_l + 2L^2l \left(\frac{2p_{i,y}p_{j,y}}{[p_ip_j]} + 1 \right) \nonumber \\
     & + 2L^2 k \frac{p_{i,x}p_{j,y} + p_{j,x}p_{i,y}}{[p_ip_j]}\nonumber \\
    	 & + 2L^2 m \frac{p_{i,y}p_{j,z} + p_{j,y}p_{i,z}}{[p_ip_j]}, \label{Eq:ddijdl}
\end{align}
\begin{align}
	& \frac{\partial }{\partial m} d_{ij}^2(\bar \tau_{ij}) = \nonumber \\
 &\underbrace{\frac{2L}{[p_i p_j]}[(x_{0i} - x_{0j})(p_{i,z}p_j + p_{j,z}p_i)] - 2L(x_{0i,z} - x_{0j,z})}_{=: a_m} \nonumber \\
 & - \frac{2L}{[p_ip_j]}[s(p_{i,z}p_j + p_{j,z}p_i)] + 2L^2m \nonumber \\	
	= & a_m + 2L^2m \left(\frac{2p_{i,z}p_{j,z}}{[p_ip_j]} + 1 \right) \nonumber \\
     & + 2L^2 k \frac{p_{i,x}p_{j,z} + p_{j,x}p_{i,z}}{[p_ip_j]}\nonumber \\
    	 &+ 2L^2 l \frac{p_{i,y}p_{j,z} + p_{j,y}p_{i,z}}{[p_ip_j]}. \label{Eq:ddijdm}
\end{align}
For the algorithm discussed below it is advantageous to keep one variable fixed and find the minimum in that plane. Keeping $m$ fixed, for example, one has to set \Eqs{Eq:ddijdk} and \eqref{Eq:ddijdl} to zero and solve for $k$ and $l$. This yields
\begin{align}
	l_0^{(m)} = & \frac{2p_{i,x}p_{j,x} + [p_ip_j]}{(p_{i,0}p_{j,z} - p_{i,z}p_{j,0})^2} \nonumber \\
 & \times\Bigg\{ \frac{[p_ip_j]}{2L^2}\left(\frac{p_{i,x}p_{j,y} + p_{j,x}p_{i,y}}{2p_{i,x}p_{j,x} + [p_ip_j]} a_k - a_l\right) \nonumber \\
	&  + m\bigg( \frac{(p_{i,x}p_{j,y} + p_{j,x}p_{i,y})(p_{i,x}p_{j,z} + p_{j,x}p_{i,z})}{2p_{i,x}p_{j,x} + [p_ip_j]} \nonumber \\
 & - (p_{i,y}p_{j,z} + p_{j,y}p_{i,z})\bigg) \Bigg\},  \label{Eq:k0m} \\
	k_0^{(m)} = & \frac{1}{2p_{i,x}p_{j,x} + [p_ip_j]} \bigg( -  \frac{[p_ip_j]}{2L^2} a_k \nonumber \\
 &- l_0^{(m)} (p_{i,x}p_{j,y} + p_{j,x}p_{i,y}) \nonumber \\
 &- m(p_{i,x}p_{j,z} + p_{j,x}p_{i,z}) \bigg) \,.
\end{align}
Similarly, one gets for fixed $l$
\begin{align}
	m_0^{(l)} = & \frac{2p_{i,x}p_{j,x} + [p_ip_j]}{(p_{i,0}p_{j,y} - p_{i,y}p_{j,0})^2} \nonumber \\
 & \times\Bigg\{ \frac{[p_ip_j]}{2L^2}\left(\frac{p_{i,x}p_{j,z} + p_{j,x}p_{i,z}}{2p_{i,x}p_{j,x} + [p_ip_j]} a_k - a_m\right)   \nonumber \\
	&  + l\bigg( \frac{(p_{i,x}p_{j,y} + p_{j,x}p_{i,y})(p_{i,x}p_{j,z} + p_{j,x}p_{i,z})}{2p_{i,x}p_{j,x} + [p_ip_j]} \nonumber \\
    & - (p_{i,y}p_{j,z} + p_{j,y}p_{i,z})\bigg) \Bigg\} , \\
	k_0^{(l)} = & \frac{1}{2p_{i,x}p_{j,x} + [p_ip_j]} \bigg( -  \frac{[p_ip_j]}{2L^2} a_k \nonumber \\
 & - l (p_{i,x}p_{j,y} + p_{j,x}p_{i,y}) \nonumber \\
 & - m_0^{(l)}(p_{i,x}p_{j,z} + p_{j,x}p_{i,z}) \bigg) \,,
\end{align}
and for fixed $k$
\begin{align}
	m_0^{(k)} = & \frac{2p_{i,y}p_{j,y} + [p_ip_j]}{(p_{i,0}p_{j,x} - p_{i,x}p_{j,0})^2} \nonumber \\
 & \times \Bigg\{ \frac{[p_ip_j]}{2L^2}\left(\frac{p_{i,y}p_{j,z} + p_{j,y}p_{i,z}}{2p_{i,y}p_{j,y} + [p_ip_j]} a_l - a_m\right)  \nonumber \\
	&  + k\bigg( \frac{(p_{i,x}p_{j,y} + p_{j,x}p_{i,y})(p_{i,y}p_{j,z} + p_{j,y}p_{i,z})}{2p_{i,y}p_{j,y} + [p_ip_j]} \nonumber\\
 & - (p_{i,x}p_{j,z} + p_{j,x}p_{i,z})\bigg) \Bigg\} , \\
	l_0^{(k)} = & \frac{1}{2p_{i,y}p_{j,y} + [p_ip_j]} \bigg( -  \frac{[p_ip_j]}{2L^2} a_l \nonumber\\
 & - k (p_{i,x}p_{j,y} + p_{j,x}p_{i,y}) \nonumber \\
 &- m_0^{(k)}(p_{iy}p_{j,z} + p_{j,y}p_{i,z}) \bigg) \,. \label{Eq:m0k}
\end{align}

Now the last missing ingredient is the point at which the line of vanishing invariant distance at closest approach intersects a plane of constant invariant time of closest approach. To find this one inserts \Eq{Eq:m0plane} into \Eq{Eq:dij2} and finds the minimum w.r.t. to $k$ and $l$. The partial derivatives of \Eq{Eq:dij2} then read
\begin{align}
	\frac{\partial}{\partial k} d_{ij}^2(\bar \tau_{ij}) = & 2 a - b' + 2L^2 c k + 2 L^2 d l \label{Eq:ddij2dk_2}, \\
	\frac{\partial}{\partial l} d_{ij}^2(\bar \tau_{ij}) = & 2 A - B' + 2L^2 d k + 2 L^2 C l \label{Eq:ddij2dl_2}
\end{align}
with
\begin{align}
	a = & [\Delta x_{ij,0} \frac{\partial s}{\partial k}]
	 - 4 L \lambda [p_ip_j] \frac{p_{i,x}-p_{j,x}}{(p_{i,z}-p_{j,z})^2}(\tau_0 - \bar \tau_{ij,0}), \\
	b' = & \frac{2L}{[p_ip_j]} \frac{p_{i,x}p_{j,z} -
		 p_{i,z}p_{j,x}}{p_{i,z}-p_{j,z}} b,\\
	b =  & [\Delta x_{ij,0} (p_i+p_j)]
	- 4 \lambda [p_ip_j] \frac{p_{i,z}+p_{j,z}}{p_{i,z}-p_{j,z}}(\tau_0 - \bar \tau_{ij,0}), \\
	c = & 1 + \left(\frac{p_{i,x}-p_{j,x}}{p_{i,z}-p_{j,z}}\right)^2 
	   \nonumber \\
    & + \frac{2}{[p_ip_j]} \left(\frac{p_{i,x}p_{j,z}-p_{i,z}p_{j,x}}{p_{i,z}-p_{j,z}}\right)^2,\\
	d = & \frac{1}{(p_{i,z}-p_{j,z})^2} \Bigg( (p_{i,x}-p_{j,x})(p_{i,y}-p_{j,y})
	 \nonumber \\
  &+ \frac{2}{[p_ip_j]}(p_{i,x}p_{j,z} - p_{i,z}p_{j,x}) (p_{i,y}p_{j,z} - p_{i,z}p_{j,y}) \Bigg),\\
	 A = & [\Delta x_{ij,0} \frac{\partial s}{\partial l}]
	 - 4 L \lambda [p_ip_j] \frac{p_{i,y}-p_{j,y}}{(p_{i,z}-p_{j,z})^2}(\tau_0 - \bar \tau_{ij,0}), \\
	B' = & \frac{2L}{[p_ip_j]} \frac{p_{i,y}p_{j,z} - 
		p_{i,z}p_{j,y}}{p_{i,z}-p_{j,z}} b,\\
	C = & 1 + \left(\frac{p_{i,y}-p_{j,y}}{p_{i,z}-p_{j,z}}\right)^2 
	   \nonumber \\
    & + \frac{2}{[p_ip_j]} \left(\frac{p_{i,y}p_{j,z}-p_{i,z}p_{j,y}}{p_{i,z}-p_{j,z}}\right)^2 \,,
\end{align}
where
\begin{equation}
	\Delta x_{ij,0} = 2 \lambda p_i(\bar \tau_{ij,0} - \tau_{0i}) - 2 \lambda p_j(\bar \tau_{ij,0} - \tau_{0j}) + x_{0i} - x_{0j} \,.
\end{equation}
Setting \Eqs{Eq:ddij2dk_2} and \eqref{Eq:ddij2dl_2} equal to zero and solving for $k$ and $l$ yields
\begin{align}
	k_0 & = \frac{b'-2a}{2L^2c} \nonumber \\
 & \quad - \frac{d}{2L^2(cC-d^2)}\left( -2A + B' + \frac{d}{c}(2a-b') \right), \\
	l_0 & = \frac{c}{2L^2(cC-d^2)} \left( -2A + B' + \frac{d}{c}(2a-b') \right) \,.
\end{align}
Inserting this into \Eq{Eq:m0plane} one has all three coordinates of the point of vanishing $d_{ij}^2(\bar \tau_{ij})$ for a given $\bar \tau_{ij} = \tau_0$.

\smallskip

The strategy for finding the next scattering of a pair of particles is now to start at this point, choosing $\tau_0$ to be the current invariant time $\tau$ of the evolution. From this point one then moves along the line of vanishing  $d_{ij}^2(\bar \tau_{ij})$ in the direction of increasing $\bar \tau_{ij}$. For the points with integer $k$, $l$ and $m$ lying closest to the line it is checked whether they lead to a scattering, i.e.\ whether $d_{ij}(\bar \tau_{ij}) < \sqrt{\sigma/\pi}$. The first such point is the sought after next scattering\footnote{In practice there is no guarantee that $\bar \tau_{ij}$ increases strictly from point to point that gets checked, so it is advisable to check also the next few points after a scattering has been found to make sure that there is no scattering with smaller $\bar \tau_{ij}$.}. In practice, the checking of nearby points is done by imagining the $(k,l,m)$ space to be divided into cubes the corners of which are the points with integer $k$, $l$ and $m$. One can then move from cube to cube by calculating through which sides the line enters and leaves the cube using \Eqs{Eq:k0m} -- \eqref{Eq:m0k}. The corners of the cubes intersected by the line are the points that have to be checked.

\section{Splitting/merging rates}
\label{appendix:gamma}

The splitting/merging rates $\gamma^a_{bc}$ are products of the spin averaged Altarelli-Parisi splitting functions and a function $\mu^2$. For the possible splitting/merging processes one gets
\begin{align}
    \gamma_{gg}^g(p;xp,&(1-x)p)  = \frac{\sqrt{2}d_AC_A\alpha_s}{(2\pi)^4} \nonumber \\ 
    & \times \left[\frac{1+x^4+(1-x)^4}{x^2(1-x)^2}\right] \mu^2\left(\eta, x;A,A,A \right),  \\
    \gamma_{gq}^q(p;xp,&(1-x)p)  = \frac{\sqrt{2}d_FC_F\alpha_s}{(2\pi)^4}\nonumber \\ 
    & \times \left[\frac{1+(1-x)^2}{x^2(1-x)}\right]\mu^2\left(\eta, x;F,A,F \right),  \\
    \gamma_{qq}^g(p;xp,&(1-x)p)  = \frac{\sqrt{2}d_FC_F\alpha_s}{(2\pi)^4}\nonumber \\ 
    & \times \left[\frac{x^2+(1-x)^2}{x(1-x)}\right]\mu^2\left(\eta, x;A,F,F \right)\,,
\end{align}
where $\eta = x(1-x)\lambda T_* p/m_g^2$ with the coupling constant given by $\lambda = g^2N_c = 4 \pi \alpha_s N_c$. The collinear splitting/joining rates are dictated by 2-dimensional quantum mechanics in the transverse plane of the splitting particle. The rate splitting rate is related to $\mu^2(\eta,x;,s_1,s_2,s_3) = \sqrt{2}(4\pi)m_g^2\mathrm{Im}(f(0))$ where $f$ is the solution to the radially symmetric two-dimensional Schrödinger equation
\begin{align}
    0 &  = \left(\delta_b^2 + \frac{3}{b}\delta_b - \hat{M}^2(s_1,s_2,s_3) \right)f(b) \nonumber \\
     &+ i\eta\left\{ C_{s_2s_3}^{s_1}\mathcal{C}(b) + C_{s_3s_1}^{s_2}\mathcal{C}(xb) + C_{s_1s_2}^{s_3}\mathcal{C}((1-x)b) \right\}
\end{align}
where $C_{s_2s_3}^{s_1} = \frac{C_{s_2}+C_{s_3}-C_{s_1}}{C_A}$ and the boundary conditions are given by
\begin{align}
    f(b) & \xrightarrow[b\to 0]{} \frac{1}{\pi}\frac{1}{b^2}, \\
    f(b) & \xrightarrow[b\to \infty]{} 0 \,.
\end{align}
The effective mass appearing in the equation is the mass difference\footnote{For the interesting special case of $d_A = 8$, $C_A = 3$, $N_f = 3$, $C_F = 4/3$, i.e.\ QCD with three flavours of quarks, the ration of the effective masses of quarks and gluons in independent of the phase space density and given by $m_g^2/m_q^2 = 9/4$.}
\begin{align}
    \hat{M}^2(s_1,s_2,s_3)m_g^2 & = x(1-x)\Bigg(\frac{1}{x(1-x)}m_{s_1}^2 \nonumber \\
    & - \frac{1}{x}m_{s_2}^2 - \frac{1}{(1-x)}m_{s_3}^2 \Bigg)\,.
\end{align}
The collision kernel $\mathcal{C}$ appearing in the Schr\"odinger equation is given by
\begin{align}
    \mathcal{C}(b) & = \int \frac{\d \mathbf{q}_\perp}{(2 \pi)^2} \mathcal{C}(\mathbf{q}_\perp) \left[ 1 - e^{i(\mathbf{q}_\perp/m_g)\cdot \mathbf{b}}\right] \nonumber \\
    & = \frac{1}{2\pi} \left[ K_0 \left(b \frac{m_D}{m_g} \right) + \gamma_E + \log \left(b\frac{m_D}{2m_g}\right) \right]
\end{align}
with
\begin{equation}
    \mathcal{C}(\mathbf{q}_\perp) = \frac{1}{q_\perp^2} - \frac{1}{q_\perp^2 + m_D^2} \,,
\end{equation}
where $m_D$ is the Debye mass.

\begin{figure}
    \centering
    \includegraphics[width=1\linewidth]{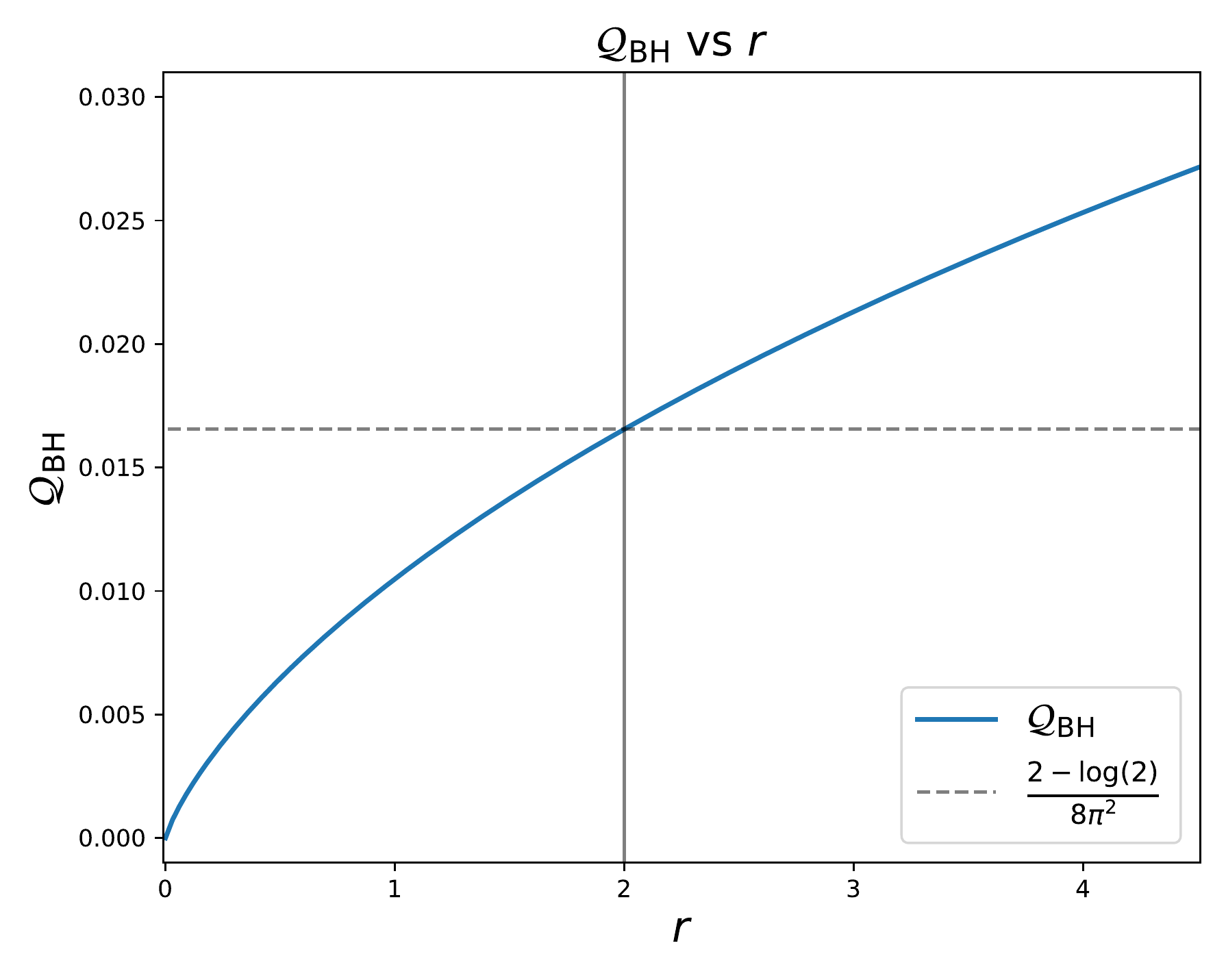}
    \caption{Function $\mathcal{Q}_{\mathrm{BH}}(r)$ appearing in the solution of the splitting/merging rates in the Bethe Heitler limit.}
    \label{fig:QBH}
\end{figure}

The quantity $\mu$ can be easily evaluated in two limits, that of small $\eta/|\hat M^2|$ corresponding to the Bethe-Heitler limit, and the large $\eta$ limit corresponding to the deep LPM limit~\cite{Arnold:2008zu}. The expression in the Bethe-Heitler limit  read\footnote{The Green's function of the free equation is $G_M(b,b_0) = \frac{I_1(Mb)}{b} \frac{K_1(Mb_0)}{b_0} \Theta(b_0-b) + \frac{I_1(Mb_0)}{b_0} \frac{K_1(Mb)}{b} \Theta(b-b_0)$. Using the Green's function the small $\eta$ solution can be written as $\Im(f_0) = \frac{1}{\pi} i \eta b^3 \int_0^\infty \d b\, [C_{s_3s_1}^{s_2} C(bx/M) + C_{s_1s_2}^{s_3} C(b(1-x)/M) + C_{s_2s_3}^{s_1} C(b/M)] \times \frac{K_1(b)}{b}G_1(b,0)$.}
\begin{align}
    \frac{\mu^2_{\text{BH}}}{m_g^2} =  & \eta \frac{4\pi}{\sqrt{2}} \Bigg\{ C_{s_2s_3}^{s_1} \mathcal{Q}_{\mathrm{BH}} \left(\frac{m_D^2}{m_g^2\hat{M}^2}\right) \nonumber \\
    & + C_{s_3s_1}^{s_2} \mathcal{Q}_{\mathrm{BH}} \left(\frac{x^2m_D^2}{m_g^2\hat{M}^2}\right) \nonumber \\
    & + C_{s_1s_2}^{s_3} \mathcal{Q}_{\mathrm{BH}} \left(\frac{(1-x)^2m_D^2}{m_g^2\hat{M}^2} \right)\Bigg\}
\end{align}
with
\begin{align}
    \mathcal{Q}_{\mathrm{BH}}(r) = & \frac{1}{8\pi^2} \Bigg[\frac{i(r-2)[\mathrm{Li}_2(r_-) - \mathrm{Li}_2(r_+)]}{\sqrt{(4-r)r}} \nonumber \\
    & - \log(r) + 2 \Bigg] \,,
\end{align}
where $r_\pm = 1 - \frac{r}{2} \pm \frac{i}{2}\sqrt{(4-r)r}$. $\mathcal{Q}_{\mathrm{BH}}$ is shown in \Fig{fig:QBH}. 
Expanding the expression further for small $x$ gives a particularly simple form that can easily be used in analytical estimates
\begin{align}
    \mathcal{Q}_{\mathrm{BH}}(0) & = 0, \\
    \mathcal{Q}_{\mathrm{BH}}(2) & = \frac{2-\log(2)}{8\pi^2}, \\
    \mathcal{Q}_{\mathrm{BH}}(4/9) & \approx 0.0271432 \,.
\end{align}

In the limit $|\hat M^2| \to 0$, to the next to leading logarithmic order\footnote{ The expression here differs from that of~\cite{Arnold:2008zu} by $\log(x) \to \log(x + 1)$. The finite shift of the argument of the log does not affect the
Next to Leading Logarithmic order but makes the expression well behaved at small values of $\eta$.},
\begin{align}
    \frac{\mu^2_{\text{LPM}}}{m_g^2}  = & \eta^{1/2}\left(\frac{1}{2\pi} \right)^{1/2}\Bigg[  C_{s_2s_3}^{s_1}\log{\left(\frac{\xi\mu^2}{m_g^2}+1\right)} \nonumber \\
    & + C_{s_3s_1}^{s_2}x^2\log{\left(\frac{\xi\mu^2}{x^2m_g^2}+1\right)} \nonumber\\ 
    &+ C_{s_1s_2}^{s_3}(1-x)^2\log{\left(\frac{\xi\mu^2}{(1-x)^2m_g^2}+1\right)}  \Bigg]^{\frac{1}{2}}
\end{align}
with $\xi = e^{2-\gamma_E+\pi/4} \approx 9.09916$. The transcendental depends on $\mu$ on both sides, and needs to be solved iteratively.

\begin{figure*}
    \centering
    \includegraphics[width=.60\linewidth]{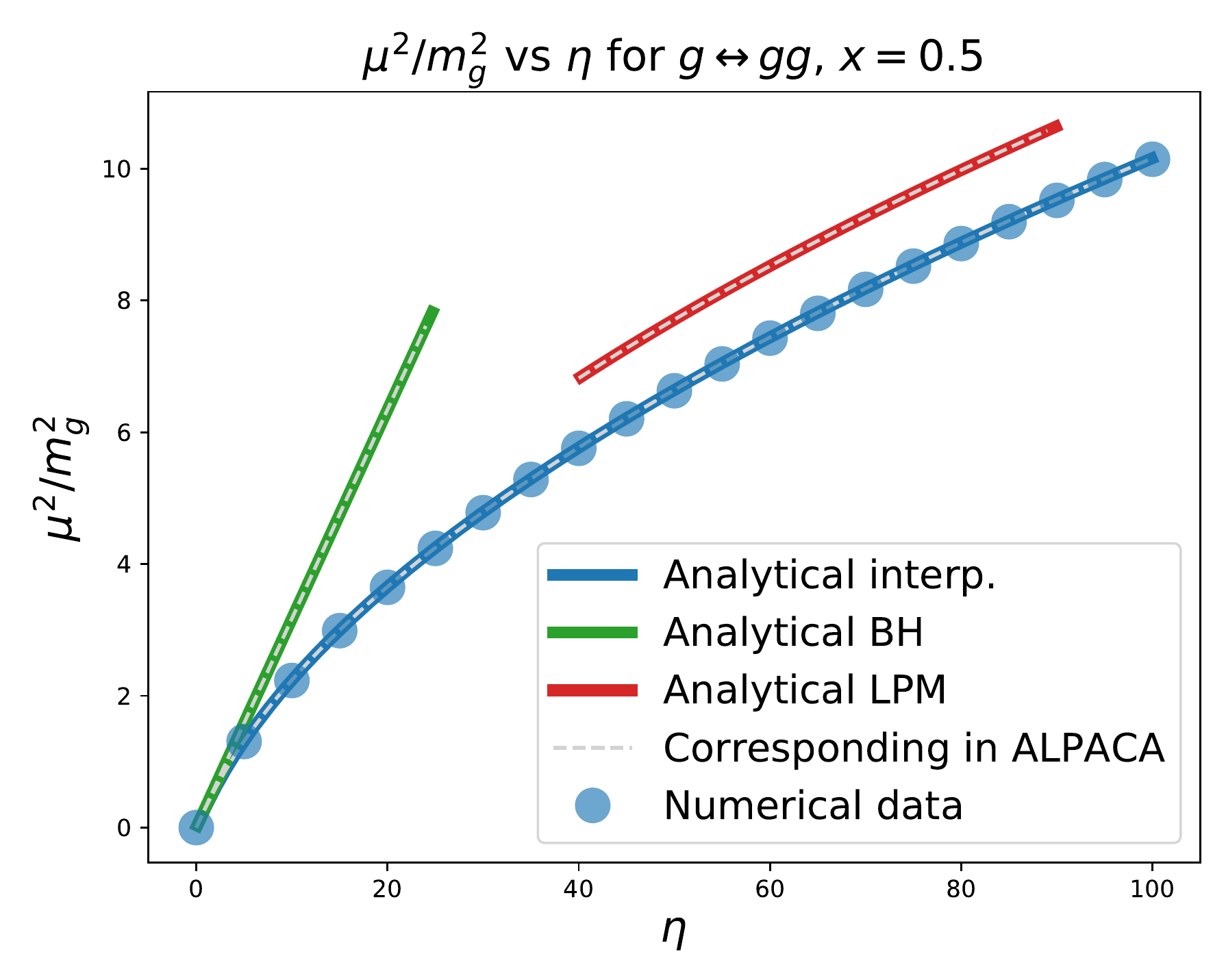}\par
    \includegraphics[width=.48\linewidth]{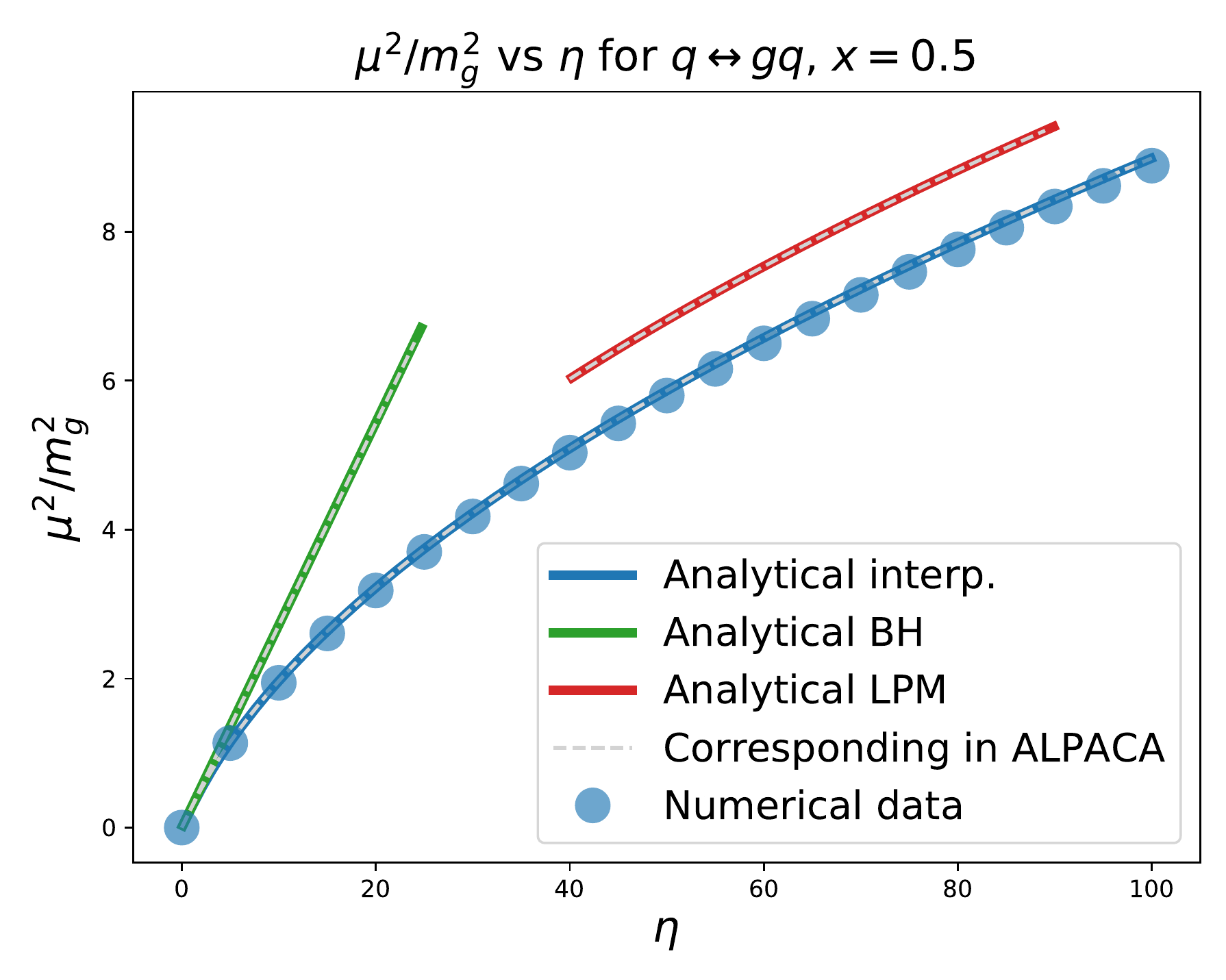}
    \includegraphics[width=.48\linewidth]{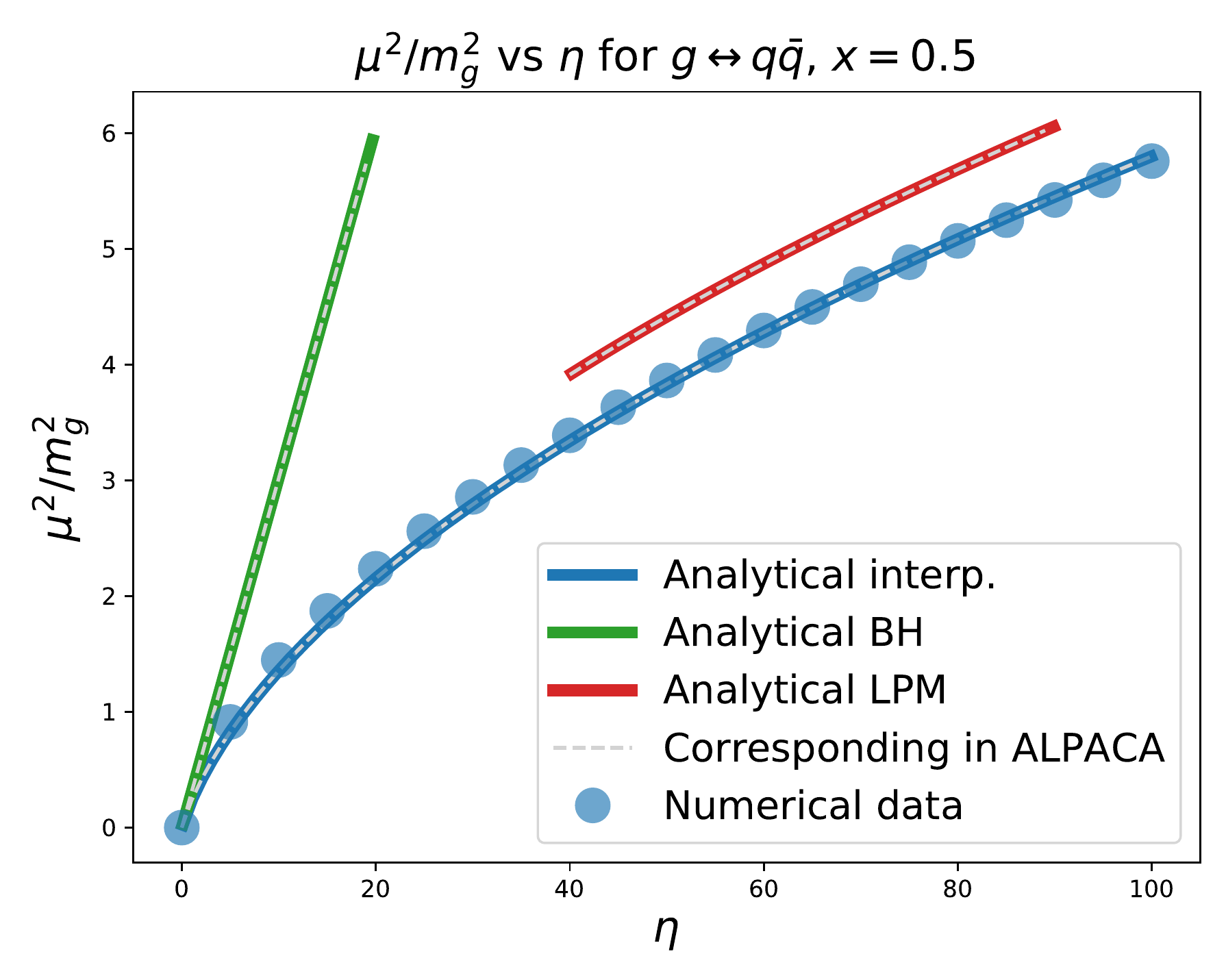}
    \caption{Top: Scale $\mu^2(\eta, x; A, A, A)$ entering the splitting rate of gluons for QCD. The dots represent numerical evaluation of $\mu^2$. At small values of $\eta$ the scale is well described by the Bethe-Heitler approximation (green line) whereas at large values the curve approaches (logarithmically) the deep-LPM approximation of~\cite{Arnold:2008zu} (purple line). The interpolation function constructed from the two limits describes the numerical data adequately at all values of $\eta$ and $x$ for $N_f = 3$ QCD. Black lines are extracted from \textsc{Alpaca}, while other lines are analytical values, with an almost complete overlap between the two. Bottom row: same as top for $\mu^2(\eta, x; F, A, F)$ (left) and $\mu^2(\eta, x; A, F, F)$ (right).}
    \label{fig:mu2}
\end{figure*}

For values of $\eta$ and $|\hat M $| for which neither of the limits gives a good approximation, the equation can be solved numerically using
We find that the numerical result is well approximated by the following interpolating function, which is indeed what we use in the simulation:
\begin{equation}
    \mu_{\mathrm{interp}}^2 = \left[\sqrt{\eta+1} - 1 \right]\left\{ \frac{2}{1+\eta} \frac{\mu_{\mathrm{BH}}^2}{\eta} + \frac{\eta}{1+\eta}\frac{\mu_{\mathrm{LPM}}^2}{\sqrt{\eta}} \right\}.
\end{equation}
\Fig{fig:mu2} shows the numerical values of $\mu$ along with the two limits and the interpolating function for the representative cases of $\mu^2(\eta, x = 0.5; A, A, A)$, $\mu^2(\eta, x = 0.5; F, A, F)$ and $\mu^2(\eta, x = 0.5; A, F, F)$, which corresponds to all different splitting processes at $x = 0.5$. The interpolating function gives a good approximation of the numerical solution.

\section{Cross sections and rates}
\label{appendix:formulae}

\subsection{Elastic scattering matrix elements}
\label{appendix:formulae_elasticmatrixelements}

Purely in terms of Mandelstam variables the total cross section becomes 
\begin{equation}
    \sigma^{ab}_{2\rightarrow 2} = \frac{\pi}{s^2} \sum_{cd}\int dt |\mathcal{M}^{ab}_{cd}|^2[1\pm f(\pp)][1\pm f(\kp)]
\end{equation}
and we overestimate this as

\begin{equation}
    \tilde{\sigma}^{ab}_{2\rightarrow 2} = 4 \frac{\pi}{s^2} \sum_{cd}\int dt |\mathcal{M}^{ab}_{cd}|^2.
\end{equation}

Below the different matrix elements used, their overestimates and integrated expressions are presented. The prefactors $\zeta_g = e^{5/3}/4$ and $\zeta_q = e^2/4$ are discussed in \Sect{subsection:alpaca_elastic_general}. It is assumed that $\nu_{\bar{q}} = \nu_q$. Note that the equations for $q\bar{q}\to gg$ are not symmetric, to get the corresponding expressions for $gg\to q\bar{q}$ one has to exchange the degeneracy factors in front as $\nu_q^2\to\nu_g^2$.

\begin{align}
     \frac{\nu_q^2}{g^4} & |\mathcal{M}^{q_iq_i}_{q_iq_i}|^2  = 8\frac{d_F^2C_F^2}{d_A}\left(\frac{s^2+u^2}{(t-\zg m_g^2)^2} + \frac{s^2+t^2}{(u-\zg m_g^2)^2} \right) \nonumber \\
     & + 16d_FC_F\left(C_f-\frac{C_A}{2}\right)\frac{s^2}{(t-\zg \mg)(u-\zg \mg)} \nonumber\\
     & \leq 32\left(\frac{s^2}{(t-\zg \mg)^2} + \frac{s^2}{(-s-t-\zg \mg)^2} \right),
\end{align}
        
\begin{align}
    &\frac{\nu_q^2}{g^4}\int |\mathcal{M}^{q_iq_i}_{q_iq_i}|^2 dt =  \frac{16}{3(s+\zg \mg)^2(\zg \mg-t)} \nonumber \\
    & \times \Bigg[ 6(s+\zg \mg)\log\left(\frac{\zg \mg-t}{s+t +\zg \mg}\right) \nonumber \\
    & + \frac{2s^2}{s+2\zg \mg}\log\left(\frac{\zg \mg-t}{s+t+\zg \mg} \right) + 6(t-\zg \mg) \nonumber\\
    & + 3(\zeta_g^2 m_g^4+2\zg \mg s + 2s^2) \left( \frac{1}{\zg \mg-t} - \frac{1}{s+t+\zg \mg} \right) \Bigg],
\end{align}

\begin{align}
     \frac{\nu_q^2}{g^4} |\mathcal{M}^{q_i\bar{q}_i}_{q_i\bar{q}_i}|^2 & = 8\frac{d_F^2C_F^2}{d_A}\left(\frac{s^2+u^2}{(t-\zg m_g^2)^2} + \frac{t^2+u^2}{(s+\zg m_g^2)^2} \right) \nonumber \\
     & + 16d_FC_F\left(C_f-\frac{C_A}{2}\right)\frac{u^2}{(s+\zg \mg)(t-\zg \mg)} \nonumber\\
     & \leq 32s^2\left[\frac{1}{(t-\zg m_g^2)^2} + \frac{4\zg \mg+s}{3\zg \mg(s+\zg \mg)^2} \right],
\end{align}

\begin{align}
     \frac{\nu_q^2}{g^4}\int & |\mathcal{M}^{q_i\bar{q}_i}_{q_i\bar{q}_i}|^2 dt =  \frac{16}{3(\zg \mg+s)^2(\zg \mg-t)} \nonumber \\
    & \times \Big[ -2 \zeta_g^4 m_g^8 + 3\zeta_g^3 m_g^6(s+2t) \nonumber\\
    & +\zeta_g^2 m_g^4(15s^2 + 9st -2t^2) \nonumber \\
    & +\zg \mg(18s^3+8s^2t +2st^2+3t^3) \nonumber\\
    & +4(\zg \mg+s)^3(\zg \mg-t)\log(\zg \mg-t) \nonumber \\
    & - 2(-3s^4+s^2t^2+st^3+t^4) \Big],
\end{align}

\begin{align}
    \frac{\nu_q^2}{g^4}|\mathcal{M}^{q_i\bar{q}_i}_{q_j\bar{q}_j}|^2 & = 8\frac{d_F^2C_F^2}{d_A}\left(\frac{t^2+u^2}{(s+\zg \mg)^2}\right) \nonumber\\
    & \leq 32\frac{s^2}{(s+\zg \mg)^2},
\end{align}

\begin{align}
    \frac{\nu_q^2}{g^4}\int |\mathcal{M}^{q_i\bar{q}_i}_{q_j\bar{q}_j}|^2 dt = &  \frac{16t(3s^2+3st+2t^2)}{3(s+\zg \mg)^2},
\end{align}

\begin{align}
    \frac{\nu_q^2}{g^4}|\mathcal{M}^{q\bar{q}}_{gg}|^2 & = 8d_FC_F^2\left(\frac{u^2+t^2}{(t-\zq \mf)(u-\zq \mf)} \right) \nonumber \\
    & - 8d_FC_FC_A\left(\frac{t^2+u^2}{(s+\zg \mg)^2} \right) \nonumber\\
    & \leq \frac{256}{3}\frac{s^2}{(t-\zq \mf)(u-\zq \mf)},
\end{align}

\begin{align}
    & \frac{\nu_q^2}{g^4}\int |\mathcal{M}^{q\bar{q}}_{gg}|^2 dt = \frac{32}{3}\Bigg[ \frac{4((s+\zq \mf)^2 + \zeta_q^2 m_q^4)}{2\zq \mf+s} \nonumber\\
    & \times \log\left(\frac{s+t+\zq \mf}{\zq \mf-t} \right)  \nonumber \\
    & -\frac{t( 8(s+\zg \mg)^2+9s^2+9ts+6t^2)}{(s+\zg \mg)^2}\Bigg],
\end{align}

\begin{align}
    \frac{\nu_q^2}{g^4}|\mathcal{M}^{q_iq_j}_{q_iq_j}|^2 & = 8\frac{d_F^2C_F^2}{d_A}\left(\frac{s^2+u^2}{(t-\zg m_g^2)^2} \right) \nonumber\\
    & \leq 32\frac{s^2}{(t-\zg m_g^2)^2},
\end{align}

\begin{align}
    \frac{\nu_q^2}{g^4}\int |\mathcal{M}^{q_iq_j}_{q_iq_j}|^2 dt = & 16\Bigg[ \frac{\zeta_g^2 m_g^4 + 2\zg \mg s + 2s^2}{\zg \mg-t}  \nonumber \\
    & +  2(s+\zg \mg)\log(\zg \mg-t) + t \Bigg],
\end{align}
    
\begin{align}
    \frac{\nu_q\nu_g}{g^4}&|\mathcal{M}^{gq}_{gq}|^2  = -8d_FC_F^2\left(\frac{s^2+u^2}{(s+\zq \mf)(u-\zq \mf)} \right) \nonumber \\
    & + 8d_FC_FC_A\left(\frac{s^2+u^2}{(t-\zg m_g^2)^2} \right) \nonumber\\
    & \leq \frac{256}{3}\frac{s^2}{\zq\mf(s+\zq \mf)} + 192\frac{s^2}{(t-\zg m_g^2)^2},
\end{align}

\begin{align}
    &\frac{\nu_q\nu_g}{g^4}\int |\mathcal{M}^{gq}_{gq}|^2 dt =  \frac{32}{3}\Bigg[ \frac{9(\zeta_g^2 m_g^4+2\zg \mg s +2s^2)}{\zg \mg-t} \nonumber \\
    & +18(s+\zg \mg)\log(\zg \mg - t) +\frac{t(5\zq \mf+13s)}{s+\zq \mf} \nonumber \\
    & +\frac{2t^2+4(s^2+\zeta_q^2 m_q^4)\log(s+t+\zq \mf)}{s+\zq \mf} \Bigg],
\end{align}
         
\begin{align}
    \frac{\nu_g^2}{g^4}|\mathcal{M}^{gg}_{gg}|^2 & = 16d_AC_A^2\Bigg(3-\frac{su}{(t-\zg m_g^2)^2} \nonumber \\
    & -\frac{st}{(u-\zg m_g^2)^2}-\frac{tu}{(s+\zg \mg)^2} \Bigg) \nonumber\\
    & \leq 1152\Bigg(3+\frac{s^2}{(t-\zg m_g^2)^2} \nonumber \\
    & -\frac{st}{(-s-t-\zg m_g^2)^2}\Bigg),
\end{align}

\begin{align}
    & \frac{\nu_g^2}{g^4}\int |\mathcal{M}^{gg}_{gg}|^2 dt =  1152\Bigg[ s\ln\left(\frac{t-\zg \mg}{\zg \mg+s+t} \right) \nonumber \\
    & + 3t +\frac{2t^3+3st^2}{6(s+\zg \mg)^2} \nonumber \\ 
    & +s(\zg \mg+s)\left(\frac{1}{\zg \mg-t} - \frac{1}{\zg \mg+s+t}\right) \Bigg].
\end{align}

\subsection{Numerical solution of $C[f_{\mathrm{thermal}}]$}
\label{appendix:numerical_solution_of_C}
    In order to verify the elastic scattering rates of \textsc{Alpaca} in the thermal case with a non-constant cross section, we have numerically integrated the collision kernels given in AMY and compared the results to the scattering rates of \textsc{Alpaca} through the relation found in \Eq{eq:total_scattering_rate}. We have looked separately at $C[f]^+$ and $C[f]^-$, which corresponds to the gain and loss terms in the collision kernel. The notation below for the coefficients is as follows, $2_i$ is due to the particles in the initial state being indistinguishable, $2_d$ is due to double-counting of the incoming particles, $2_b$ is due to quark and anti-quark both contributing and $3_f$ (or $2_f$) is due to the three flavours currently in \textsc{Alpaca}. Hence, for particles scattering out of $\p$ we have for gluons that the loss is
    
    \begin{align}
        \label{eq:appendix_I_g_loss}
         I_g^- & = \int\frac{d^3p}{2(2\pi)^3} C_g^{-,2\leftrightarrow2}[f,\p] \nonumber \\
        & =  \frac{1}{2\cdot2_{i,d}\cdot\nu_g} \int_{\p, \k,\pp, \kp}(2\pi)^4 \nonumber \\
        & \delta^{(4)}(P+K-P'-K') \Big[ |\mathcal{M}^{gg\leftrightarrow gg}|^2 \nonumber \\
        & \times f_g(\p)f_g(\k)(1+f_g(\pp))(1+f_g(\kp)) \nonumber \\
        & \quad  +2_d\cdot 2_b\cdot3_f\cdot|\mathcal{M}^{gq\leftrightarrow gq}|^2\nonumber \\
        & \times f_g(\p)f_q(\k)(1+f_g(\pp))(1-f_q(\kp)) \nonumber \\
        & \quad  +2_d\cdot3_f\cdot|\mathcal{M}^{gg\leftrightarrow q\bar{q}}|^2\nonumber \\
        & \times f_g(\p)f_g(\k)(1-f_q(\pp))(1-f_{\bar{q}}(\kp))\Big] \nonumber \\ 
        & = I_g^{-, gg\leftrightarrow gg} + I_g^{-, gq\leftrightarrow gq} +I_g^{-, gg\leftrightarrow q\bar{q}}
    \end{align}
    and the gain is
    
    \begin{align}
        \label{eq:appendix_I_g_gain}
        I_g^+ & = \int\frac{d^3p}{2(2\pi)^3} C_g^{+,2\leftrightarrow2}[f,\p] \nonumber \\
        & =  \frac{1}{2\cdot2_{i,d}\cdot\nu_g} \int_{\p, \k,\pp, \kp}(2\pi)^4  \nonumber \\
        &   \times \delta^{(4)}(P+K-P'-K') \Big[ |\mathcal{M}^{gg\leftrightarrow gg}|^2 \nonumber \\
        & \times f_g(\pp)f_g(\kp)(1+f_g(\p))(1+f_g(\k)) \nonumber \\
        &      +2_d\cdot 2_b\cdot3_f\cdot|\mathcal{M}^{gq\leftrightarrow gq}|^2 \nonumber \\
        & \times f_g(\pp)f_q(\kp)(1+f_g(\p))(1-f_q(\k)) \nonumber \\
        &      + 2_d\cdot3_f\cdot|\mathcal{M}^{q\bar{q}\leftrightarrow gg}|^2\nonumber \\
        & \times f_q(\pp)f_{\bar{q}}(\kp)(1+f_g(\p))(1+f_g(\k))\Big].
    \end{align}
    The factor of $2_b$ in front of $gq\leftrightarrow gq$ is due to that full row being equivalent to $g\bar{q}\leftrightarrow g\bar{q}$, assuming $f_q=f_{\bar{q}}=f_g$. For a quark of species $s$ we have that the loss is
    
    \begin{align}
        \label{eq:appendix_I_q_loss}
        I_q^- & = \int\frac{d^3p}{2(2\pi)^3} C_{q,s}^{-,2\leftrightarrow2}[f,\p] \nonumber \\
        & =  \frac{1}{2\cdot2_{i,d}\cdot\nu_g} \int_{\p, \k,\pp, \kp}(2\pi)^4 \delta^{(4)}(P+K-P'-K') \nonumber \\
        & \times \quad \Big[ \big\{2_d\cdot 2_b\cdot2_f\cdot|\mathcal{M}^{q_iq_j\leftrightarrow q_iq_j}|^2+|\mathcal{M}^{q_iq_i\leftrightarrow q_iq_i}|^2 \nonumber \\
        & \quad\quad +2_d\cdot|\mathcal{M}^{q_i\bar{q}_i\leftrightarrow q_i\bar{q}_i}|^2 + 2_d\cdot 2_f\cdot|\mathcal{M}^{q_i\bar{q}_i\leftrightarrow q_j\bar{q}_j}|^2 \big\} \nonumber \\
        & \quad       \times f_q(\p)f_q(\k)(1-f_q(\pp))(1-f_q(\kp)) \nonumber \\
        & \quad       +2_d\cdot|\mathcal{M}^{qg\leftrightarrow qg}|^2f_q(\p)f_g(\k) \nonumber \\
        & \times (1-f_q(\pp))(1+f_g(\kp)) \nonumber \\
        & \quad       +  |\mathcal{M}^{q\bar{q}\leftrightarrow gg}|^2f_q(\p)f_{\bar{q}}(\k) \nonumber \\
        & \times (1+f_g(\pp))(1+f_g(\kp))\Big] \nonumber \\
        & = I_q^{-, q_iq_j\leftrightarrow q_iq_j} + I_q^{-, q_iq_i\leftrightarrow q_iq_i} + I_q^{-, q_i\bar{q}_i\leftrightarrow q_i\bar{q}_i}  \nonumber \\
        & \quad + I_q^{-, q_i\bar{q}_i\leftrightarrow q_j\bar{q}_j} + I_q^{-, qg\leftrightarrow qg} + I_q^{-, q\bar{q}\leftrightarrow gg}
    \end{align}
    and the gain is
    
    \begin{align}
        \label{eq:appendix_I_q_gain}
        I_q^+ & = \int\frac{d^3p}{2(2\pi)^3} C_{q,s}^{+,2\leftrightarrow2}[f,\p] \nonumber \\
         & =  \frac{1}{2\cdot2_{i,d}\cdot\nu_g} \int_{\p, \k,\pp, \kp}(2\pi)^4 \delta^{(4)}(P+K-P'-K')  \nonumber \\
        & \times \Big[ \big\{2_d\cdot 2_b \cdot 2_f\cdot|\mathcal{M}^{q_iq_j\leftrightarrow q_iq_j}|^2+|\mathcal{M}^{q_iq_i\leftrightarrow q_iq_i}|^2 \nonumber \\
        & \quad +2_d\cdot|\mathcal{M}^{q_i\bar{q}_i\leftrightarrow q_i\bar{q}_i}|^2 + 2_d\cdot 2_f\cdot |\mathcal{M}^{q_i\bar{q}_i\leftrightarrow q_j\bar{q}_j}|^2 \big\} \nonumber \\
        &        \times f_q(\pp)f_q(\kp)(1-f_q(\p))(1-f_q(\k)) \nonumber \\
        &        +2_d\cdot|\mathcal{M}^{qg\leftrightarrow qg}|^2f_q(\pp)f_g(\kp)(1-f_q(\p))(1+f_g(\k)) \nonumber \\
        &        + |\mathcal{M}^{gg\leftrightarrow q\bar{q}}|^2f_g(\pp)f_g(\kp)(1-f_q(\p))(1-f_{\bar{q}}(\k))\Big]
    \end{align}
    where the factor $2_b$ on the first matrix element is due to the fact that $q_iq_j\leftrightarrow q_iq_j$ is equivalent to $q_i\bar{q}_j \leftrightarrow q_i\bar{q}_j$. Lastly, $C_q^{-,2\leftrightarrow2}[f,\p] = C_{\bar{q}}^{-,2\leftrightarrow2}[f,\p]$ and $C_q^{+,2\leftrightarrow2}[f,\p] = C_{\bar{q}}^{+,2\leftrightarrow2}[f,\p]$.

    The integrals above have been solved numerically, separately from \textsc{Alpaca}, using Monte Carlo methods with stratified sampling, and compared to the elastic scattering rate in \textsc{Alpaca} through the relation shown in \Eq{eq:total_scattering_rate}. The results of the full integration are shown in \Sect{subsection:alpaca_elastic_thermal}.

\subsection{Splitting probability overestimate}
\label{appendix:splitting_probability_overestimate}
To find an overestimate $\tilde{h}_a(x)\geq h_a(x)$, with $h_a(x)$ as defined in \Eq{eq:splitting_h}, we note the inequality relation

\begin{align}
    h_a(x) & \leq \frac{(2\pi)^3}{2|\p|\nu_a}  \sum_{b,c} \gamma^a_{bc}(\p;xp\phat,(1-x)p\phat)\hat{f}_{bc} \nonumber \\
    & = \frac{(2\pi)^3}{2|\p|\nu_a} \sum_{b,c} P^a_{bc} \mu \hat{f}_{bc} \leq \frac{(2\pi)^3}{2|\p|\nu_a} \sum_{b,c} P^a_{bc} \mu^2_{\mathrm{LPM}} \hat{f}_{bc}
\end{align}
where $P^a_{bc}$, $\mu^2$ and $\mu^2_{\mathrm{LPM}}$ are defined in \app{appendix:gamma}, and $\hat{f}_{bc} \geq [1\pm f_b][1\pm f_c]$ is an overestimate of the Bose/Pauli factors. It also holds that

\begin{align}
    \mu^2_{\text{LPM}}  & \leq m_g^2\sqrt{\frac{\lambda T_* p}{2\pi m_g^2}x(1-x)} \nonumber \\
    & \quad \times \Bigg[C_{s_2s_3}^{s_1}\left(\frac{\xi\mu^2_{\text{LPM}}}{m_g^2}\right) + C_{s_3s_1}^{s_2}x^2\left(\frac{\xi\mu^2_{\text{LPM}}}{x^2m_g^2}\right) \nonumber \\
    & \quad + C_{s_1s_2}^{s_3}(1-x)^2\left(\frac{\xi\mu^2_{\text{LPM}}}{(1-x)^2m_g^2}\right)  \Bigg]^{\frac{1}{2}} \nonumber \\
    & = \mu_{\text{LPM}}\sqrt{\frac{\lambda T_* p}{2\pi}x(1-x)} \nonumber \\
    & \quad \times \left[\xi(C_{s_2s_3}^{s_1} + C_{s_3s_1}^{s_2} + C_{s_1s_2}^{s_3}) \right]^{\frac{1}{2}}
\end{align}
and so

\begin{align}
    \mu^2_{\text{LPM}} & \leq \frac{\lambda T_* p \xi x(1-x) }{2\pi}\left(C_{s_2s_3}^{s_1} + C_{s_3s_1}^{s_2} + C_{s_1s_2}^{s_3}\right) \nonumber \\
    & = \tilde{\mu}^2_{\text{LPM}}.
\end{align}
Note also that for all splitting channels it holds that $P^a_{bc}(x)\leq P^a_{bc}(x_{\mathrm{min}})$, with the condition that $x_{\text{max}}=1-x_{\text{min}}$. Combining all of the inequalities presented above we end up with the three following overestimates,

\begin{align}
    h^{g\rightarrow gg}(x) & \leq \tilde{h}^{g\rightarrow gg} \nonumber \\
    & = \frac{(2\pi)^3}{2|\p|\nu_g} P^g_{gg}(x_\mathrm{min}) \tilde{\mu}^2_{\text{LPM}} \hat{f}_{gg} \nonumber \\
    & = \frac{\sqrt{2}d_AC_A\alpha_s}{2(2\pi)^2\nu_g}\left[\frac{1+x_{\text{min}}^4+(1-x_{\text{min}})^4}{x_{\text{min}}(1-x_{\text{min}})}\right] \nonumber \\
    & \times \lambda T_* \xi\left(C_{s_2s_3}^{s_1} + C_{s_3s_1}^{s_2} + C_{s_1s_2}^{s_3}\right)\hat{f}_{gg},
\end{align}

\begin{align}
    h^{q\rightarrow gq} (x) & \leq \tilde{h}^{q\rightarrow gq} \nonumber \\
    & = \frac{(2\pi)^3}{2|\p|\nu_q} P^q_{gq}(x_\mathrm{min}) \tilde{\mu}^2_{\text{LPM}} \hat{f}_{gq} \nonumber \\
    & = \frac{\sqrt{2}d_FC_F\alpha_s}{2(2\pi)^2\nu_q}\left[\frac{1+(1-x_{\text{min}})^2}{x_{\text{min}}}\right] \nonumber \\
    & \times \lambda T_* \xi\left(C_{s_2s_3}^{s_1} + C_{s_3s_1}^{s_2} + C_{s_1s_2}^{s_3}\right)\hat{f}_{gq},
\end{align}

\begin{align}
    h^{g\rightarrow q\bar{q}}(x) & \leq  \tilde{h}^{g\rightarrow q\bar{q}} \nonumber \\
    & = \frac{(2\pi)^3}{2|\p|\nu_q} P^g_{q\bar{q}}(x_\mathrm{min}) \tilde{\mu}^2_{\text{LPM}} \hat{f}_{q\bar{q}} \nonumber \\
    & = \frac{\sqrt{2}d_FC_F\alpha_s}{2(2\pi)^2\nu_q}\left[x_{\text{min}}^2+(1-x_{\text{min}})^2\right]  \nonumber \\
    & \times  \lambda T_* \xi \left(C_{s_2s_3}^{s_1} + C_{s_3s_1}^{s_2} + C_{s_1s_2}^{s_3}\right) \hat{f}_{q\bar{q}}.
\end{align}

\subsection{Relation between $|\mathcal{M}^c_{ab}|^2$ and $\gamma^c_{ab}$}
\label{section:relation_between_gamma_M}

To find the relation between the effective (i.e. including multiple soft scattering during the formation time) merging matrix element $|\mathcal{M}^c_{ab}|^2$ and the  merging rate $\gamma^c_{ab}$ we start by looking at the term in the collision kernel which corresponds to a particle with momentum $\p$ merging with a particle with momentum $\k$ to produce a particle with momentum $\pp$,

\begin{align}
    \label{eq:C2to1}
    C_a^{2\leftrightarrow 1} =&  \frac{1}{2|\p|\nu_a}\sum_{b,c}\int_{\k,\pp}\nu_a\nu_b|\mathcal{M}^c_{ab}(\pp;\p,\k)|^2(2\pi)^4 \nonumber \\
    & \times \delta^{(4)}(P-P'-K') f_a(\p)f_b(\k)[1\pm f_c(\pp)].
\end{align}
Using the integral identity
\begin{equation}
    \int_{\k,\pp}(2\pi)^4\delta^{(4)}(P+K-P')  = \int_0^\infty dk \int_0^\infty dp' \frac{\delta(p+k-p')}{8\pi|\p|}
\end{equation}
with an on-shell condition for $\pp$ implicitly in the last row, the collision kernel simplifies to

\begin{align}
    &C_a^{2\leftrightarrow 1} = \frac{(2\pi)^3}{|\p|^2\nu_a}\sum_{b,c}\int_0^\infty dk dp'\delta(p + k-p') \nonumber \\
    & \times\left[ \frac{\nu_a \nu_b}{8(2\pi)^4}|\mathcal{M}^c_{ab}(\pp;\p,\k)|^2 \right]f_a(\p)f_b(\k)[1\pm f_c(\pp)].
\end{align}
Comparing to the corresponding collinear expression given in~\cite{Arnold:2002zm},

\begin{align}
    \label{eq:C2to1_collinear}
    C_a^{2\leftrightarrow 1} & =  \frac{(2\pi)^3}{|\p|^2\nu_a}\sum_{b,c}\int_0^\infty dkdp'\delta(p+k-p') \nonumber \\
    & \times\Big[ \gamma^c_{ab} (p'\phat;\p,k\phat)\Big] f_a(\p)f_b(k\phat)[1\pm f_c(p'\phat)]
\end{align}
we see that $\gamma^c_{ab}$ is related to $\mathcal{M}^c_{ab}$ through
\begin{equation}
    \gamma_{ab}^c = \frac{\nu_a \nu_b}{8(2\pi)^4}|\mathcal{M}_{ab}^c|^2_.
\end{equation}




\end{document}